\documentclass[12pt]{article}

\usepackage{setspace,graphicx,epstopdf,amsmath,amsfonts,amssymb,amsthm}
\usepackage{graphicx,verbatim,array,multicol}
\usepackage{amssymb}
\usepackage{amsfonts}
\usepackage{amsmath}
\usepackage{epsfig}
\usepackage{eurosym}
\usepackage{latexsym}
\usepackage{amssymb}
\usepackage{color}
\usepackage{amscd}
\usepackage{multirow}
\usepackage{graphicx}
\usepackage{lscape}
\usepackage{bm}
\usepackage{url}
\usepackage{graphics}
\usepackage{placeins}
\usepackage{amsthm}
\usepackage{verbatim}
\usepackage{setspace} 
\usepackage{tabularx}
\usepackage{natbib}

\newtheorem{definition}{Definition}
\newcommand\independent{\protect\mathpalette{\protect\independenT}{\perp}}
\def\independenT#1#2{\mathrel{\rlap{$#1#2$}\mkern2mu{#1#2}}}

\begin{document}

\title{Designating market maker behaviour in Limit Order Book markets}

 
 \author{ Efstathios Panayi$\dag$ $\ast$ \thanks{The support of the Economic and Social Research Council (ESRC) in funding the Systemic Risk Centre is gratefully acknowledged [grant number ES/K002309/1]. \newline Corresponding author. Email: efstathios.panayi.10@ucl.uk}\and Gareth W. Peters$\ddag$ $\star$ $\ast$ \and Jon Danielsson$\ast$ \and Jean-Pierre Zigrand$\ast$\\
 {\small{
$\dag$ Department of Computer Science, University College London}}\\
{\small{
$\ddag$ Department of Statistical Science, University College London}} \\
{\small{
$\star$ Associate Fellow, Oxford Mann Institute, Oxford University}}\\
{\small{
$\ast$ Systemic Risk Center, London School of Economics and Political Science}}\\
}

\date{\today}
\maketitle


\vspace*{1in}

\centerline{\bf Abstract}
\medskip

Financial exchanges provide incentives for limit order book (LOB) liquidity provision to certain market participants, termed designated market makers or designated sponsors. While quoting requirements typically enforce the activity of these participants for a certain portion of the day, we argue that liquidity demand throughout the trading day is far from uniformly distributed, and thus this liquidity provision may not be calibrated to the demand. We propose that quoting obligations also include requirements about the speed of liquidity replenishment, and we recommend use of the Threshold Exceedance Duration (TED) for this purpose. We present a comprehensive regression modelling approach using GLM and GAMLSS models to relate the TED to the state of the LOB and identify the regression structures that are best suited to modelling the TED. Such an approach can be used by exchanges to set target levels of liquidity replenishment for designated market makers.

\noindent \textbf{Keywords:} Limit Order Book, liquidity, resilience, GLM, GAMLSS

\noindent \textbf{JEL codes:} C41, C52, D47

\newpage 

\section{Introduction}

Financial exchanges have different modes of operation, or market models, for different assets. This is often determined by an asset's liquidity in the prevailing period, and an exchange will endeavour to choose a market model that facilitates trading in the asset. As an example, the electronic trading system Xetra, operated by Deutsche B{\"o}rse, offers continuous trading for the most liquid assets, and the same mode of operation is offered for the second most liquid category of assets, but supplemented with a `Designated Sponsor', who has market-making obligations. Other securities, such as structured products, feature a single market maker, while the less liquid assets are traded instead in `continuous auction' mode, which features a specialist. 

The classification of assets in most electronic exchanges is performed according to their liquidity which is averaged over a particular period of time (typically quarterly). For assets which feature a Designated Sponsor (termed a Designated Market Maker in other exchanges), there are requirements regarding the maximum spread, minimum quote size and the effective trading time. In return for fulfilling their quoting obligations, Designated Sponsors receive a full reimbursement of the transaction fees incurred. 

In this paper, we argue that in order to ensure high-frequency liquidity provision, exchanges need to consider not only the average liquidity over time, but also the time required for liquidity to be replenished, which we will explain and quantify as an indication of liquidity resilience. This is because large orders are increasingly being partitioned by execution algorithms into multiple smaller tranches, and traders take advantage of liquidity replenishment to improve execution\footnote{\citep{chlistalla2011high} notes that the average order size is one-eighth of that of fifteen years ago, in terms of number of shares, and one-third in dollar value.}. 

Such replenishment is swift when market liquidity is resilient, and the effect of resilience on e.g. optimal execution has been considered in the past in the models of \citet{obizhaeva2012optimal} and \citet{alfonsi2010optimal}. However, these models generally considered resilience to be constant or have a very simple parametric form. Thus, they failed to attribute the resilience characteristics to interpretable features of the limit order book structure.

The model of \citet{panayi2014market} instead introduced a new notion of resilience explicitly measuring the time for liquidity to return to a previously-defined threshold level. This approach was agnostic to the particular class of liquidity measure considered, and could therefore accommodate volume-based, price-based and cost-of-return-trip-based measures. They showed that resilience was not constant, but was instead related to the state of the LOB. This allowed them to understand the effect of different LOB structural explanatory variables on the resilience metric constructed, and as part of this, they considered a regression based specification. In particular, they considered simple log-linear regression structures to relate the response (the duration of liquidity droughts) to instantaneous and lagged limit order book structural regressors intra-daily. 

Using Level 2 LOB data from the multi-lateral trading facility Chi-X, we have access to the state variables considered by \citet{panayi2014market}, and can therefore consider this notion of resilience further in the study undertaken in this manuscript. We significantly extend their resilience modelling framework to allow for additional structural features, as well as a greater class of distribution model types to better explain and capture the liquidity resilience features of a range of assets intra-daily. In particular, we consider two classes of regression models which allow for more general resilience model dynamics to be captured and more flexible distributional features to be explored, ultimately improving the fit and forecast performance of the models. Firstly, we have Generalised Linear Models, or GLMs, which typically assume a conditionally specified exponential family of distributions for observation assumptions for the response, in our case the exceedance times over a liquidity threshold. The second class is that of Generalised Additive Models for Location, Shape and Scale, or GAMLSS, which relax this assumption and can consider a wider, general distribution family with the limit order book regressors entering not just into the location/mean relationship through a link function, but also into the shape and scale parameters directly. This informs the skewness and kurtosis of the liquidity profiles and the resilience of the liquidity in settings of liquidity leptokurticity and platokurticity. 

It is critical to develop these new modelling approaches, as they provide a directly interpretable modelling framework to inform exchanges and market making participants of the influence different structural features of the LOB for a given asset will have on affecting instantaneously within a trading day the local liquidity resilience. They thus provide insights into how best to manage and design market making activities to improve resilience in markets. Our results reveal that considering the more flexible Generalised Gamma distribution assumption within a GAMLSS framework, with multiple link functions to relate the LOB covariates to the different distribution parameters, improves the explanatory performance of the model. On the other hand, the simpler Lognormal specification also achieves respectable explanatory power and its estimation is more robust.

We also statistically assess the significance of the explanatory variables in greater detail, and across datasets for companies from 2 different countries. We find that, in agreement with empirically observed market features, a larger deviation of the liquidity from a given resilience threshold level would be associated with a longer deviation from that level of liquidity (liquidity drought). On the other hand, a larger frequency of such deviations from a liquidity threshold level would be associated with swifter returns to that level (shorter duration liquidity droughts). Using the proposed liquidity resilience modelling framework we can also determine the regimes under which we are likely to see different structural features in the resilience behaviour.

Our results indicate that resilience considerations should also be a factor when deciding the quoting requirements for exchange-designated liquidity providers, such as the aforementioned Designated Sponsors. That is, along with the requirements for maximum spread and minimum volumes, they should be subject to additional requirements for liquidity replenishment, ensuring that throughout the trading day, the LOB returns swiftly back to normal levels. As we have shown that liquidity resilience is dependent on the state of the LOB, exchanges can use the modelling approaches we have proposed, in order to determine the appropriate level of liquidity replenishment requirements, given prevailing market conditions. In addition, liquidity providers may use the model to determine the best response to a liquidity drought.

The remainder of this paper is organised as follows: In Section \ref{sec:incentives} we discuss incentives for liquidity provision in the limit order book and other market structures. Section \ref{sec:background} introduces existing concepts of liquidity resilience, as well as the TED metric analysed in this paper. Section \ref{sec:regmodels} outlines the regression model structures of increasing complexity employed in our analysis of liquidity resilience. Section \ref{sec:data} describes the data used in this study and section \ref{sec:results} presents the results in terms of importance of individual covariates for explaining resilience, and the explanatory power of the models with different regression structures and different distributional assumptions for the response. Section \ref{sec:conc} concludes with proposals about altering current incentive schemes for liquidity provision.

\section{Ensuring uninterrupted liquidity provision via exchange incentives}
\label{sec:incentives}

In many modern financial markets and across different asset classes, a large part of liquidity provision originates from high-frequency traders. Indicatively, for the equities market, a typical estimate of at least about 50\% of total volume is contributed by such market participants, see details in the report by the \cite{securities2010concept}. However, these firms have no legal obligations to provide continuous access to liquidity, and may (and indeed do\footnote{\cite{kirilenko2014flash} note that during the 2010 `Flash crash', the activity of high frequency traders accounted for a much lower share of overall activity, compared to the preceeding days.}) reduce their activity in times of distress. For this reason, and in order to also ensure access to liquidity for younger, smaller cap, or more volatile stocks, exchanges provide incentives to firms to facilitate liquidity provision. These market making obligations have been found to improve liquidity for these assets, and by extension, also improve year-on-year returns \citep{venkataraman2007value,menkveld2013designated}. \cite{benos2012role} summarise the impact of introducing designated market makers into a stock market.    

Both the incentive structure and the obligations differ across exchanges, and in particular, they may be applicable only for certain market structures. For example, in London Stock Exchange's hybrid SETS market, Designated Market Makers must maintain an executable quote for at least 90\% of the trading day, as well as participate in the closing auctions, and they are also subject to maximum spread and quote size requirements, which vary across stocks. In return, they incur no trading fees, and are allowed to ask for the suspension of trading of an asset when prices are volatile \citep{benos2012role}.  

As an example of specific exchange considerations for classifying assets and incentivizing liquidity provision we present details for the German electronic trading system Xetra, originally developed for the Frankfurt stock exchange. Xetra offers a number of different trading models adapted to the needs of its various trading groups, as well as the different assets classes. The models differ according to\footnote{Xetra trading models, accessed 25/05/2015, available at \url{http://www.xetra.com/xetra-en/trading/trading-models}}:
\begin{itemize}
\item Market type (e.g. number of trading parties);
\item The transparency level of available information pre- and post-trade;
\item The criteria of the order prioritisation;
\item Price determination rules;
\item The form of order execution.
\end{itemize}

For equity trading, the following trading models are supported\footnote{Xetra Market Model Equities Release, accessed 25/05/2015, available at  \url{http://www.deutsche-boerse-cash-market.com/blob/1193332/8b79d504d5aaf80be8853817a6152ecd/data/Xetra-Market-Model-Equities-Release-15.0.pdf}}:

\begin{itemize}
\item Continuous trading in connection with auctions (e.g. opening and closing auction, and possibly, one or more intra-day auctions);
\item Mini-auction in connection with auctions;
\item One or more auctions at pre-defined points in time. 
\end{itemize}

We will focus on the first model, which is the market model that reflects the activity considered here, i.e. in the context of the LOB. For many of the most well-known assets (such as those in the main indices), there is sufficient daily trading interest, such that one should be able to execute their orders without much delay and without causing a significant price shift (although Xetra also offers a price improvement service, termed Xetra BEST\footnote{Xetra continuous trading with best executor, available at \url{http://www.xetra.com/xetra-en/trading/handelsservices/continuous-trading-with-best-executor}}). However, there are also less frequently traded assets, for which Xetra tries to ensure uninterrupted liquidity provision, by offering incentives to trading members to provide quotes throughout the trading day.  

Xetra defines liquid equities as those in which the Xetra Liquidity Measure (XLM)\footnote{The Xetra Liquidity Measure is a Cost-of-Round-Trip measure, quantifying the cost to buy and immediately sell an amount 25000 EUR of an asset} does not exceed 100 basis points and daily order book turnover is higher that or equal EUR 2.5m on a daily average in the preceeding four month period\footnote{Designated sponsor guide, accessed 25/05/2015, available at \url{http://www.deutsche-boerse-cash-market.com/blob/1193330/215d37772fbec9fbc39391cbc7c5821c/data/Designated-Sponsor-Guide.pdf}}. Assets for which this is not true require liquidity provision from trading members, termed `Designated Sponsors' for continuous trading to be offered, otherwise the assets are traded under a continuous auction model, with a specialist. 

Designated Sponsors have to adhere to strict quoting obligations, which are verified daily. In return for meeting them, they have the transaction fees they generate fully reimbursed. These quoting requirements depend on the liquidity of the asset in the preceeding 3 month period. Table \ref{tab:quoting} shows both the basis of determination of an asset's liquidity class, as well as the quoting requirements for Designated Sponsors for each of these classes.

\begin{table}[h]
\begin{tabular}{llll}
\hline
\multicolumn{4}{c}{\textbf{Liquidity class determination}}                                                                                     \\ \hline
\textbf{Liquidity class}          & \multicolumn{1}{c}{LC1}                     & \multicolumn{1}{c}{LC2}                     & \multicolumn{1}{c}{LC3}                     \\\textbf{XLM}                      & $\leq$ 100 basis points            & $\leq$ 500 basis points            & $\leq$ 500 basis points            \\ 
                                  &                                             &                                             &                                            \\ \hline
\multicolumn{4}{c}{\textbf{General quoting requirements}}                                                                                   \\ \hline
       & \multicolumn{1}{c}{LC1}                     & \multicolumn{1}{c}{LC2}                     & \multicolumn{1}{c}{LC3}                     \\\textbf{Minimum quote size} & \euro20,000                             & \euro15,000                             & \euro10,000                             \\
\textbf{Maximum spread}           &                                    &                                    &                                    \\
$\geq$EUR 8.00                   & 2.5 \%                             & 4 \%                               & 5\%                                \\
 $<$EUR 8.00                      & $\min \left\{\text{\euro} 0.20;10.00\% \right\}$ & $\min \left\{\text{\euro} 0.32;10.00\% \right\}$ & $\min \left\{\text{\euro} 0.40;10.00\% \right\}$ \\
 $<$EUR 1.00                      & \euro 0.10                                & \euro 0.10                                & \euro 0.10                                \\
                                  &                                             &                                             &                                            \\  \hline
\multicolumn{4}{c}{\textbf{Minimum requirements in continuous trading: Quotation duration}}                                                    \\ \hline
\multicolumn{4}{c}{90\%}                                                                                                                        
\end{tabular}
\caption{Quoting requirements for Designated Sponsors on the 3 liquidity classes of Xetra (reproduced from the Xetra Designated Sponsor guide), accessed 28/05/2015. }
\label{tab:quoting}
\end{table}

\subsection{Limitations of current incentive schemes}

Designated Sponsors can select the time within the trading day for which they wish to be active, as long as it exceeds 90\% of the day on average. We argue that a more useful quoting requirement would also reflect the intra-day trading patterns, i.e. considering also the variation in trading activity throughout the trading day. If the 10\% of the day for which the Designated Sponsor is not active corresponds to a significant proportion of daily activity (e.g. close to the beginning and end of the trading day), then 90\% activity in calendar time does not correspond to 90\% in participation over the day. 

\begin{figure}[ht!]
	\begin{center}
	\includegraphics[width=0.49\textwidth]{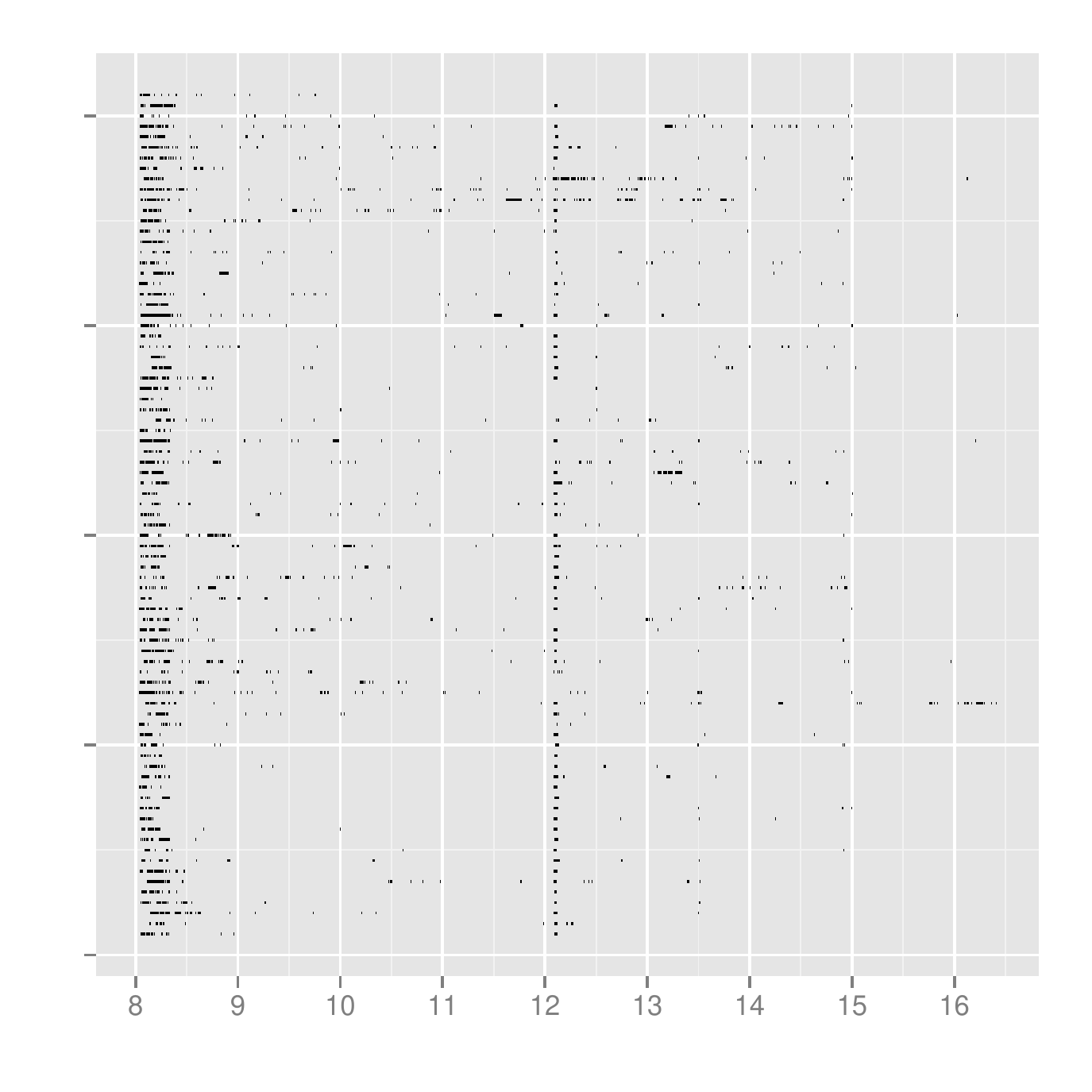}
	\includegraphics[width=0.49\textwidth]{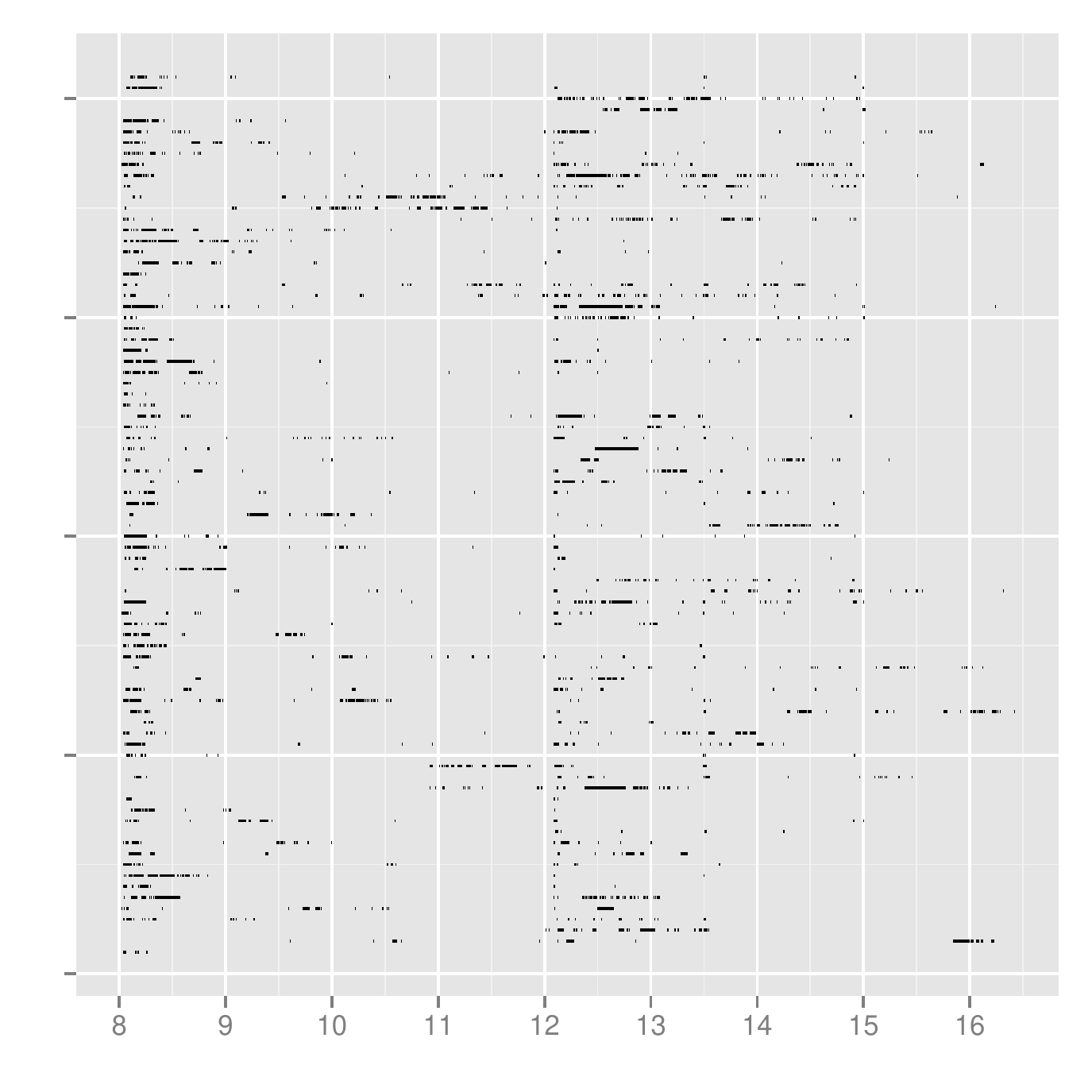}
	\caption{Every line corresponds to a trading day, and the shaded regions represents periods for which the spread (left) and XLM (right) is in the top quintile for the day for stock Sky Deutschland. }
	\label{fig:intradayTED}
	\end{center}
\end{figure}

Indeed, an empirical analysis of intra-day liquidity behaviour shows that liquidity demand throughout the trading day is far from uniformly distributed, and thus the quoting requirements above may not have the desired effect. Figure \ref{fig:intradayTED} shows the proportion of the trading day for which the spread for Sky Deutschland on the Chi-X exchange was in the top quintile for the day. \cite{panayi2014market} explained that the presence of larger spreads at the start of the trading day can be explained by the uncertainty of market makers about what the fair price should be for the asset, while the second concentration can be explained by the release of certain economic announcements. 

This variation in intra-day liquidity demand is prevalent in the equities class across different markets and different industries. However, as we have seen in this section, it is not currently considered when determining the liquidity class of an asset, and for an extended period of time near the start of the trading day or around important economic announcements, we may have extended periods of low liquidity.  We therefore argue that both the determination of the liquidity class and the quoting requirements for Designated Market Makers could be adjusted, in order to reflect both the absolute level of liquidit as well as the speed of order replenishment after a shock (e.g. a large market order). Specifically, given the remit of Designated Sponsors, one would expect that they should not only provide 2-way markets in periods in which there is no trading interest, but also swiftly replenish liquidity in more illiquid periods, such as the ones indicated above. 

To determine whether this is the case, we propose the use of resilience in market liquidity, as measured through the Threshold Exceedance Duration (TED) \citep{panayi2014market}, which we introduce in the next section. We acknowledge that enforcing swift replenishment of limit orders may lead to an increase in adverse selection costs for liquidity providing operators. We therefore also present a comprehensive modelling approach to identify the most informative determinants of resilience, and thus aid market participants in understanding how their behaviour can affect resilience of liquidity in the limit order book.     

\section{Concepts of market liquidity resilience}
\label{sec:background}

\subsection{Liquidity resilience introduction}
\label{sec:liqres}

The seminal paper of \citet{kyle1985continuous} acknowledged the difficulty in capturing the liquidity of a financial market in a single metric, and identified tightness, depth and resiliency as three main properties that characterize the liquidity of a limit order book. Tightness and depth have been mainstays of the financial literature (and indeed, are easily captured through common liquidity measures such as the spread and depth, respectively) and there has been substantial literature in studying the intra-day variation\footnote{\citet{chan1995market} found a declining intra-day spread for NASDAQ securities (an L-shaped pattern), while \citet{wood1985investigation} and \citet{abhyankar1997bid} found a U-shape pattern (with larger spreads at the beginning and at the end of the day), for the NYSE and LSE respectively. \citet{brockman1999analysis} found an inverted U-shaped pattern for the depth, which mirrors the U-shaped spread pattern (in that the peak of the depth and the trough of the spread both correspond to higher levels of liquidity).} and commonality\footnote{There is a rich literature studying the cross-sectional commonality in liquidity in the equity markets through the principal components of individual asset liquidity, starting with the work of \cite{hasbrouck2001common} and extended by \cite{korajczyk2008pricing} and \cite{karolyi2012understanding}. More recent work by \cite{panayi2015liquidity}, however, have identified weaknesses in the PCA and PCA regression approaches for quantifying commonality and suggests caution when heavy-tailed features are present in liquidity. } in these measures. However, resilience has received decidedly less attention. 

\cite{panayi2014market} provided a review of the state of the art in liquidity resilience and noted that the extant definitions seemed to be divided into two categories: In the first, definitions provided by  \citet{kyle1985continuous} and \citet{obizhaeva2012optimal} were related to price evolution, and specifically to the return of prices to a steady state. The second category of definitions, proposed by \citet{garbade1982securities} and \citet{harris2002trading} was concerned with liquidity replenishment.

\begin{figure}[ht!]
	\begin{center}
	\includegraphics[width=0.9\textwidth]{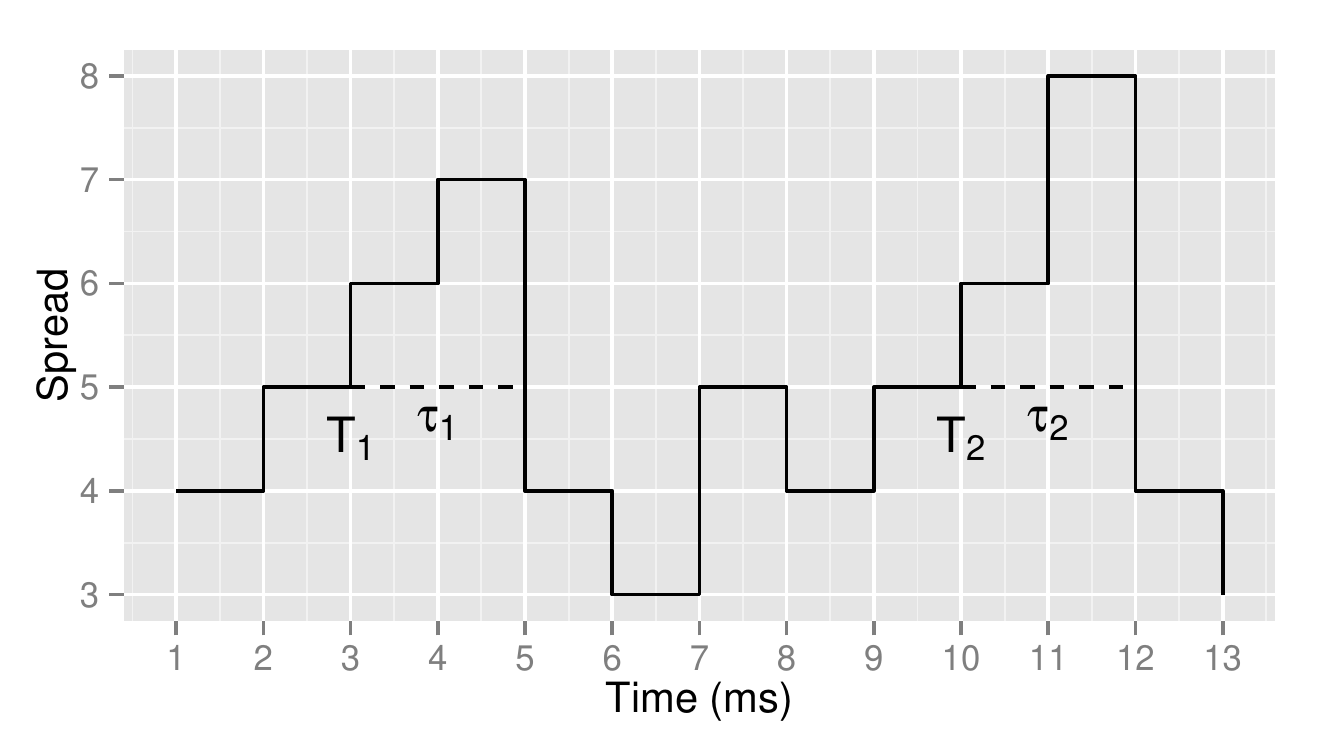}
	\caption{An example of the duration of the exceedance over a spread threshold. The spread threshold is $c$ is 5 cents, $T_i$ denotes the $i$-th time instant that the spread exceeds the threshold, and $\tau_i$ is the duration of that exceedance.}
	\label{spreadDeviation}
	\end{center}
\end{figure}

The liquidity resilience notion introduced in \citep{panayi2014market} is a member of the latter category, and was the first to explicitly define resilience in terms of any of the possible liquidity measures and in terms of a liquidity threshold at which resilience is measured against. Hence, the concept of resilience of the liquidity measure was converted to a notion of relative resilience for a given operating liquidity threshold that a user may specify. This was important, since as discussed in \cite{lipton2013trade}, there are several different market participants in modern electronic exchanges and their liquidity demands and requirements differ depending on their mode of operation. In particular, this will mean that they would likely care about relative liquidity resilience characteristics at different liquidity thresholds, which may also depend on the type of liquidity measure being considered. All such characteristics are then easily accommodated by the framework developed in \cite{panayi2014market}, where the central concept is that liquidity is considered `replenished' by the market or market maker when a (user-specified) liquidity measure returns to a (again, user-specified) threshold. In a financial market where liquidity is resilient, one would expect that the time required for this liquidity replenishment would be low. This replenishment time was captured by \cite{panayi2014market} through the idea of the threshold exceedance duration (TED):

\begin{definition}
The threshold exceedance duration (TED) is the length of time between the point at which a liquidity measure deviates from a threshold liquidity level (in the direction of less liquidity), and the point at which it returns to at least that level again.

Formally, we have
\begin{equation}
\label{eq:tedspread}
\tau_i=\inf\left\{\tau~:~L_{T_i+\tau} \leq c, \; T_i+\tau > T_{i} \right\}.
\end{equation}
\end{definition}
where $L_t$ denotes the level of liquidity at time $t$ and $T_i$ is the $i$-th time in the trading day where liquidity deviates from the threshold level $c$. This notation is explained in detail in Section \ref{sec:notation}.

This definition was designed to intentionally allow the flexibility for it to be utilised for any measure of liquidity of choice, be it price based, volume based or some combination. It also allowed for the use of different threshold liquidity levels, and the setting of a very low liquidity threshold (e.g. a high level of the spread), which meant that one could model the duration of low liquidity regimes, which would be of interest in a regulation setting. In the high-frequency liquidity provision setting we are considering in this paper, an exchange would be interested in modelling the time for return to a `normal' liquidity level, which one could consider to be the median intra-day liquidity level. 

\subsection{Regression structures for the modelling of liquidity resilience}

In this paper, we employ a number of regression structures in order to model the TED liquidity resilience metric. Such regression structures provide an interpretable conditional dependence specification between liquidity resilience, for a desired liquidity measure, given structural observed attributes or features of a given asset's limit order book, or other important market based intra-day trading volume/price/activity indicators. In addition, structural features such as known reporting times and announcement features can also be incorporated into the model explicitly to see their influence on core aspects of the resilience in the liquidity of the asset. 


Although regression models based on simple linear structures linking the mean of a response to a linear functional of the covariates have been in widespread use for over 200 years, it is only relatively recently that such structures have begun to be significantly generalized. Innovations in the class of parameteric regression relationships available have included the incorporation of non-linear structures, random effects, functional covariates and relationships between not just the mean (location of the response variable) and the linear model, but also direct relationships between covariates and variance/covariance, skewness, kurtosis, shape, scale and other structure features of a number of distributions for the response variable.  

The beginning of this revolution in regression modelling was heralded by the highly influential family of Generalized Linear Models (GLMs), see for instance \citet{nelder1972generalized}, which was intended to unify the extant regression approaches of the time. Up to this point, only parametric mean regression models were considered, i.e. only the mean of the distribution was related to the explanatory variables. Shortly after, the variance of the distribution was also modelled as a function of explanatory variables, in the case of normal models \citep{harvey1976estimating}. Parametric variance regression models for different assumptions of the response (e.g. for the exponential family) followed, see further discussion regarding the adaptation to different contexts in \citet{rigby2013discussion}.

GAMLSS, introduced by \citet{rigby2005generalized}, enabled the modelling of a response variable from a general parametric distribution. The explanatory variables of the model are then related to each distribution parameter through a link function, which can have both linear and non-linear components. One can explicitly see if the distribution of resilience in liquidity under a given market regime, on a given day, is likely to affect the mean, variance, skewness or kurtosis of the liquidity resilience. More importantly, which LOB characteristics are most likely to be influential in affecting these attributes of the resilience. By identifying these one can then devise market making strategies to improve resilience in a given market regime.

It is clear, therefore, that such a formulation is very general, encompassing previous approaches such as GLMs. Indeed, the supporting R package used in this paper \citep{stasinopoulos2007generalized} enables the modelling of more than 80 parametric distributions. We can see, therefore, that incorporating GAMLSS into our approach would result in a highly desirable and flexible framework for modelling TED duration data, and the next section will detail precisely how one can define different link function to relate the distribution parameters of the TED response to the LOB explanatory variables.

\section{Hierarchy of regression models for liquidity resilience in the LOB}
\label{sec:regmodels}

The framework we employ here aims to explain the variation in the TED random variables as a function of independent explanatory covariates obtained from the state of the LOB at the point of exceedance. We start first with a very brief description of log-linear regression structures, in order to introduce concepts before moving on to Generalised Linear Models (GLM) and Generalised Additive Models for Location Shape and Scale (GAMLSS). 

\subsection{Log-linear regression structures}

For the TED random variables, since we are modelling positive random variables, we can consider a log-linear formulation, incorporating model covariates as follows
\begin{equation}
\label{eq:ln}
\ln(\tau_i)=\bm{x}_i^{T}\boldsymbol{\beta}+\varepsilon_i,
\end{equation}
where $\varepsilon_i \sim \mathcal{N}(0,\sigma)$ is a random error term. Under the assumption that $\tau_i | \bm{x}_i \independent \tau_j |\bm{x}_j$ we can relate the expected log response to the covariates

\begin{align*}
\mathbb{E} \left[ \ln(\tau_i) | \bm{x}_i\right]=\bm{x_i}^{T}\boldsymbol{\beta}.
\end{align*}

One can see that a unit change in, say, $x_i^{(k)}$ to $x_i^{(k)} + 1$ will have a multiplicative effect of $e^{\beta_k}$ on the response $\tau_i$, i.e. the Threshold Exceedance Duration. The sign of the coefficient for a given covariate indicates the direction of the partial effect of this variable.   

\subsection{GLM}

While convenient, the log-linear regression structure is also restrictive in the model's expressive power and flexibility, and we may therefore opt to consider instead a GLM. The construction of such a parametric regression model requires the specification of three key components:

\begin{enumerate}
\item A distribution for the response random variables. In this context, this is the conditional distribution of the TED, given the covariates constructed from the LOB structure. 
\item The conditional mean structure of the TED, which links the linear regression model comprised of the independent observed explanatory covariates to the response (typically through a transformation known as the link function). 
\item A specification of the variance function, perhaps also as a function of the mean. 
\end{enumerate}

GLMs enable us to fit models when the response variable belongs to the exponential family of distributions, i.e.
\begin{equation}
\label{eq:exp}
f(\tau|\theta,\phi)=\exp \left( \frac{\tau \theta - b(\theta)}{a(\phi)}+c(\tau,\phi)\right)
\end{equation}
where $\theta$ is the location parameter and $\phi$ the scale parameter and where $a(\phi), b(\theta), c(\tau,\phi)$ are known functions defining particular subfamilies.

A GLM relates the expected response $\mu = \mathbb{E} \left[ \tau \right]$ to a linear predictor $\bm{x}^T \bm{\beta}$ through a link function $g(\cdot)$, i.e. $g(\mu)=\bm{x}'_t \bm{\beta}$. When this link function is the identify function, it is equivalent to a standard linear model. If $g(\mu)=\theta$ then the function $g(\cdot)$ is called the \textit{canonical link} and we have $\theta=\bm{x}'_t \bm{\beta}$. 

Let us consider two members of the exponential family, which are widely used for analysing duration data, for the distributions of our response. The Gamma distribution
\begin{equation}
f(\tau|\alpha,\beta)=\frac{\beta^{\alpha} \tau^{\alpha -1}}{\Gamma(\alpha)}\exp \left( -\beta \tau \right), \tau \geq 0, \alpha \geq 0, \beta \geq 0,
\end{equation} 
which we can see is in the exponential family with $\theta=\frac{\beta}{\alpha}$, $\phi= \frac{1}{\alpha}$,  $a(\phi)=-\frac{1}{\alpha}$ and $b(\theta)=\log(\theta)$. We use the reparameterisation 
\begin{equation}
f(\tau|\mu,\sigma)=\frac{1}{\left(\sigma^2 \mu \right)^{1/ \sigma^2}} 
\frac{\tau ^{\frac{1}{\sigma^2} -1} \exp \left(-\tau/(\sigma^2 \mu) \right) }{\Gamma(1/\sigma^2)},\tau \geq 0, \mu \geq 0, \sigma \geq 0,
\end{equation} 
so that $\mathbb{E} [\tau]=\mu$. Then $\sigma^2=\frac{1}{\alpha}$ and $\mu=\frac{\alpha}{\beta}$ and we see that the linear predictor $\bm{x}^T \bm{\beta}$ would be related to both parameters. However, if the shape parameter $\alpha$ is fixed, only the scale parameter $\beta$ varies with the linear predictor.    

We also consider a second member of the exponential family, namely the Weibull distribution, given by
\begin{equation}
f(\tau|\mu,\sigma)=\frac{\sigma}{\beta} \left( \frac{\tau}{\beta} \right)^{\sigma-1} \exp \left\{-\left( \frac{\tau}{\beta}\right)^{\sigma} \right\},\tau \geq 0, \sigma \geq 0, \beta \geq 0,
\end{equation} 
where $\beta=\mu/\Gamma(\frac{1}{\sigma}+1)$. We can see that this is also a member of the exponential family (for fixed $\sigma$), with $a(\phi)=\phi=1$, $\theta=\frac{-1}{\mu^\sigma}$, and $b(\theta)=\sigma\ln \mu$. As in the case of the gamma distribution, if the shape parameter is fixed, only the scale parameter varies with the linear predictor.

In order to ensure that $\mu$ is positive, in both cases we use the log link, i.e. $g(\mu)=\log(\mu)$. For the exponential family of distributions one can obtain the conditional expectation of the response as

\begin{equation}
\mathbb{E}\left[\tau | \bm{x} \right]=b'(\theta)=\mu
\end{equation}
and the variance as 
\begin{equation}
Var\left[\tau | \bm{x} \right]=a(\phi) b''(\theta)=a(\phi) V(\mu)
\end{equation}
see the derivation in \cite{mccullagh1989generalized}. It is clear then that the formulation of the exponential model above allows one to model cases when the response variables are of the same distributional form and they are independent, but not identically distributed, in that they may have different mean and variance. In this case, the variance varies as a function of the mean. 

\cite{de2008generalized} suggests that there are many functions $V(\mu)$ that cannot arise from an exponential family distribution. In addition, one may also want to consider a distribution which is more flexible in terms of skewness and kurtosis. In this case, we propose the flexible generalised gamma distribution class of models given by
\begin{equation} \label{EqnGGD}
f_{\tau}(\tau; b,a,k) = \frac{b}{\Gamma(k)}\frac{\tau^{b k - 1}}{a^{b k}}\exp\left(-\left(\frac{\tau}{a}\right)^{b} \right), k > 0, a >0, b > 0
\end{equation}
which was first introduced by \cite{stacy1962generalization} and considered further in a reparameterised form by \cite{lawless1980inference}. This distribution has the additional advantage that it has a closed form expression for the quantile function, which we will see in Section \ref{sec:quantile}. This means we can also explicitly study the relationship between quantiles of the TED and certain LOB covariates, i.e. also interpreting the resulting regression as a quantile regression \citep{noufaily2013parametric}. As the generalised gamma distribution is not a member of the exponential family of distributions, it cannot be modelled using a GLM. We will thus show in the next subsection how one can model this using the GAMLSS framework.

\subsection{GAMLSS}

GAMLSS requires a parametric distribution assumption for the response variable, but this can be of the general distribution family, rather than only in the exponential distribution family, as in the case of the GLM. It differs from the GLM additionally, in that where the former assume that only the expected response is related to the predictor through a link function, GAMLSS has separate link functions relating each of the distribution parameters to the explanatory variables. As such, it enables one to capture features such as overdispersion or positive and negative skew in the response data. We will base the formulation presented on that of \cite{rigby2005generalized}.

Let $\bm{\tau}'=\left( \tau_1,\ldots,\tau_n \right)$ be the vector of the response and let us assume a density $f(\tau_i|\bm{\theta}_i)$ where $\bm{\theta}_i=\left( \theta_{1,i},\theta_{2,i},\theta_{3,i},\theta_{4,i} \right)=\left( \mu_i,\sigma_i, \kappa_i, \nu_i \right)$, where $\mu_i,\sigma_i, \kappa_i, \nu_i$ are the distribution parameters, with the two first relating to location and scale and others typically relating to shape. Let also $\bm{X}_k$ be a fixed known design matrix containing the covariates at the point of exceedance $T$ for each observed TED random variable. In this setting, we can define the following link functions to relate the $k$-th distribution parameter $\theta_k$ to the vectors of explanatory variables $\bm{X}_k$ and $\left\{\bm{Z}_{jk}\right\}_{j=1}^{J_k}$

\begin{align}
g_k(\bm{\theta}_k)= \bm{X}_k \bm{\beta}_k + \sum_{j=1}^{J_k} \bm{Z}_{jk} \gamma_{jk} 
\end{align} 
where $\left\{\bm{Z}_{jk}\right\}_{j=1}^{J_k}$ are optional components for the incorporation of random effects.  

GAMLSS are referred to as semi-parametric regression type models, as they assume a parametric distribution for the response, and they allow for non-parametric smoothing functions in the response. To see this, let $\bm{Z}_{jk}=\bm{I}_n$, the $n \times n$ identity matrix, and let $\bm{\gamma}_{jk} = \bm{h}_{jk}=h_{jk}(\bm{x}_{jk})$. Then the semi-parametric form can be be obtained according to 

\begin{equation}
g_k(\bm{\theta}_k)= \bm{X}_k \bm{\beta}_k + \sum_{j=1}^{J_k}h_{jk}(\bm{x}_{jk}).
\end{equation} 

Hence, under the GAMLSS framework, the parameters of the distribution can be modelled using linear functions as well as flexible smoothing functions (e.g. cubic splines) of the explanatory variables, in addition to random effects. A parametric linear model can be recovered when linear functions of explanatory variables are considered, i.e. in the case of the identity link function
  \begin{equation}
g_k(\bm{\theta}_k)= \bm{X}_k \bm{\beta}_k.
\end{equation} 

Using this specification, we can consider very flexible multiple-parameter distributions, such as the Generalised Gamma distribution (hereafter g.g.d.). We assume that the TED random variables are conditionally independent, given the LOB covariates, i.e. $\tau \stackrel{i.i.d}{\sim} F\left(\tau;k,a,b\right)$, with the density given in Equation \ref{EqnGGD}. The g.g.d. family includes as sub-families several popular parametric models: the exponential model $(b=k=1)$, the Weibull distribution (with $k=1$), the Gamma distribution (with $b=1$) and the Lognormal model as a limiting case (as $k \rightarrow \infty$).

We now wish to relate this statistical model assumption to a set of explanatory variables (covariates) from lagged values of the LOB. In practice, to achieve this, one could work on the log scale with $\ln(\tau)$, i.e. with the log-generalized gamma distribution, as this parameterisation is known to improve identifiability and estimation of parameters. Discussions on this point are provided in significant detail in \cite{lawless1980inference}. Instead, as we employ the {\em gamlss} R package of \cite{stasinopoulos2007generalized} for estimation, we have the following reparameterisation

\begin{equation}
f'_{\tau}(\tau;\mu,\sigma,\nu)=\frac{|\nu| \theta^{\theta} \left( \frac{\tau}{\mu}\right)^{\nu \theta} \exp \left( -\theta \left( \frac{\tau}{\mu}\right)^{\nu } \right) } {\Gamma(\theta) \tau}
\end{equation}  
where $\theta=\frac{1}{\sigma^2 \nu^2}$. This corresponds to the parameterisation in Equation \ref{EqnGGD} under the transformation 
\begin{equation}
f'_{\tau}(\tau;\mu,\sigma,\nu) \equiv f_{\tau}(\tau;\nu,\mu \theta^{-\frac{1}{\nu}},\theta).
\end{equation}

The regression structure we adopt for the g.g.d. model involves a log link for the time-varying coefficient $\mu(\bm{x}_t)$:  
\begin{equation} \label{eq:EqnLink}
\ln \left(\mu(\bm{x}_t)\right) = \beta_0 + \sum_{s=1}^px_t^{(s)}\beta_s.
\end{equation}
with $p$ covariates $\bm{x}_t=\left\{x^{(s)}_t\right\}_{s=1}^p$ measured instantaneously at the point of exceedance $t=T_i$, and the link functions for parameters $\sigma$ and $\nu$ can be found in Table \ref{tab:linkfn}. Each of the covariates is a transform from the LOB for which the liquidity measure is observed, and all covariates are described in Section \ref{sec:covariates}. We note that we also considered models with interactions between the covariates, but interaction terms were not found to be significant in the majority of our models. 

Under a model with this regression structure, we observe that the conditional mean of the duration is also related directly to this linear structure where for the $i$-th exceedance of the threshold, we have

\begin{equation}\label{eq:CEF}
\mathbb{E}\left[\tau_i|\bm{x}_{T_i}\right] =  \exp \left( \beta_0 + \sum_{s=1}^p \bm{x}_{T_i}^{(s)}\beta_s \right) 
\left(\frac{1}{k}\right)^{\frac{1}{b}}\frac{\Gamma\left(k + \frac{1}{b}\right)}{\Gamma\left(k\right)}
\end{equation}
see details in \cite{lo2002econometric}. 

\subsection{Notation and definitions of LOB variables and regression model components}
\label{sec:notation}
We define a level of the LOB as a price level at which there is at least 1 resting order: The $1$-st level of the bid side is then the 1st level below the price midpoint at which there is a resting buy order.       
In addition, we utilise the following notation for a single asset, on a single trading day. 
\begin{itemize}
\item $a$ denotes the ask, $b$ denotes the bid. 

\item $P_{t}^{b,i}\in\mathbb{N}^{+}$ denotes the random variable for the limit price of the $i^{th}$ level bid at time $t$ in tick units

\item $P_{t}^{a,i}\in\mathbb{N}^{+}$ denotes the random variable for the limit price of the $i^{th}$ level ask at time $t$ in tick units

\item $\bm{V}_{t}^{b,i}\in\mathbb{N}^{n}$ denotes a column random vector of orders at the $i^{th}$ level bid at time $t$, with $n$ being the number of such orders  

\item $L_t$ denotes a random variable at time $t$ for the generic proxy for the liquidity measure. 

\item $c$ for denotes the exceedance threshold level, defined relative to the liquidity measure $L_t$. $c$ is deterministic and constant over time.

\item $T_i$ denotes the $i$-th random time instant in a trading day that the liquidity measure $L_{t}$ exceeds the threshold $c$. Formally, we define $T_{i} = \inf\left\{t: L_{t} > c, \; t > T_{i-1} , \; t>T_0  \right\}$, where $T_0$ denotes the start of the observation window (1 minute after the start of the trading day). 

\item $\tau_i$ will denote the duration of time in ms, relative to the exceedance event $T_i$, that the liquidity measure $L_{t}$ remains above the threshold $c$. These are the response random variables which correspond to the TED. 

%

\end{itemize}

\subsubsection{Model LOB Covariates}
\label{sec:covariates}
For each TED random variable $\tau_i$ we consider the corresponding contemporaneous covariates in our regression design matrix, i.e. at the times of exceedance above the specified liquidity threshold, $t=T_i$. In the following, a `level' of the LOB is defined as one in which there is at least 1 resting limit order. Thus the first 5 levels of the bid are the 5 levels closest to the quote mid-point, where there is available volume for trading.  
The covariates chosen pertain to the state of the limit-order book of one given stock.

\begin{itemize}
\item{The total number of asks in the first 5 levels of the LOB at time $t$, obtained according to $x_t^{(1)}=\sum_{i=1}^{5}\left |V_{t}^{a,i}\right |$} (where $\left | \cdot  \right |$ is the number of orders at a particular level), and is denoted $ask$ hereafter

\item{The total number of bids in the first 5 levels of the LOB at time $t$, obtained according to $x_t^{(2)}=\sum_{i=1}^{5}\left |V_{t}^{b,i}\right |$}, denoted $bid$

\item{The total ask volume in the first 5 levels of the LOB at time $t$, obtained according to $x_t^{(3)}=\sum_{i=1}^{5}TV_{t}^{a,i}$}, denoted $askVolume$

\item{The total bid volume in the first 5 levels of the LOB at time $t$, obtained according to $x_t^{(4)}=\sum_{i=1}^{5}TV_{t}^{b,i}$}, denoted $bidVolume$

\item{The number of bids $x_t^{(5)}$ in the LOB that had received price or size revisions (and were thus cancelled and resubmitted with the same order ID), denoted by $bidModified$}. 

\item{The number of asks $x_t^{(6)}$ in the LOB that had received price or size revisions, denoted by $askModified$}. 

\item{The average age (in ms) $x_t^{(7)}$ of bids in the first 5 levels at time $t$, denoted by $bidAge$.}
		
\item{The average age $x_t^{(8)}$ of asks in the first 5 levels at time $t$, denoted by $askAge$.}

\item The instantaneous value of the spread at the point at which the $i$-th exceedance occurs, which is given by $x_t^{(9)}=P_{t}^{a,1}-P_{t}^{b,1}$ and denoted as $spreads$.

\item{For the nine previously defined covariates, we also include exponentially weighted lagged versions. For example, in the case of the $x_t^{(s)}$ covariate, the respective lagged covariate value is then given by:
\begin{equation}
z_t^{(s)}=\sum_{n=1}^d w^n x^{(s)}_{t - n\Delta},
\end{equation}
where for a time $t$, we consider $w=0.75$ is the weighting factor, $d=5$ is the number of lagged values we consider and $\Delta=1s$ is the interval between the lagged values. These covariates are hereafter denoted with the `l' prefix.}

\item The number $x_t^{(10)}$ of previous TED observations in the interval $[t-\delta,t]$,  with $\delta=1s$, denoted by $prevexceed$. 

\item The time since the last exceedance, $x_t^{(11)}$, denoted by $prevexceed$.

\item The average of the last 5 TEDs, $x_t^{(12)}$, denoted by $prevTEDavg$.

\item The activity in the associated CAC40 index (in number of order additions, cancellations and executions) in the previous second, $x_t^{(13)}$, denoted by $indact$. 

\item A dummy variable indicating if the exceedance occurred as a result of a market order to buy, $x_t^{(14)}$, denoted by $mobuy$.

\item A dummy variable indicating if the exceedance occurred as a result of a market order to sell, $x_t^{(15)}$, denoted by $mosell$.

\end{itemize}

Altogether we then have 24 variates, 15 instantaneous and 9 lagged.

\section{Data description}
\label{sec:data}

We use an 82 day trading sample (January 2nd to April 27, 2012) of all order submissions, executions and cancellations in the limit order book for 20 German and French stocks traded on Chi-X\footnote{Chi-X was a pan-European multilateral trading facility (MTF) which merged with BATS in 2012. For the period under consideration, it accounted for between a quarter and a third of total trading activity in the French and German stocks considered, for more details see  \url{http://www.liquidmetrix.com/LiquidMetrix/Battlemap}.}. Information for these assets is provided in Table \ref{tab:stockinfo} and they were to chosen include both small and large cap stocks in a number of different industries. We note that while the trading hours for Chi-X are 08:00 to 16:30, we do not consider activity before 08:01 and after 16:29, in order to avoid market opening and closing effects.

There is a degree of flexibility regarding the manner in which one selects the threshold levels over which to consider exceedances. \citep{panayi2014market} suggest that these may be specified based on an interest in particular liquidity resilience scenarios to be studied, in other cases they can be specified based on historical observations of the empirical distribution of the selected proxy for a liquidity measure. The implications of each choice are discussed in \citep{panayi2014market}. For this paper we consider exceedances over the median threshold level, which could be of interest in the market making setting described in this paper, as well as exceedances over the daily 95\% threshold level.

\section{Results and Discussion}
\label{sec:results}

In the following model selection procedure we assess both the appropriateness of the various distributional assumptions one might make for the TED response, as well as the importance of the different covariates in explaining the variation in the TED, in the interest of obtaining a parsimonious model. \cite{panayi2014market} explained that the assumption of stationarity in liquidity resilience over an extended period is not supported by the data, and for this reason we also fit the model individually for each day and each asset, where daily local stationarity is reasonably assumed. We will first identify the covariates that are most frequently found to be significant in daily regressions for different assets and for most of the period under consideration, in order to obtain a parsimonious covariate subset. We will then proceed by comparing the explanatory power of the regression model for lognormal, Weibull, gamma and Generalised gamma distributional assumptions for the response random variable.  

\subsection{Model selection - covariate significance}
\label{sec:modselection}

\begin{figure}[ht!]
	\begin{center}
\includegraphics[width=0.99\textwidth]{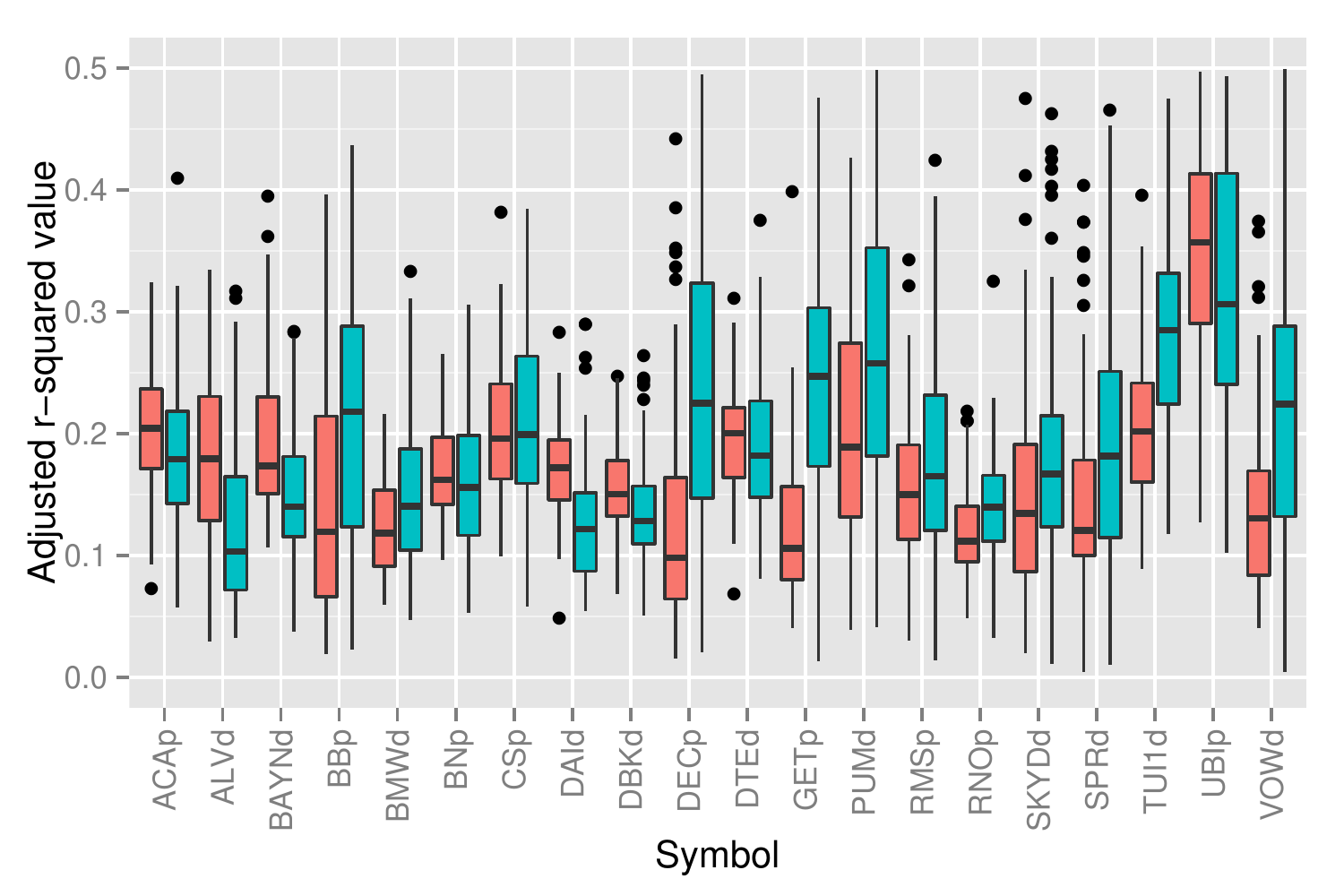}
	\caption{Boxplots of the adjusted $R^2$ value obtained from fitting the full regression model separately for each day in our dataset for both the threshold corresponding to the 5th decile of the spread (the median - red) and the 95th percentile threshold (blue).}
	\label{fig:adjr2}
	\end{center}
\end{figure}

For the empirical evaluation of the importance of the various LOB covariates in the regression, we selected a lognormal model specification. This is because the simple linear formulation of this model enables us to use existing model selection techniques, in order to identify the model structure that produces the highest explanatory power for each daily regression for every asset. We evaluate the explanatory power of this model in terms of the proportion of the variation in the TED resilience measure that can be explained by the selected model covariates, as captured by the adjusted coefficient of determination (adjusted $R^2$). 

We first fit a multiple regression model, in which all covariates explained in Section \ref{sec:covariates} are considered in each daily model for the entire 4 months of our dataset, for each asset under consideration. We also considered interactions between covariates, but these were not found to be significant in the vast majority of cases. Figure \ref{fig:adjr2} shows the adjusted $R^2$ values obtained from fitting the full model, using as a threshold either the median or the 95th percentile spread, obtained every day. We find that for the vast majority of the stocks, the median adjusted $R^2$ value is over 10\%. For some stocks we find even more remarkable median adjusted $R^2$ values of over 20\%, rising to as much as 50\% for some daily models. 

We then used the branch-and-bound algorithm implemented in the the leaps package in R \citep{lumley2004leaps}, in order to identify the best scoring model (in terms of the adjusted $R^2$ value). In this context, a model subspace is the set of all possible models containing a particular number of covariates $v$ from the LOB. For example, the full model contains all covariates and is the only model in its subspace, while the smallest model subspace is comprised of models that contain the intercept and any one of the possible covariates. Intermediate model subspaces are comprised of models with all combinations of $v=2 \ldots n-1$ covariates, where $n=p+m$ is the total number of covariates, contemporaneous $p$ plus lagged $m$ variates. There are  thus $\frac{n!}{(n-v)!}$ models in each model subspace. 

\begin{figure}[ht!]
	\begin{center}
	\includegraphics[width=0.58\textwidth]{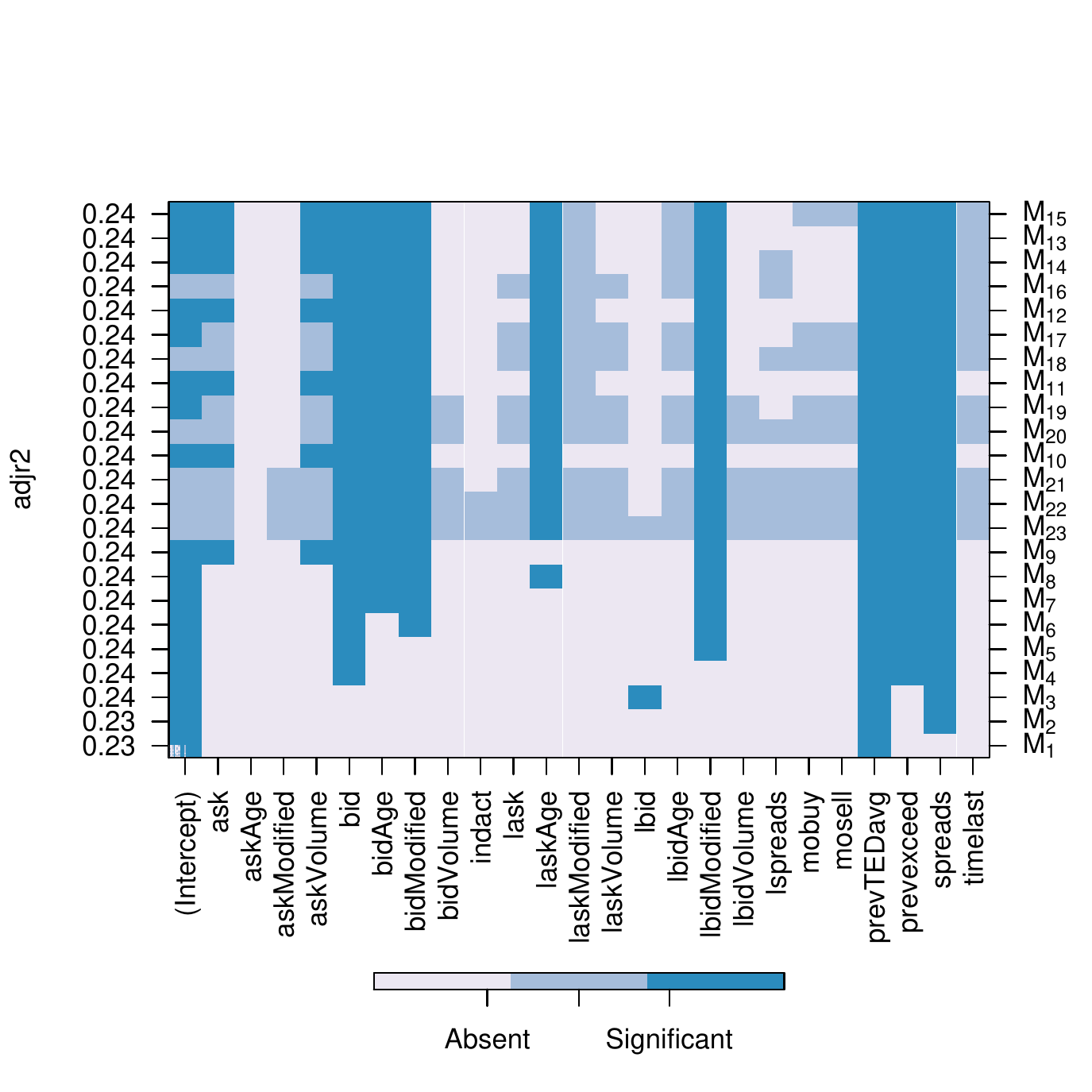}
	\includegraphics[width=0.58\textwidth]{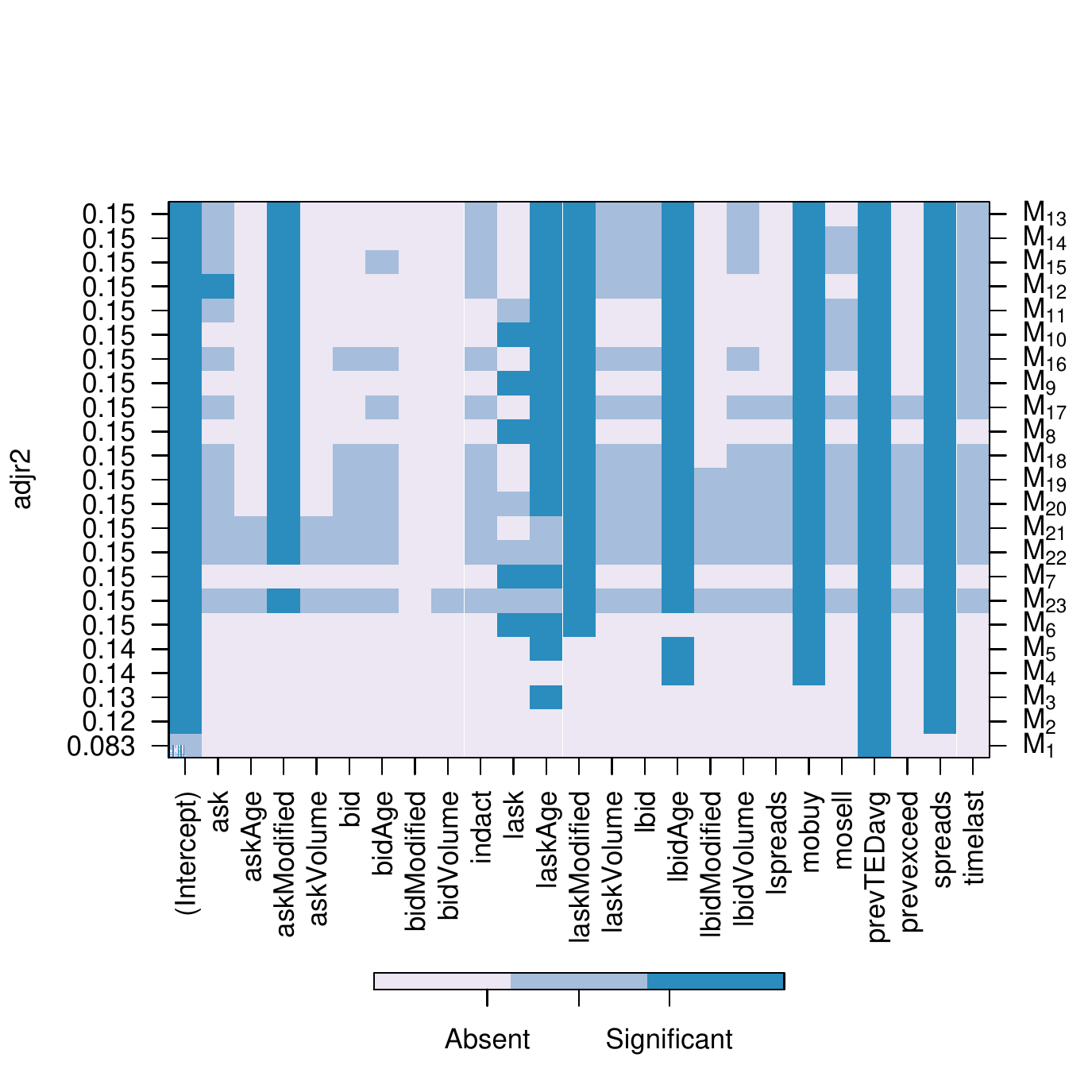}
	\caption{The adjusted $R^2$ values for models of using the best subsets of covariates (of size 1 to 24, in this case) for a single trading day (the 17th of January 2012) for stock Credit Agricole in the lognormal specification and the median spread (top) or the 95th percentile (bottom) as the threshold. Any one row corresponds to a submodel with the highlighted squares indicating whether a variate has been included in the model or not. A dark square indicates statistical significance at the 5\% level, with light squares not statistically significant at the 5\% level. For instance, row $M_3$ corresponds to a specification with the following variates: $intercept$, $ask$, $prevTEDavg$ and $spreads$.  The models are ranked by the best adjusted $R^2$ value, and we see that in this case, the best scoring model is obtained using a subset of 15 covariates for the top plot. We differentiate between covariates that were found to be significant or not, and of the 15 covariates in the best model, only 8 are found to be significant at the 5\% level.}
	\label{fig:subsetsACAp}	
	\end{center}
\end{figure}

To illustrate our findings we first present results for a given day of data for Credit Agricole in Figure \ref{fig:subsetsACAp}, where for all model subspaces, we rank the models in the subspace based on their adjusted $R^2$ score. We thus obtain the best combination of covariates, for each subspace and for each day of data. We can then identify the covariates that are consistently present as we move between model subspaces. This is interesting because it gives us a relative measure of the contribution of that covariate across different assumptions of parsimony for the model. Particularly for higher dimensional model subspaces, some of the covariates in each subset model are not significant, and we distinguish between the covariates that are significant or not, at the 5\% level of significance. 

\begin{figure}[ht!]
	\begin{center}
	\includegraphics[width=0.49\textwidth]{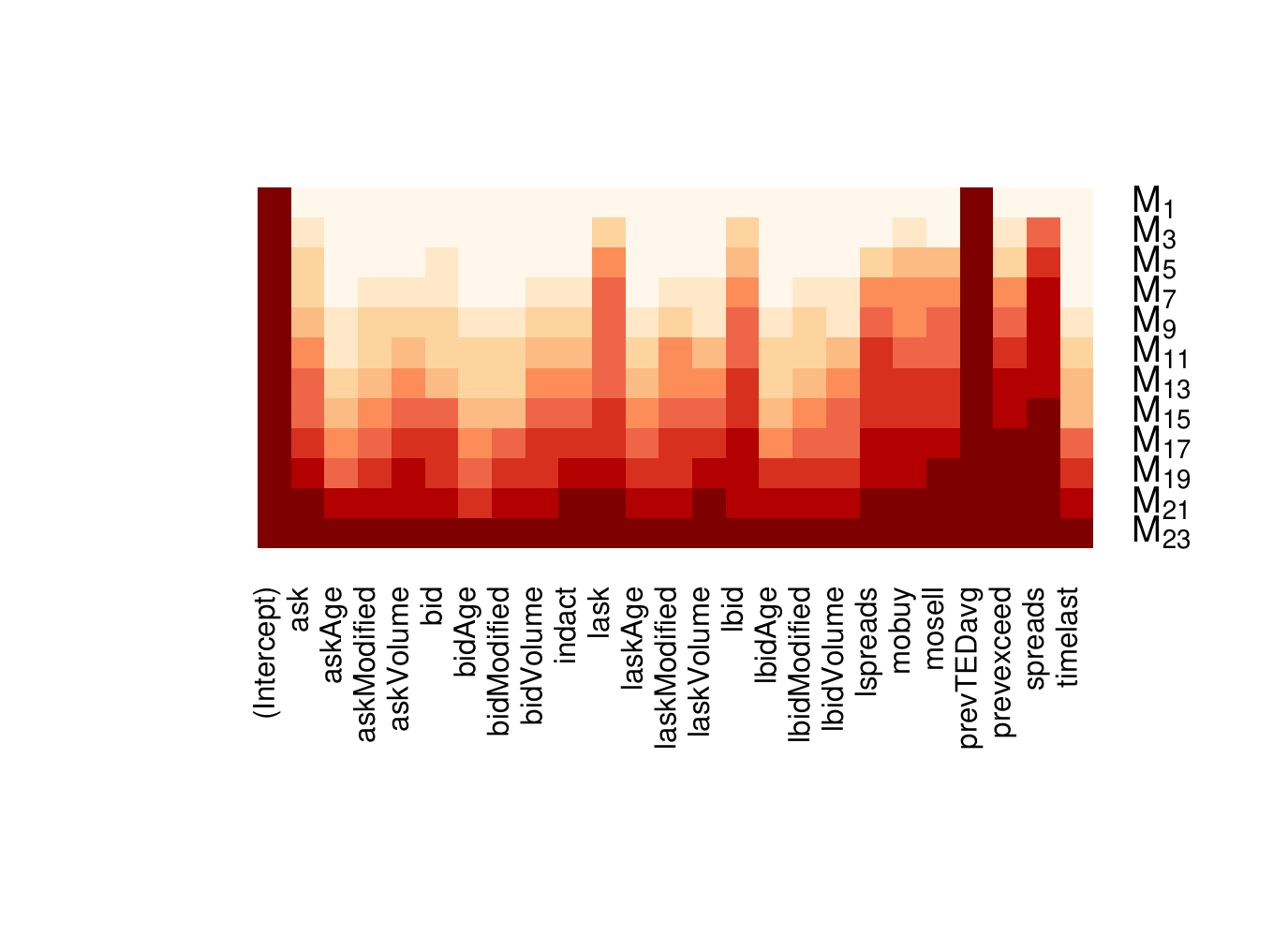}
	\includegraphics[width=0.49\textwidth]{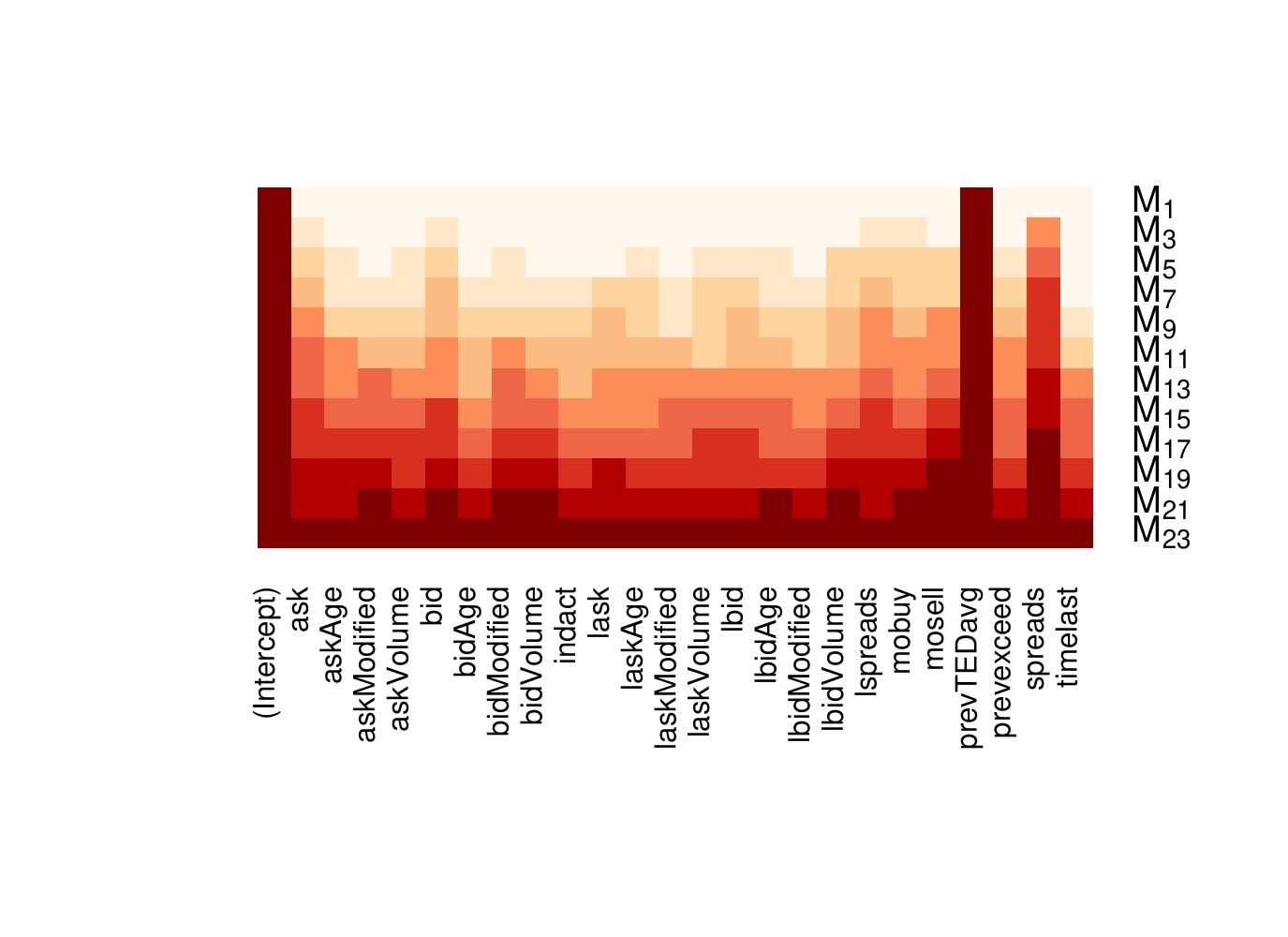}
	\caption{Heatmap of the relative frequency with which parameters appear in the best daily models of every subspace (frequency in terms of the number of daily models over the 82 day period) for the Credit Agricole dataset using the daily median (left) or the 95th percentile (right) of the spread as the threshold value.  So for instance, the element in row M11 and column {\em lask} indicates the relative frequency (in terms of the fraction of days over the 82 day period) by which the variate {\em lask} has appeared in the best model with 10 variates amongst all models with 10 variates. The bottom row is not informative since by construction all variates appeared amongst the best model having all variates.}
	\label{fig:heatmapACAp}	
	\end{center}
\end{figure}

To get an indication of the time stability of these model structures (and identify covariates that are consistently selected in the model), we illustrate the relative frequency with which parameters appear in the best models of every subset. That is, for each model subspace, we count the number of times each covariate forms part of the model with the highest adjusted-$R^2$ value over the four month period. Figure \ref{fig:heatmapACAp}  indicates that the covariates identified earlier as being important in explaining the variation in the TED for a single day ({\em prevTEDavg} and {\em spreads}) are also consistent features in models across time. However, {\em prevexceed} does not form part of the best model very frequently, except in higher model subspaces, possibly because it is less informative in the presence of the aforementioned covariates.  
 
\begin{figure}[hb]
	\begin{center}
	\includegraphics[width=0.49\textwidth]{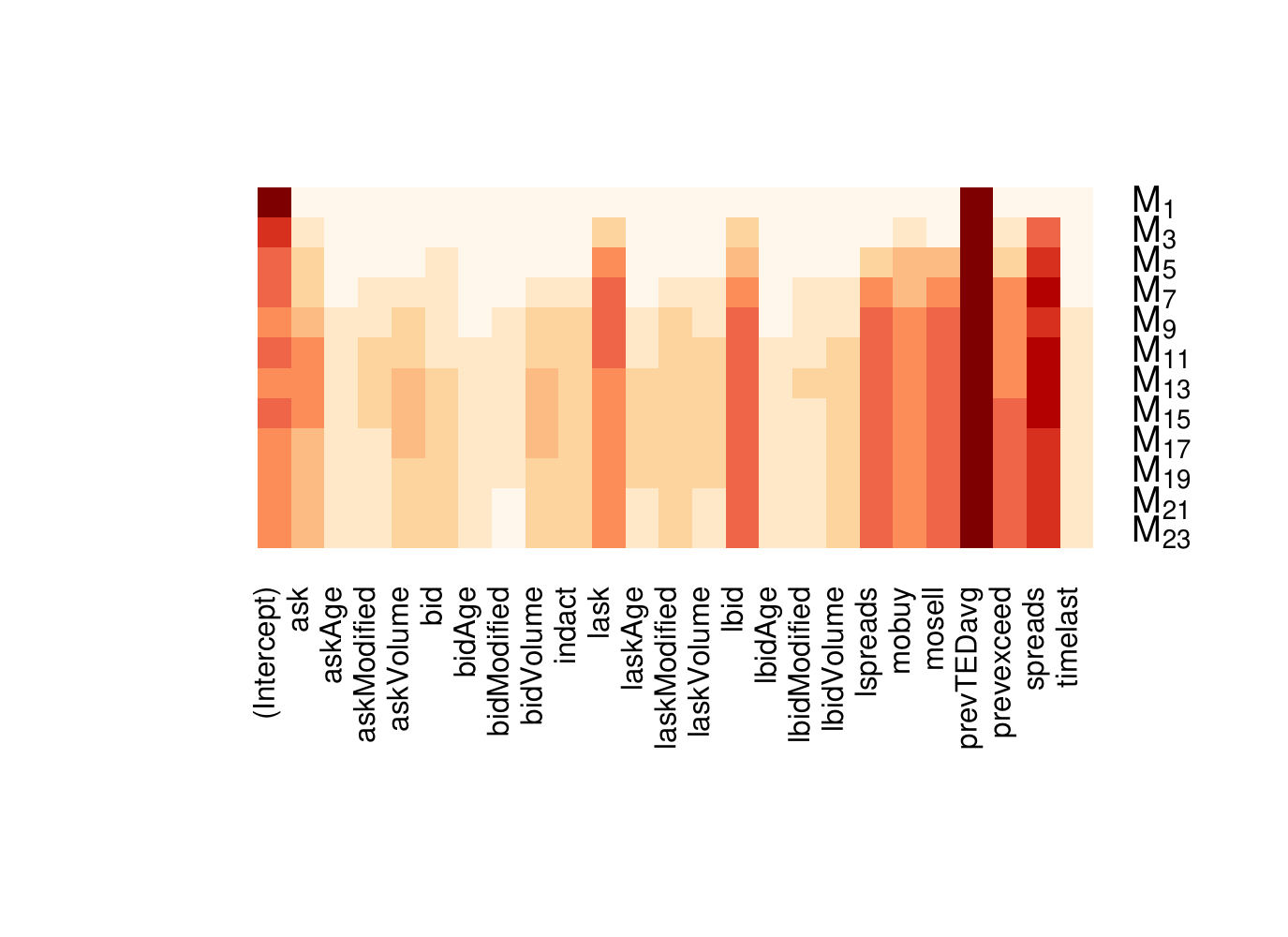}
	\includegraphics[width=0.49\textwidth]{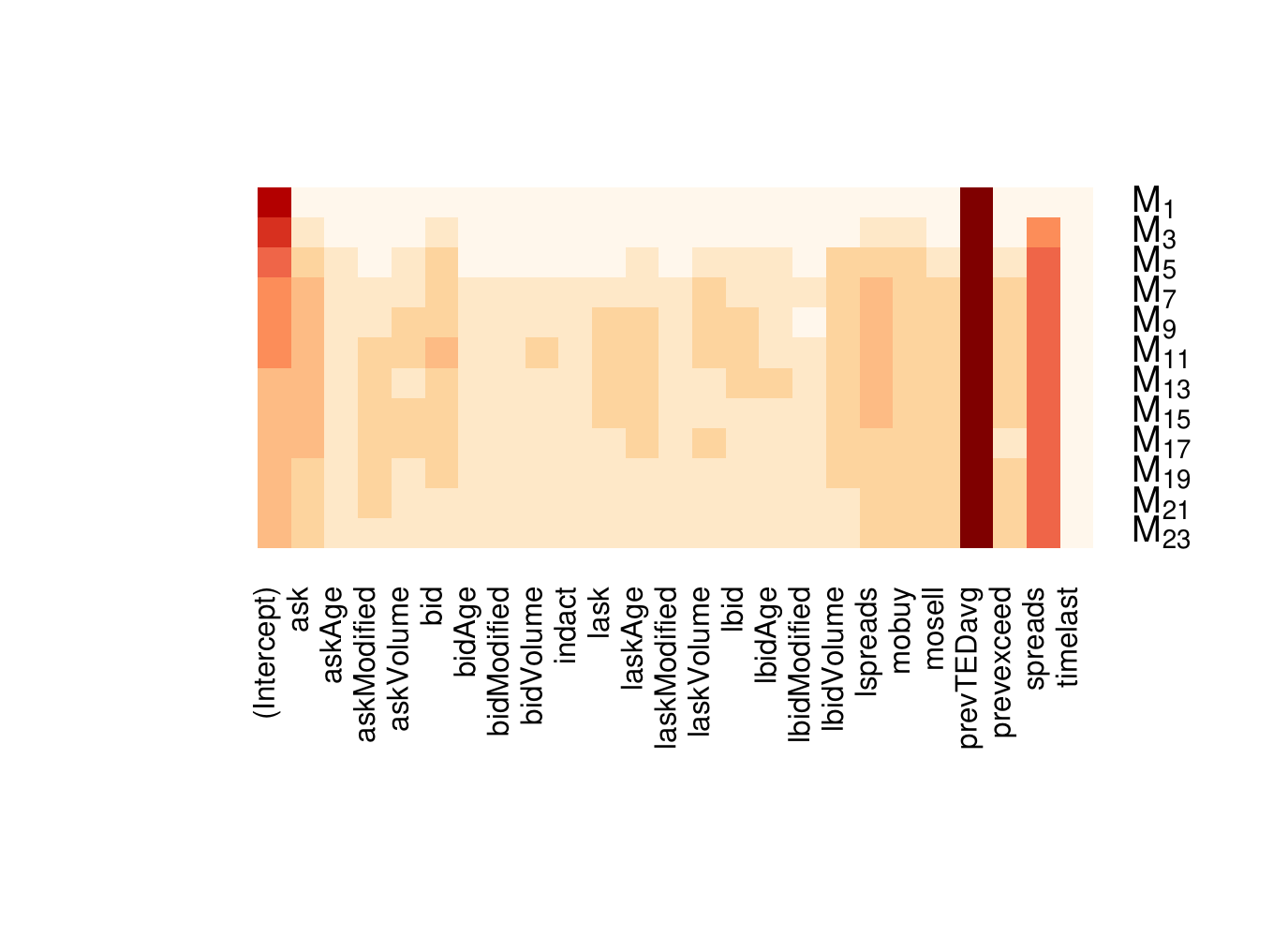}
	\caption{Heatmap of the relative frequency with which the parameters are found to be significant at the 5\% level (frequency in terms of the number of daily models over the 82 day period) for the Credit Agricole dataset, using the median and 95th percentile thresholds.}
	\label{fig:heatmapACApwsig}	
	\end{center}
\end{figure}

Besides the frequency of the presence of each covariate in the best fitting model of a given subspace, we also evaluate individual covariate significance over time via a formal partial t-test at the 5\% level in Figure \ref{fig:heatmapACApwsig}. At higher model subspaces, we find that several covariates are found to be statistically significant  (i.e. reject a null hypothesis for a partial t-test) less frequently. This is what one may expect, when covariates become less significant in the presence of other correlated covariates, i.e. collinearity in the factors of the LOB covariates takes effect. To validate this hypothesis regarding the model structures we develop, we have performed further analysis on the correlation between the covariates and the effect on our estimated coefficients, which can be provided on request.

In this analysis, we recall that under our regression framework, the sign of the coefficient for a given covariate indicates the direction of the partial effect of this variable, on the conditional probability that the resilience, as measured by the exceedance duration for a given threshold, will exceed a time $t$. Therefore we can interpret positive coefficient values as influencing the liquidity resilience of the LOB by slowing the return to a desirable level, whilst negative coefficients tend to result in a rapid return to the considered liquidity level, indicating higher resilience marginally, with respect to that covariate.  \cite{panayi2014market} provides an economic/theoretic interpretation of the significant covariates for the case of the lognormal model, and we provide in Section \ref{sec:interpretation} a discussion regarding the sign and variation in coefficients for the different model structures we considered. 

\begin{table}[hb]
\label{tab:coefs}
\centering
\begin{tabular}{rrr}
  \hline
 & \% significant & \% positive \\ 
 \hline
askAge & 48.2 & 22.4 \\ 
  askModified & 62.0 & 50.2 \\ 
  askVolume & 52.3 & 27.6 \\ 
  bid & 58.2 & 34.4 \\ 
  bidAge & 51.2 & 23.2 \\ 
  bidModified & 59.5 & 46.6 \\ 
  bidVolume & 54.6 & 27.7 \\ 
  indact & 52.6 & 35.6 \\ 
  lask & 61.0 & 12.1 \\ 
  laskAge & 49.0 & 22.3 \\ 
  laskModified & 65.6 & 7.9 \\ 
  laskVolume & 53.0 & 29.7 \\ 
  lbid & 63.5 & 12.4 \\ 
  lbidAge & 45.5 & 20.7 \\ 
  lbidModified & 63.7 & 9.3 \\ 
  lbidVolume & 53.1 & 30.0 \\ 
  lspreads & 71.6 & 58.9 \\ 
  mobuy & 84.3 & 36.5 \\ 
  mosell & 83.5 & 35.7 \\ 
  prevTEDavg & 96.3 & 96.2 \\ 
  prevexceed & 69.9 & 7.3 \\ 
  spreads & 75.2 & 70.5 \\ 
  timelast & 57.6 & 8.5 \\ 
   \hline
\end{tabular}
\caption{The percentage of daily models (for all assets) for which each covariate is found to be significant at the 5\% level, and the percentage of daily models for which the sign of the associated coefficient is positive.}
\end{table}

In Table \ref{tab:coefs} we summarise the significance and sign of the assets over time and over the different assets in our dataset. This is interesting as we can identify the regressors for which loadings are positive (negative) and thus produce marginal increases (decreases) in the expected TED under the model, and thus an associated decrease (increase) in the resilience of market liquidity. 

In the following, we will now consider a fixed subset of covariates in the regression models, which we identified in our previous model selection procedure as being most significant in the daily regressions, as well  a consistent sign. These are:

\begin{itemize}
\item {\em prevTEDavg}
\item {\em spreads}
\item {\em prevexceed}
\item {\em mobuy, mosell}
\item {\em ask, bid}
\item {\em lask, lbid}
\end{itemize}

\subsection{Model selection - distributional assumptions}

We now consider the effect of different distributional assumptions on the explanatory power of the model, using the fixed subset of covariates selected above. We will compare the explanatory power of the lognormal, Weibull, gamma and Generalised gamma regression models. We will first relate the covariates to the mean of the response, as in the GLM structure, before considering separate link functions for further distribution parameters, as in the GAMLSS structure. 

\begin{table}[h]
\label{tab:linkfn}
\begin{center}
\begin{tabular}{llll}
\hline
                  & \multicolumn{3}{c}{Link function} \\
Distribution      & $\mu $       & $\sigma$   & $\nu$       \\
\hline
Lognormal         & identity   & log      & -         \\
Gamma             & log        & log      & -         \\
Weibull           & log        & log      & -          \\
Generalised Gamma & log        & log      & identity \\
\hline
\end{tabular}
\caption{The link functions in the GAMLSS framework for each parameter for the four distributions under consideration.}
\end{center}
\end{table}

\begin{figure}[ht!]
	\begin{center}
	\includegraphics[width=0.48\textwidth]{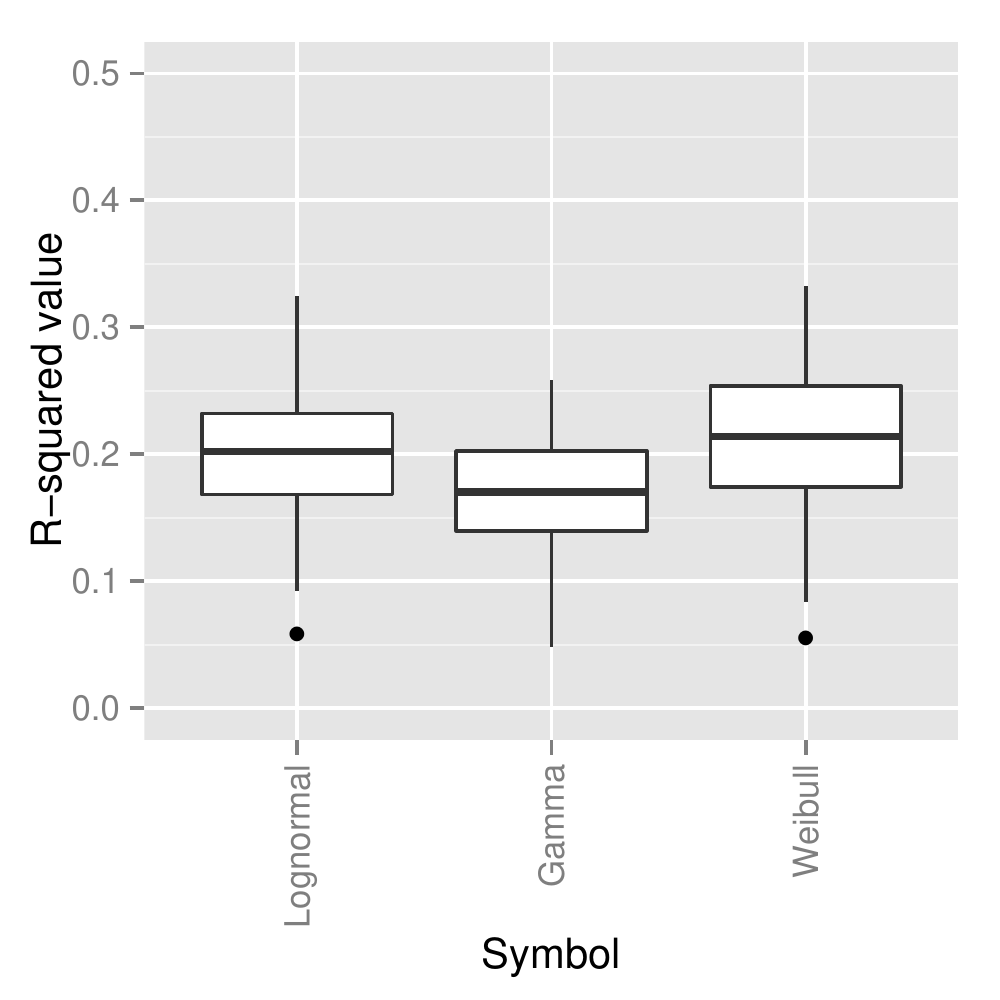}
		\includegraphics[width=0.48\textwidth]{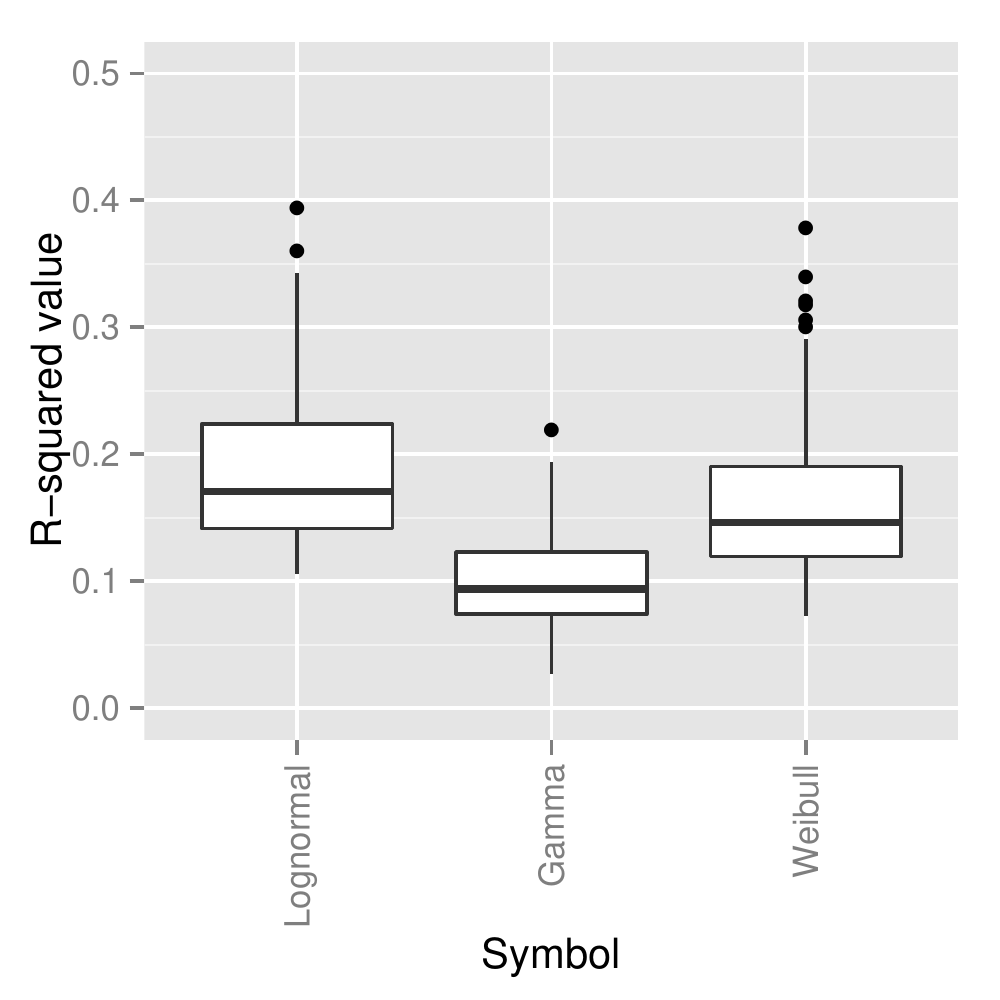}
			\includegraphics[width=0.48\textwidth]{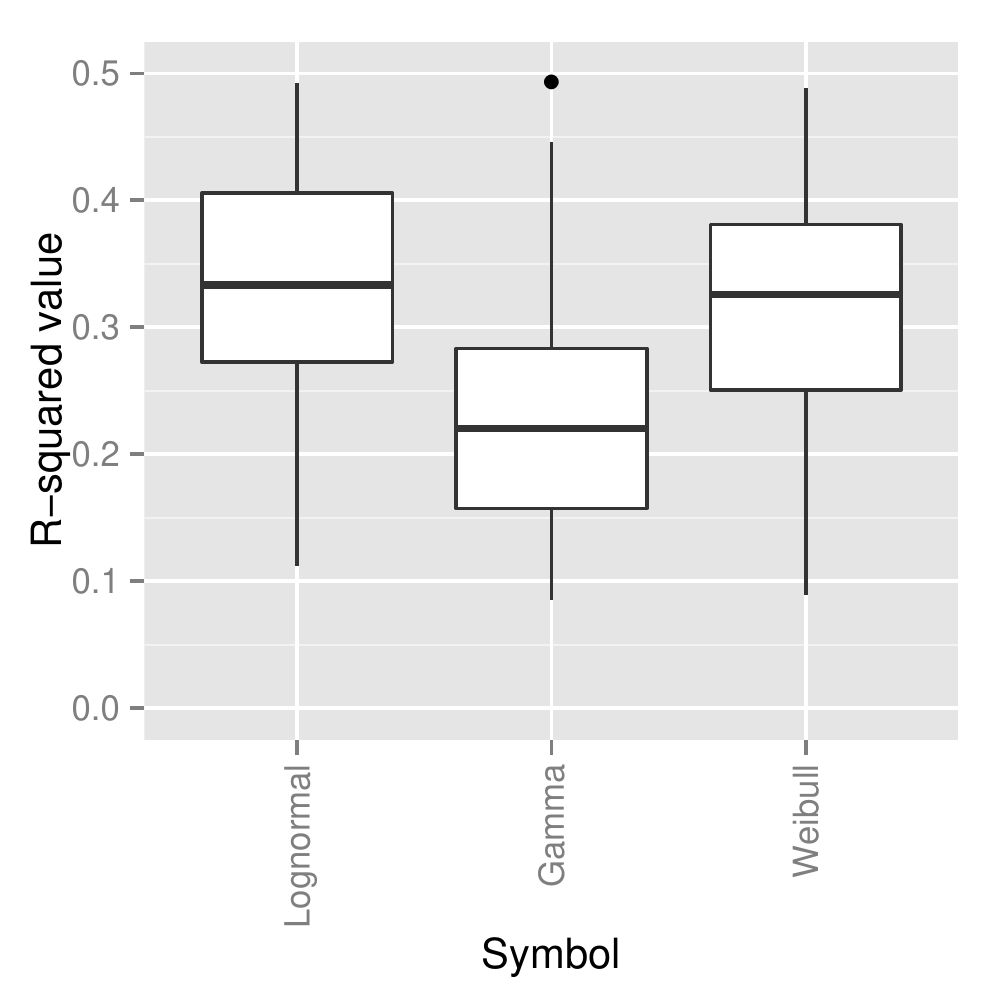}
				\includegraphics[width=0.48\textwidth]{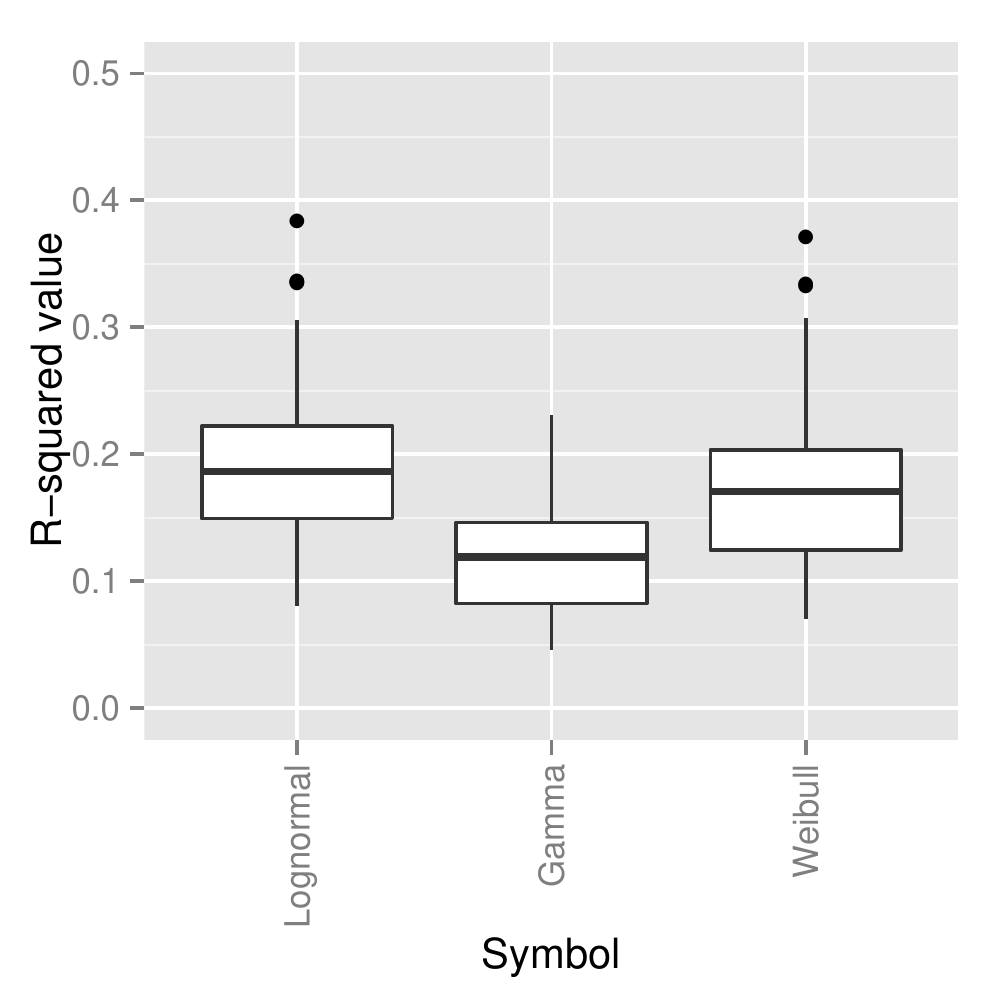}
	\caption{Boxplots of the adjusted $R^2$ value obtained from fitting the regression models with the various distributional assumptions separately for each day in our dataset for the threshold corresponding to the 5th decile of the spread }
	\label{fig:boxplotr2dist}
	\end{center}
\end{figure}

Figure \ref{fig:boxplotr2dist} shows the range of adjusted-$R^2$ values obtained from daily fits of each model over the 4-month period for the regression models with the different distributional assumptions. We note that in general, making lognormal and Weibull distributional assumptions leads to regression models where the explanatory power is comparable, whereas the explanatory power of gamma regression models is lower for the vast majority of assets. This indicates that the tail behaviour of the liquidity resilience measure tends to be better fit with moderate to heavy tailed distributions which admit more flexible skew and kurtosis features.

\begin{figure}[p]
	\begin{center}
	\includegraphics[width=\textwidth]{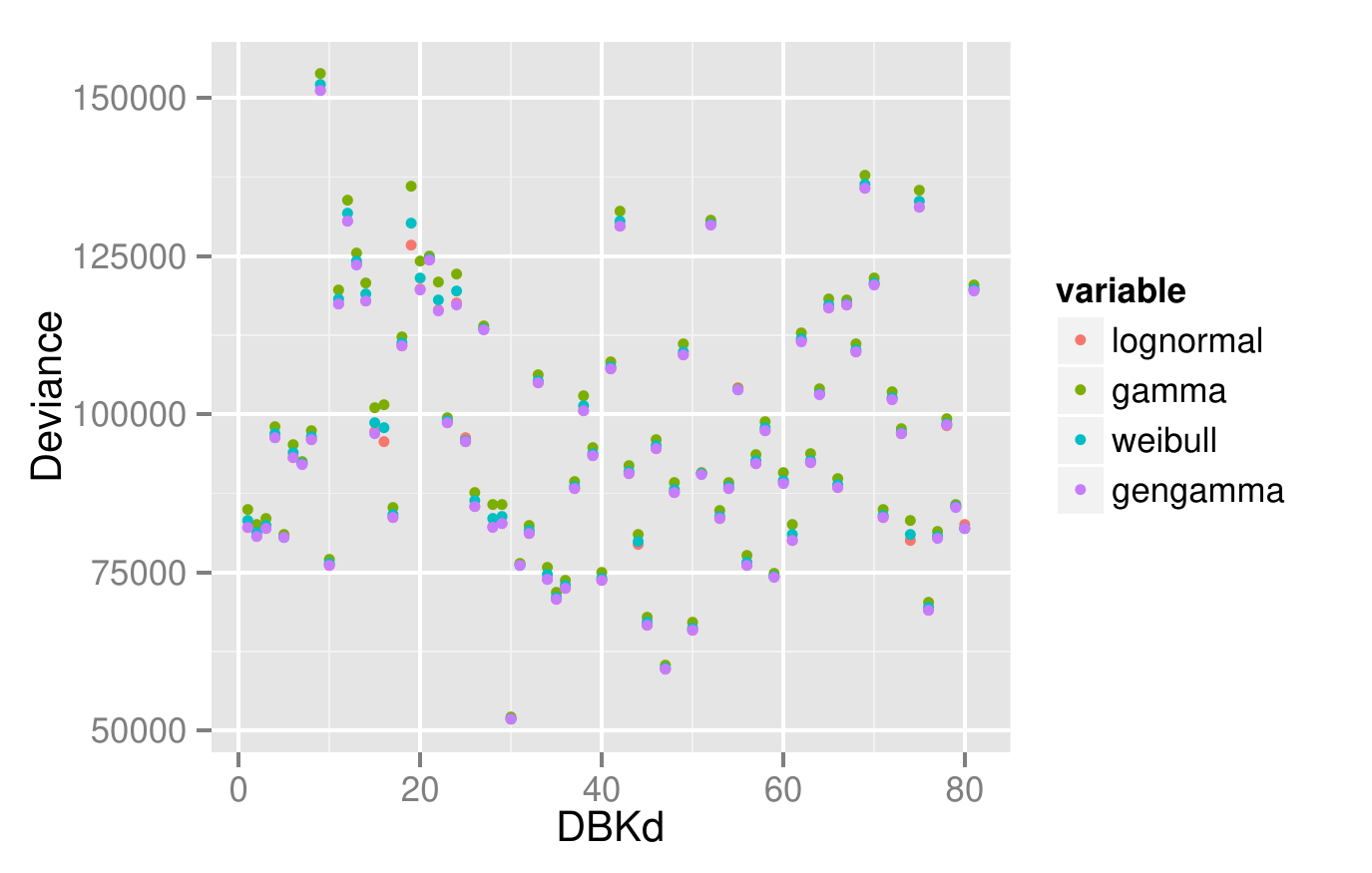}
		\includegraphics[width=\textwidth]{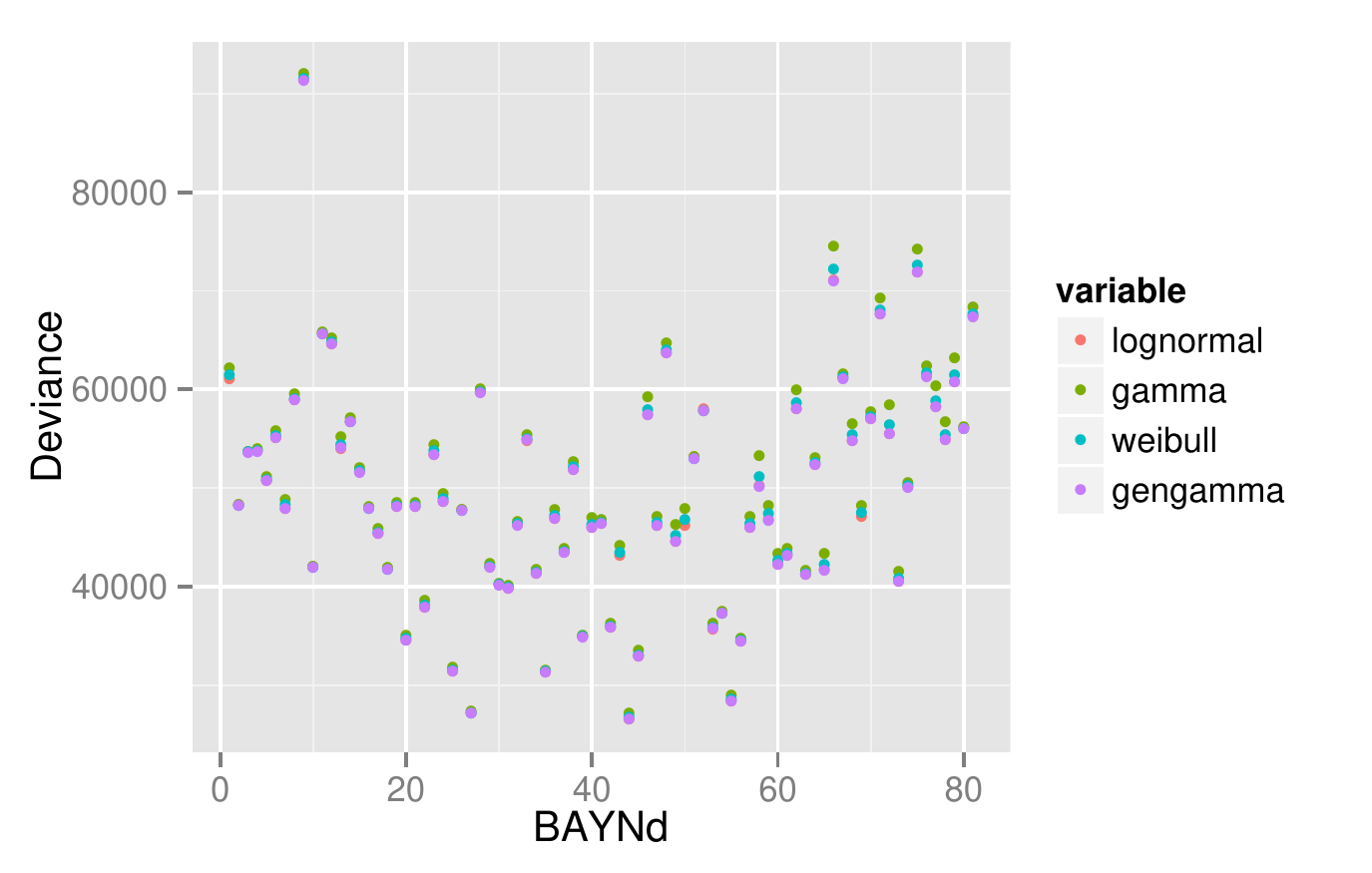}
	\caption{The deviance for every distributional assumption, with a model fit fit every day over an 81-day trading period.}
	\label{fig:deviance}
	\end{center}
\end{figure}

We present in Figure \ref{fig:deviance} for two assets, the estimated deviance of the fitted models every day, for all 4 distributional assumptions that we make. We see that in general, the Generalised Gamma produces model fits with the lowest deviance values, which is as one would expect, as it encompasses all the other distributions as special limiting cases. We also observed that there are a few days where the Generalised Gamma regression model failed to converge, and in these cases the Lognormal dominates. Over our dataset, we find that the Generalised Gamma model is the best performing model for approximately 75\% of the data, while for most of the remaining cases (which would mostly include daily datasets where the Generalised Gamma model failed to converge), the best model is the Lognormal one.    

\begin{table}[ht]
\centering
\begin{tabular}{rr}
  \hline
 & Lowest deviance \\ 
  \hline
Lognormal & 24.6\% \\ 
	Gamma & 0.0\% \\ 
  Weibull & 0.8\% \\ 
  Generalised Gamma & 74.6\% \\ 
   \hline
\end{tabular}
\caption{The percentage of daily datasets for which each model fit produced the lowest deviance.}
\end{table}

\subsection{Incorporating more flexibility through a GAMLSS framework specification}

The advantage of the GAMLSS framework is that is one is able to relate regression covariates to every distribution parameter through different link functions. Within the {\em gamlss} package, the distributions are reparameterised (as explained in Section \ref{sec:regmodels}) so that they have common parameters $\mu,\sigma$, and possibly $\nu$ and $\kappa$. For the Lognormal model the default link function for $\mu$ is the identity link 
\begin{align*}
g(\mu)&=\mu \\
&=\bm{X}_1 \bm{\beta}_1,
\end{align*}

while for most other parameters the log link is used, e.g. for $\sigma$

\begin{align*}
h(\sigma)&=\log(\sigma) \\
&=\bm{X}_2 \bm{\beta}_2,
\end{align*}

\begin{figure}[hb]
	\begin{center}
	\includegraphics[width=0.48\textwidth]{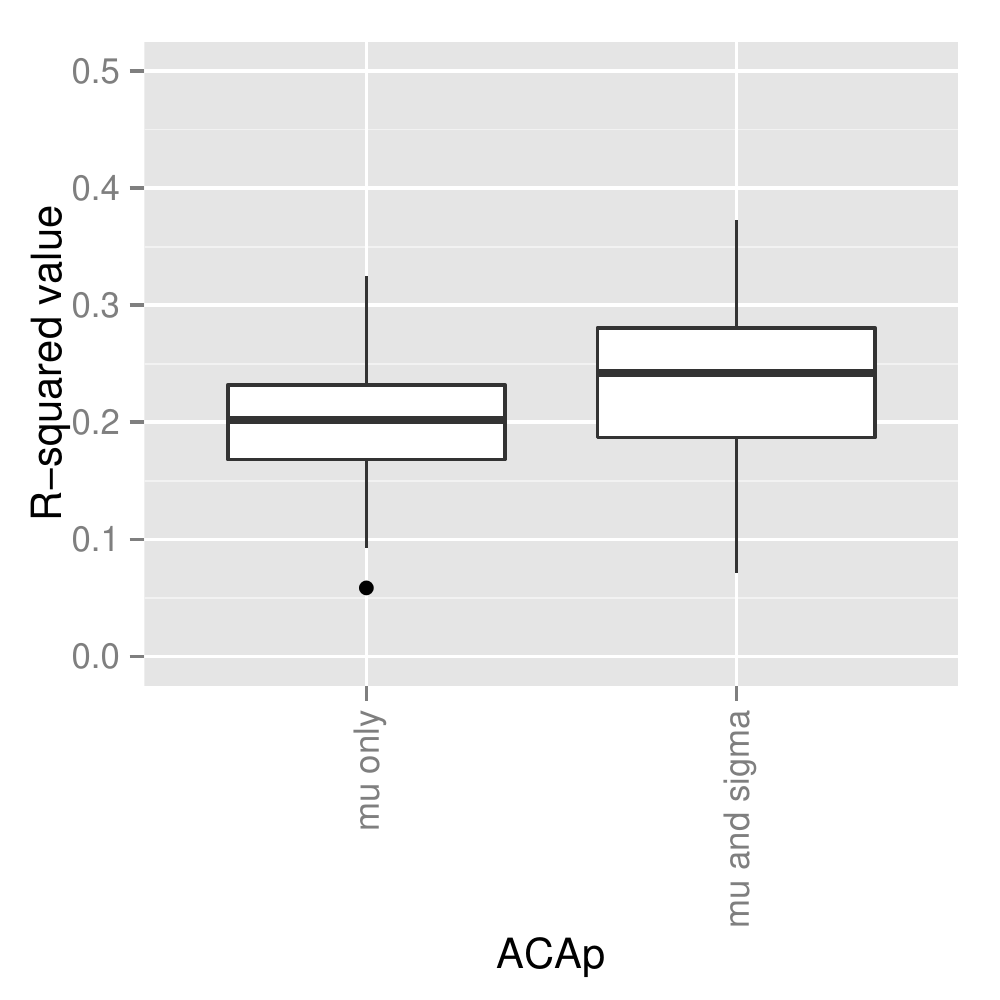}
		\includegraphics[width=0.48\textwidth]{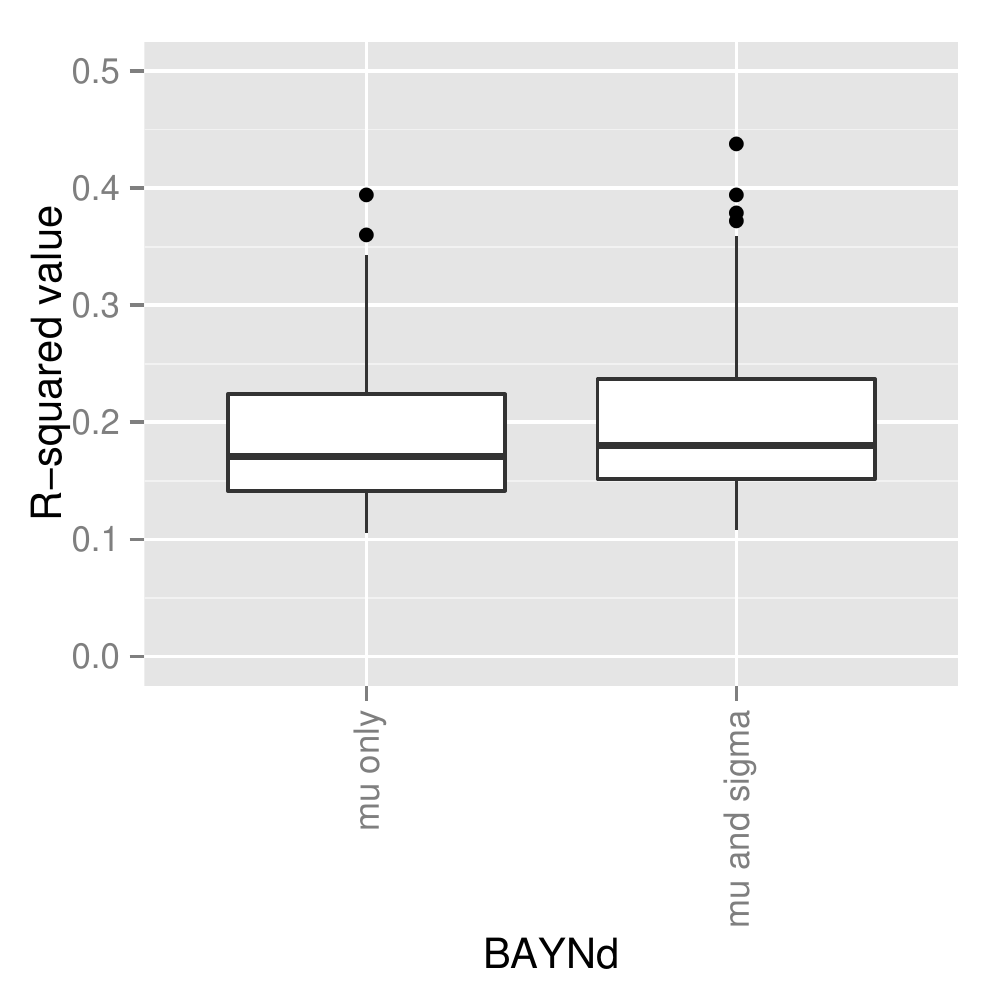}
			\includegraphics[width=0.48\textwidth]{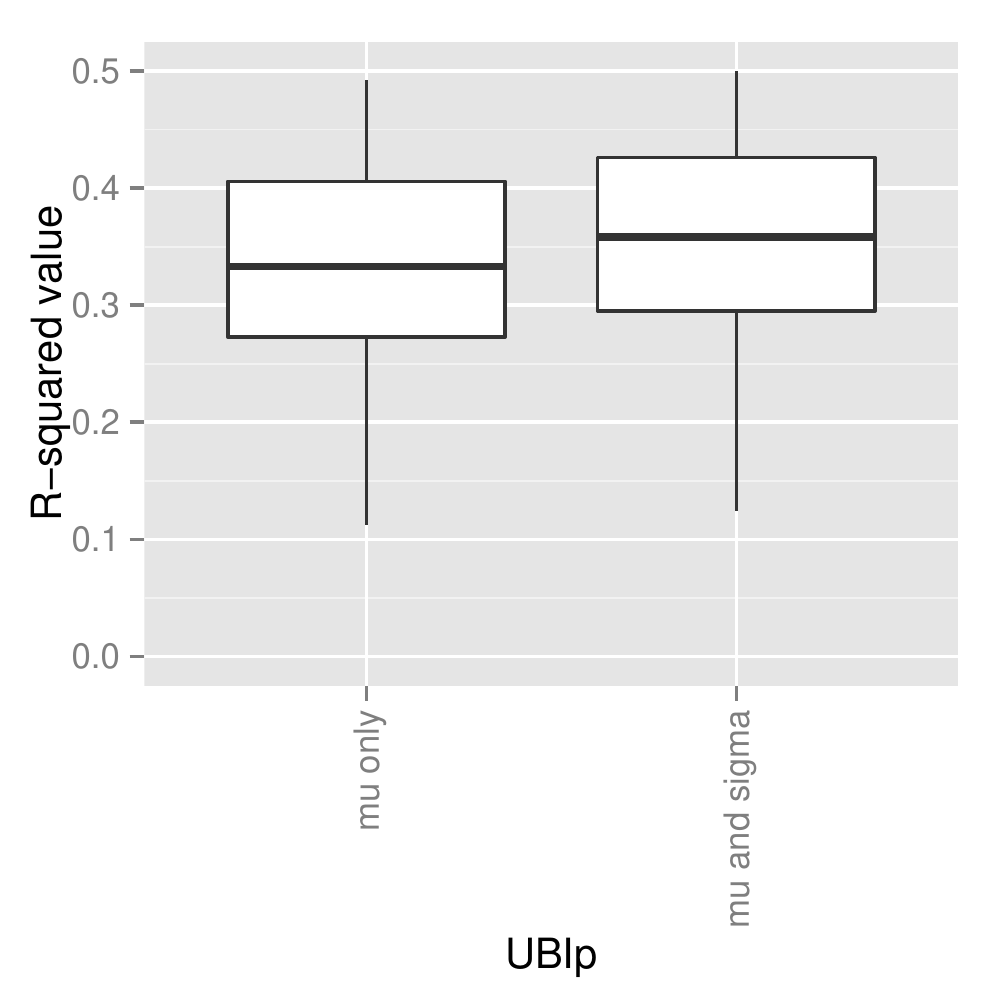}
				\includegraphics[width=0.48\textwidth]{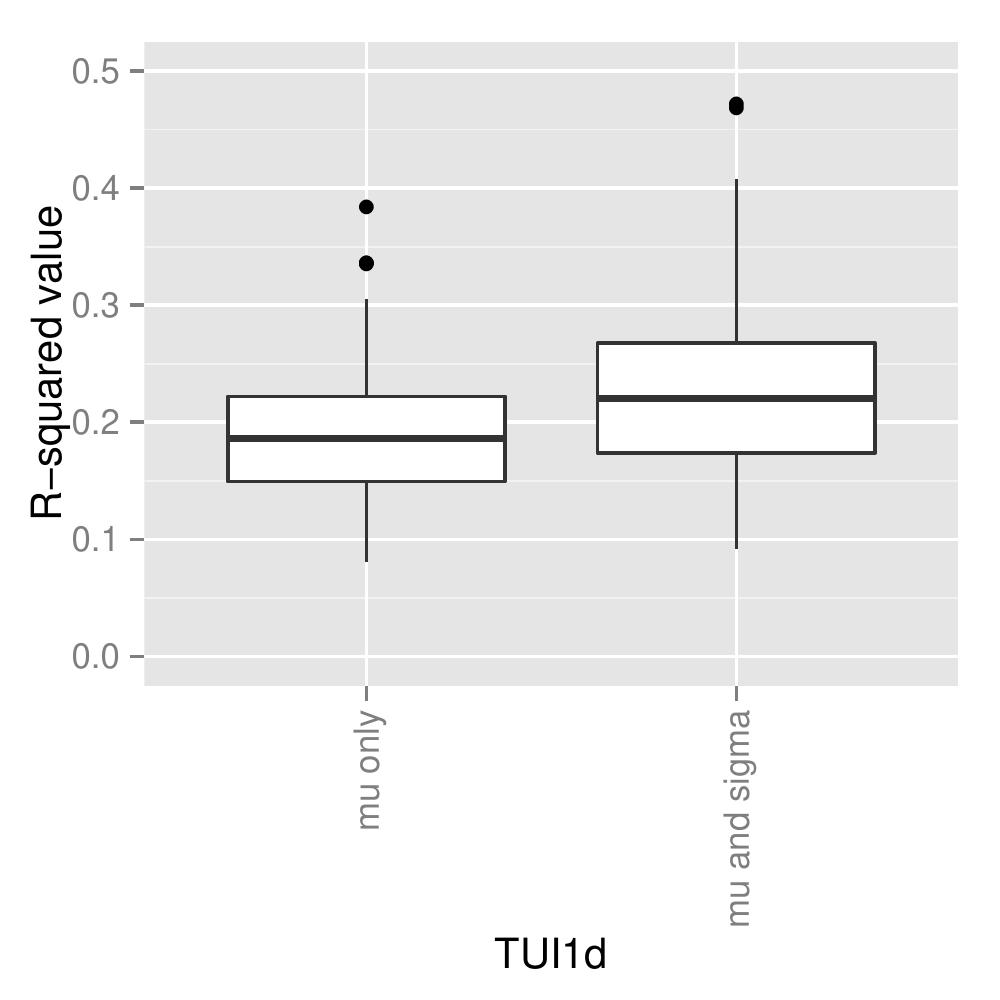}
	\caption{Boxplots of the adjusted $R^2$ value obtained from fitting the lognormal regression model, comparing the explanatory power when relating only a single distribution parameter $\mu$ to covariates, and when relating both $\mu$ and $\sigma$ in a GAMLSS framework, for four different assets. (Top left): Credit Agricole SA. (Top right): Bayer AG. (Lower left): UBISOFT Entertainment. (Lower right): TUI AG. }
	\label{fig:boxplotr2musigma}
	\end{center}
\end{figure}

As we do not know apriori whether covariates are more important in affecting one distribution parameter than another, we use a common set of covariates for each link function and therefore in the expressions for the link functions above $\bm{X}_1=\bm{X}_2=\bm{X}$. We present in Figure \ref{fig:boxplotr2musigma} the $R^2$ explanatory power of the Lognormal regression model when considering only a single link function for $\mu$, and when considering an additional link function for $\sigma$  also. We note that there is a slight increase in the median $R^2$, and this is observed across the assets in our dataset. 

\subsection{Effects of unit changes in LOB dynamics on the TED}

In this section we consider how to study the influence on the TED  arising from a unit change in the statistically most important covariates given by: {\em prevTEDavg}; {\em spreads}; {\em prevexceed}; {\em mobuy, mosell}; {\em ask, bid}; {\em lask, lbid}. This is interesting to study as it will depend on the distributional choice and model structure. We study the perturbation effect of a unit change of one covariate in the GAMLSS model, given all the other covariates on the mean and variance functions of the model. This will allow us to interpret the influence of the sign and magnitude of the coefficient loadings in the model for each covariate on the average TED (replenishment time) and the variance in the TED for an asset.

\subsubsection{Lognormal GAMLSS mean and variance functions  in the GAMLSS framework}
In the case of the Lognormal model with a single link function on $g\left(\mu(\bm{x})\right) = \bm{\beta}^T\bm{x}$ we know that the mean and variance functions are given as follows:
\begin{equation}
\begin{split}
\mathbb{E}\left[\left. \tau \right| \bm{x} \right] &= \exp\left(\mu(\bm{x}) + \sigma^2/2\right)\\
\mathbb{V}\text{ar}\left[\left. \tau \right| \bm{x} \right] &= \exp\left(\sigma^2/2 \right) \left[ \exp\left(\sigma^2/2 \right) - 1 \right]\exp\left(2 \mu(\bm{x}) \right).
\end{split}
\end{equation}
We can therefore consider the influence of a unit change in a covariate under this model by considering the partial difference in the mean and variance functions given a change in say the $j$-th covariate, which is given by
\begin{equation}
\begin{split}
&\frac{\partial}{\partial \bm{x}_j} \mathbb{E}\left[\left. \tau \right| \bm{x} \right] = \beta_j \\
\frac{\partial}{\partial \bm{x}_j} \mathbb{V}\text{ar}\left[\left. \tau \right| \bm{x} \right] &= 2 \beta_j \exp\left(\sigma^2/2 \right) \left[ \exp\left(\sigma^2/2 \right) - 1 \right]\exp\left(2 \mu(\bm{x}) \right).
\end{split}
\end{equation}
From this analysis one sees that a unit change in the $j$-th covariate $x_j$ with a negative coefficient loading will produce an increase in the mean liquidity resilience by reducing the average TED. Conversely, a positive loading will result in an decrease in the mean liquidity resilience.

In the case of a Lognormal model with two link functions, assuming both parameters are related to the same set of covariates in vector $\bm{x}$, then one has $g\left(\mu(\bm{x})\right) = \bm{\beta}^T\bm{x}$ and $h\left(\sigma(\bm{x})\right) = \log\left(\sigma(\bm{x})\right) = \bm{\alpha}^T\bm{x}$. An approximation of the log link function
\begin{equation}
\ln\mathbb{E}\left[\left. \tau \right| \bm{x} \right] \approx \left(\sum_{j} \beta_j x_j + 1 +  \sum_{k} \alpha_k x_k + O\left(\sum_{k} \alpha^2_k x_k^2\right)\right),
\end{equation}
results in the following approximate relationship for the partial derivative of a covariate $x_j$
\begin{equation}
\frac{\partial}{\partial x_j} \mathbb{E}\left[\left. \tau \right| \bm{x} \right] \approx \left(\beta_j + \alpha_j\right)\exp\left(1 + \sum_{j} \beta_j x_j + \sum_{k} \alpha_k x_k \right).
\end{equation}
Hence, if $\beta_j + \alpha_j > 0$ then a unit change in covariate $x_j$ will result in an increase in the average TED. Conversely if $\beta_j + \alpha_j<0$ then an increase in covariate $x_j$ will decrease the average TED and in the third case that $\beta_j + \alpha_j=0$, changes in the covariate have no effect on liquidity resilience, as measured by the TED.

\subsubsection{Gamma mean and variance functions in the GAMLSS framework}
In the case of the Gamma model with a link function on $g\left(\mu(\bm{x})\right) = \bm{\beta}^T\bm{x}$, the mean and variance function are given as follows:
\begin{align*}
\mathbb{E}\left[\left. \tau \right| \bm{x} \right] &= \mu(\bm{x}) = \exp\left(\sum_j \beta_j x_j\right) \\
\mathbb{V}\text{ar}\left[\left. \tau \right| \bm{x} \right] &= \mu \left(\bm{x}\right)\sigma \left(\bm{x}\right) = \exp\left(2\sum_j \beta_j x_j + 2\sum_j \alpha_j x_j \right).
\end{align*}
We can therefore consider the influence of a unit change in a covariate under this model by considering the partial difference in the mean and variance functions given a change in say the $j$-th covariate given by
\begin{align*}
\frac{\partial}{\partial \bm{x}_j} \mathbb{E}\left[\left. \tau \right| \bm{x} \right] &= \beta_j \exp\left( \sum_j \beta_j x_j \right) \\
\frac{\partial}{\partial \bm{x}_j} \mathbb{V}\text{ar}\left[\left. \tau \right| \bm{x} \right] &= 2 \left(\beta_j + \alpha_j \right) \exp\left(2\sum_j \beta_j x_j + 2\sum_j \alpha_j x_j \right).
\end{align*}
We see from this analysis that a unit change in the variable $x_j$ when $\beta_j > 0$ results in an increase in the average TED, as well as an increase in the variance in the variance of the TED $\beta_j +\alpha_j >0$.

\subsubsection{Weibull and Generalised Gamma mean and variance functions  in the GAMLSS framework}

The Weibull distribution parameterised has mean and variance functions

\begin{align*}
\mathbb{E} \left[\tau | \bm{x} \right]&=\mu(\bm{x})= \exp\left(\sum_j \beta_j x_j\right) ,\\
\mathbb{V} \left[\tau | \bm{x} \right]&= \mu^2(\bm{x}) \left\{ \Gamma\left(\frac{2}{\sigma(\bm{x})}+1 \right) \left[ \Gamma\left(\frac{1}{\sigma(\bm{x})}+1 \right) \right]^{-2} -1 \right\}.
\end{align*}

The generalised gamma distribution has mean and variance functions

\begin{align*}
\mathbb{E} \left[\tau | \bm{x} \right]&=\mu(\bm{x}) \frac{\Gamma \left( \theta + \frac{1}{\nu} \right)}{\theta^{1/v} \Gamma(\theta)},\\
\mathbb{V} \left[\tau | \bm{x} \right]&= \mu^2(\bm{x}) \frac{  \Gamma(\theta) \Gamma \left(\theta+\frac{2}{\nu}\right) - \left[ \Gamma\left( \theta + \frac{1}{\nu} \right) \right]^{2}  }{  \theta^{\frac{2}{\nu}} \left[ \Gamma(\theta) \right]^2}.
\end{align*}

We can obtain the partial difference in the mean and variance functions as in the lognormal and Gamma cases above. However, the presence of the gamma functions makes the identification of the partial contributions of unit changes in covariates to the mean and variance functions more involved.  


\subsection{Interpretation}
\label{sec:interpretation}
\begin{figure}[p]
	\begin{center}
	\includegraphics[width=0.48\textwidth]{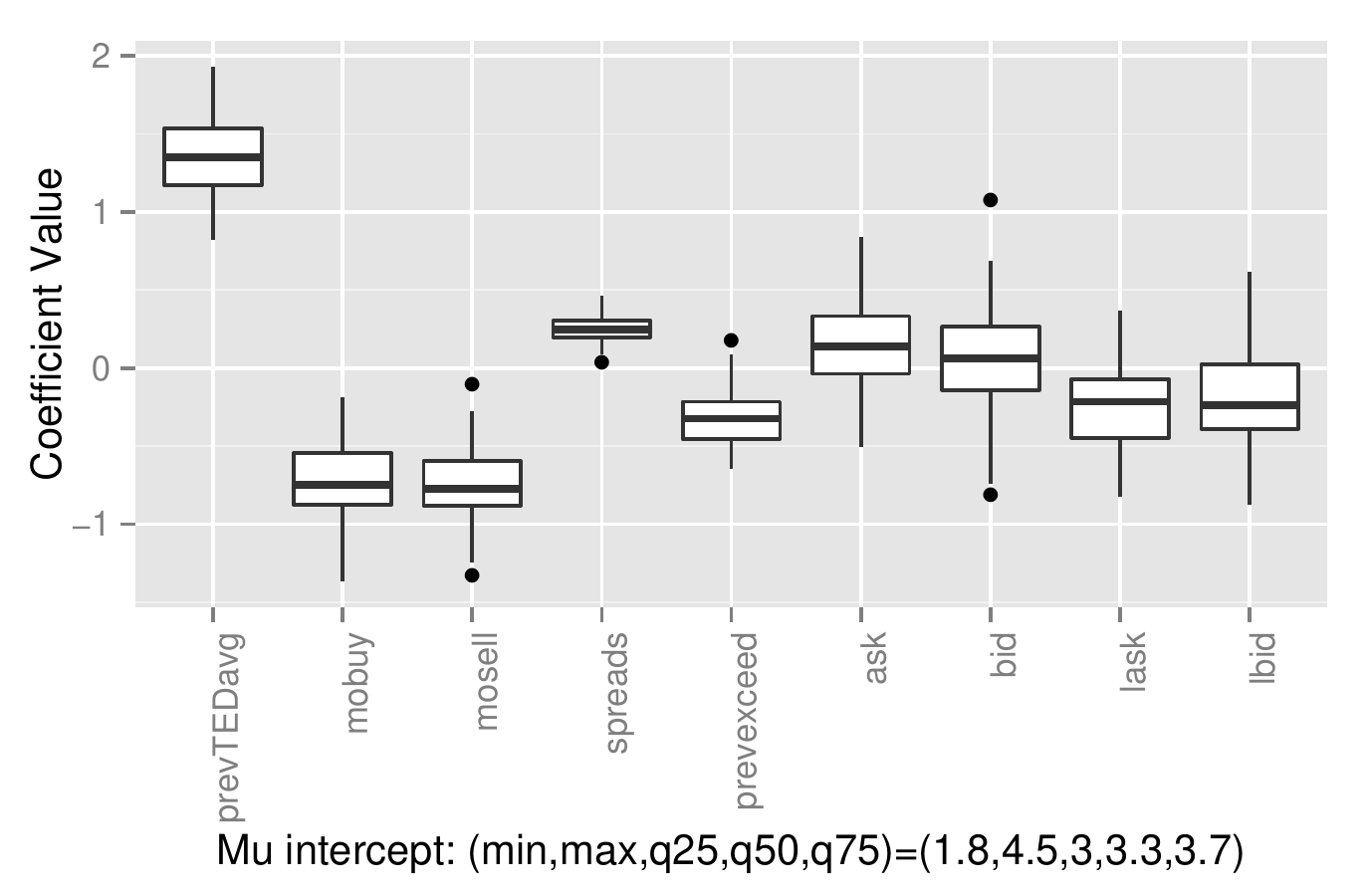}
	\includegraphics[width=0.48\textwidth]{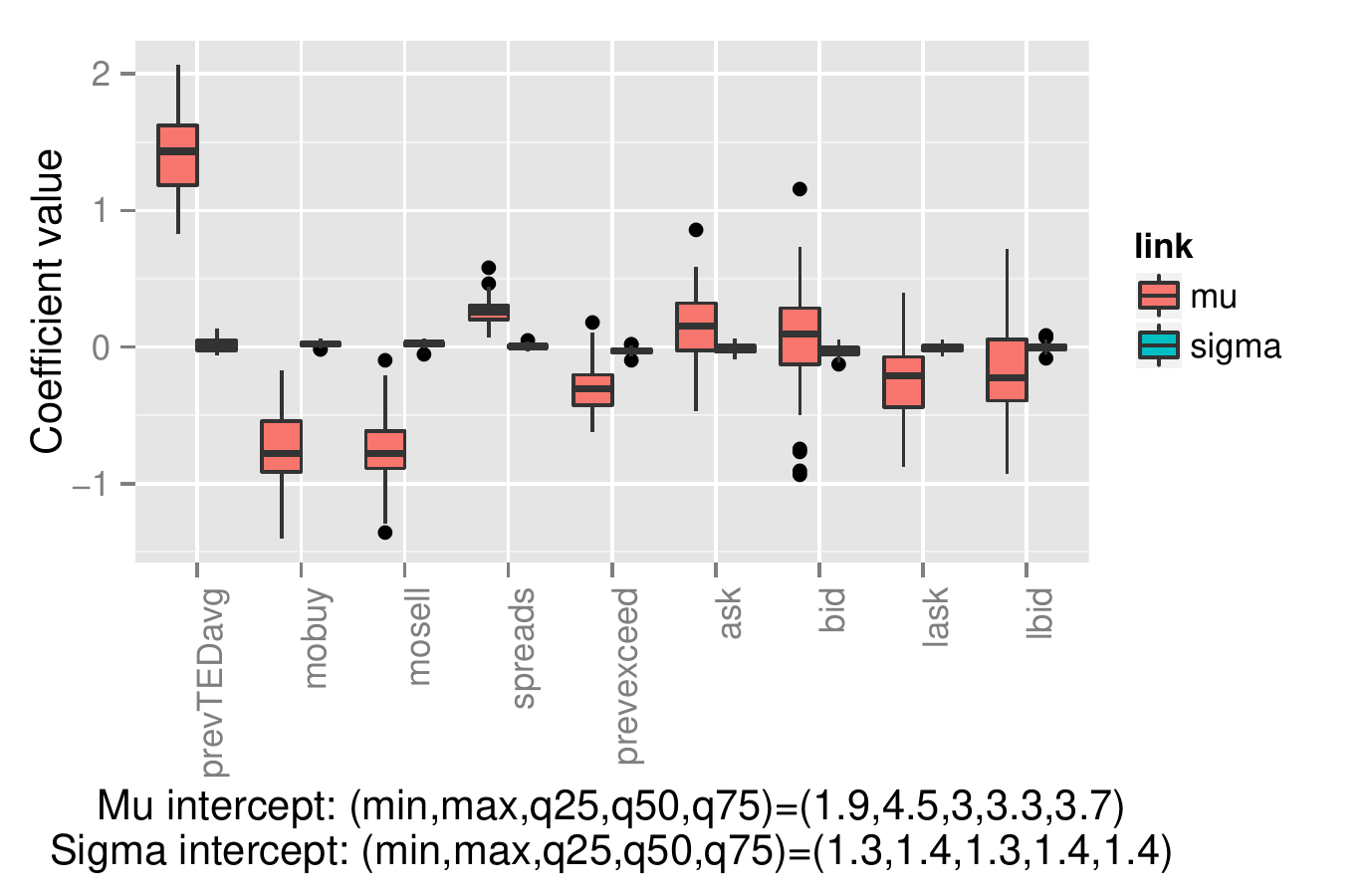}
		\includegraphics[width=0.48\textwidth]{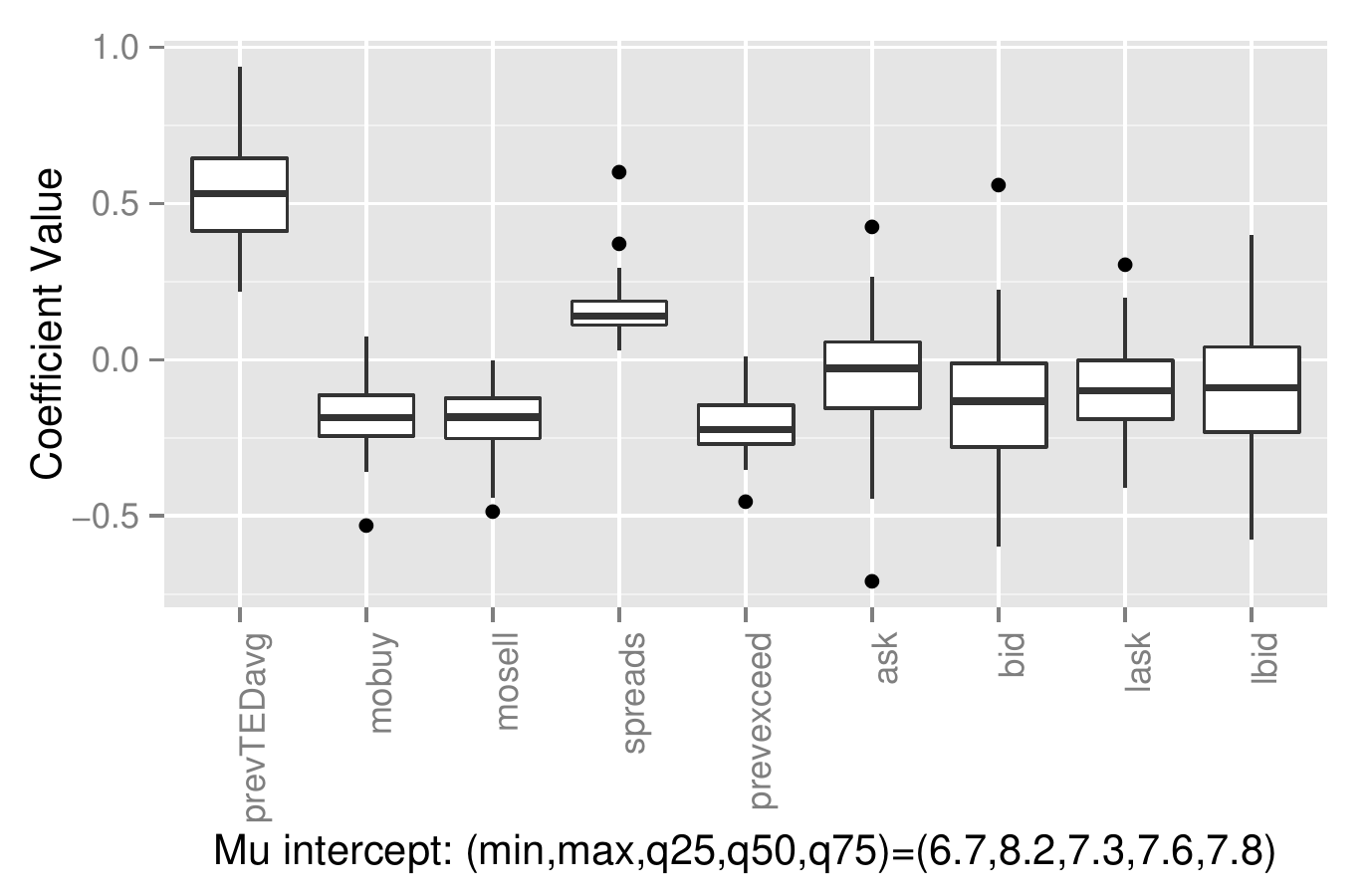}
		\includegraphics[width=0.48\textwidth]{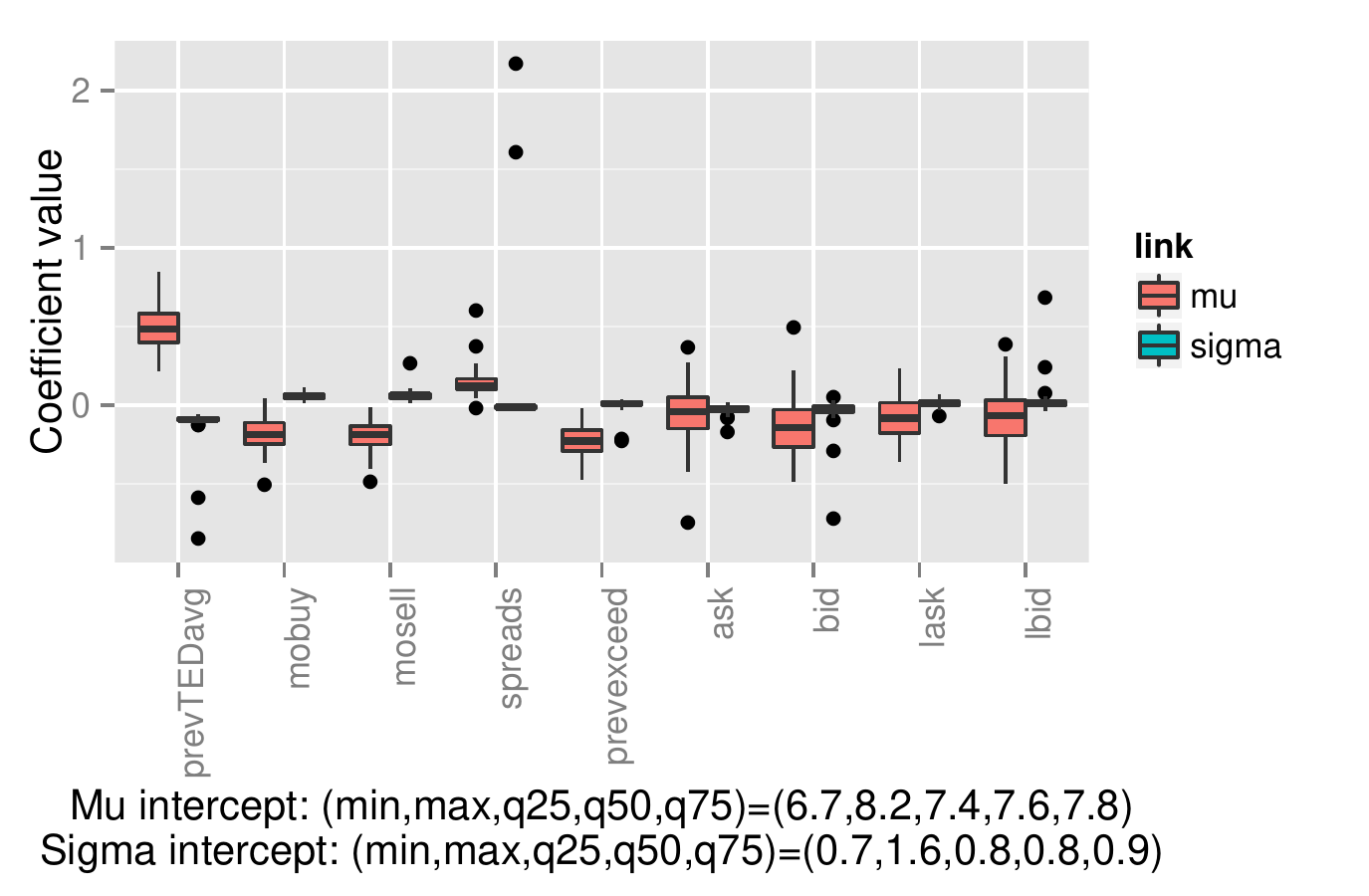}
		\includegraphics[width=0.48\textwidth]{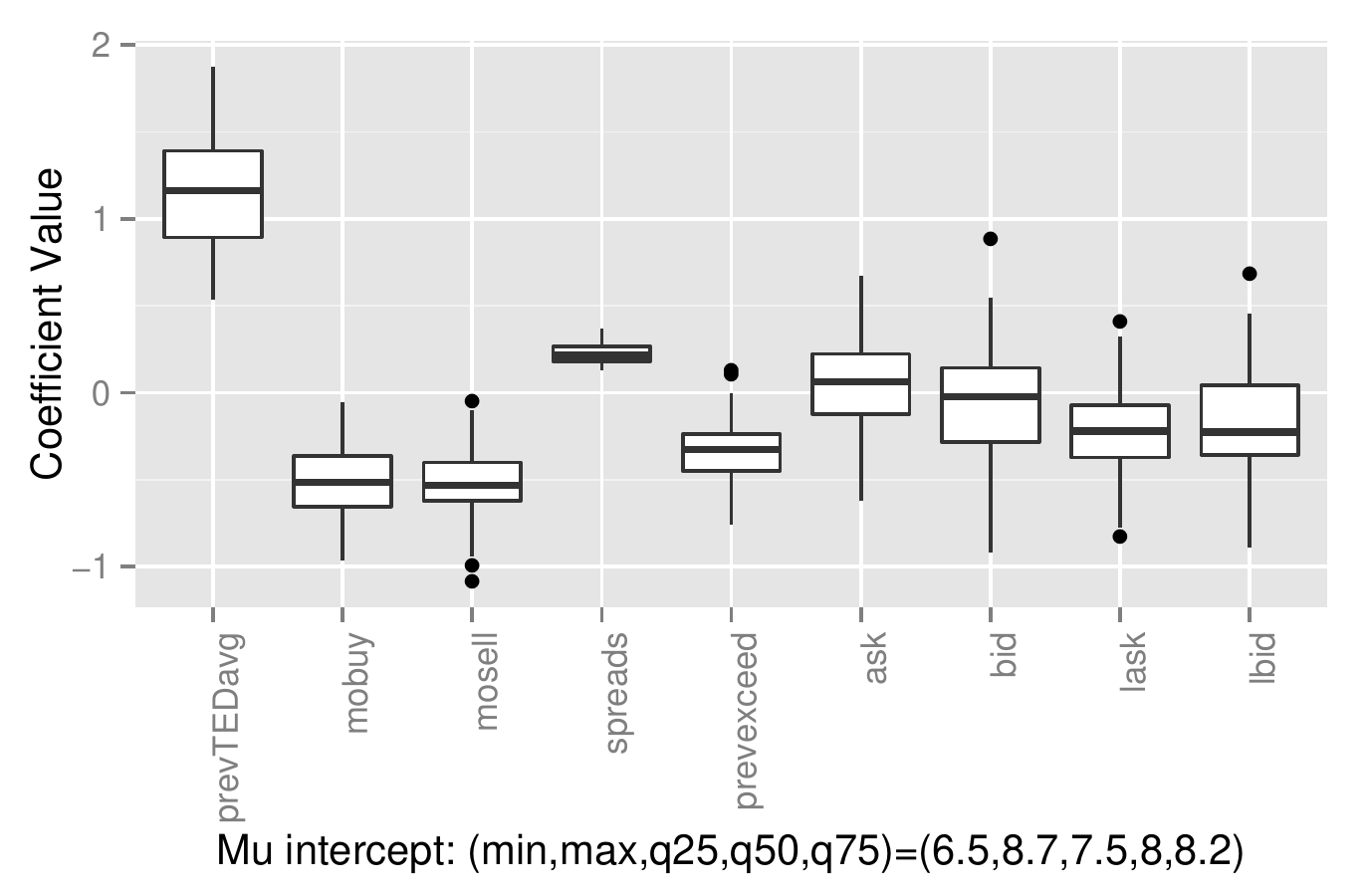}
				\includegraphics[width=0.48\textwidth]{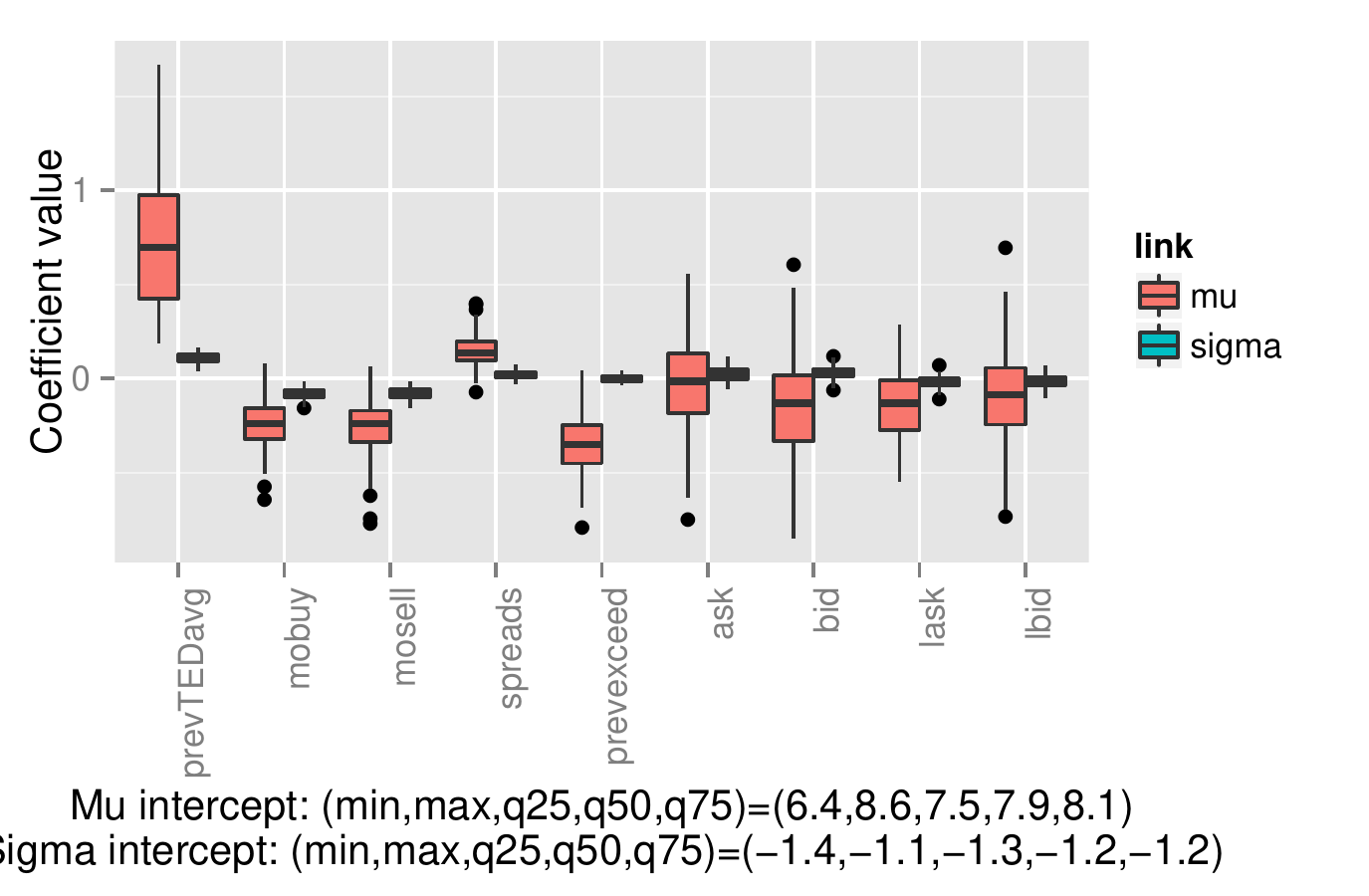}
						\includegraphics[width=0.48\textwidth]{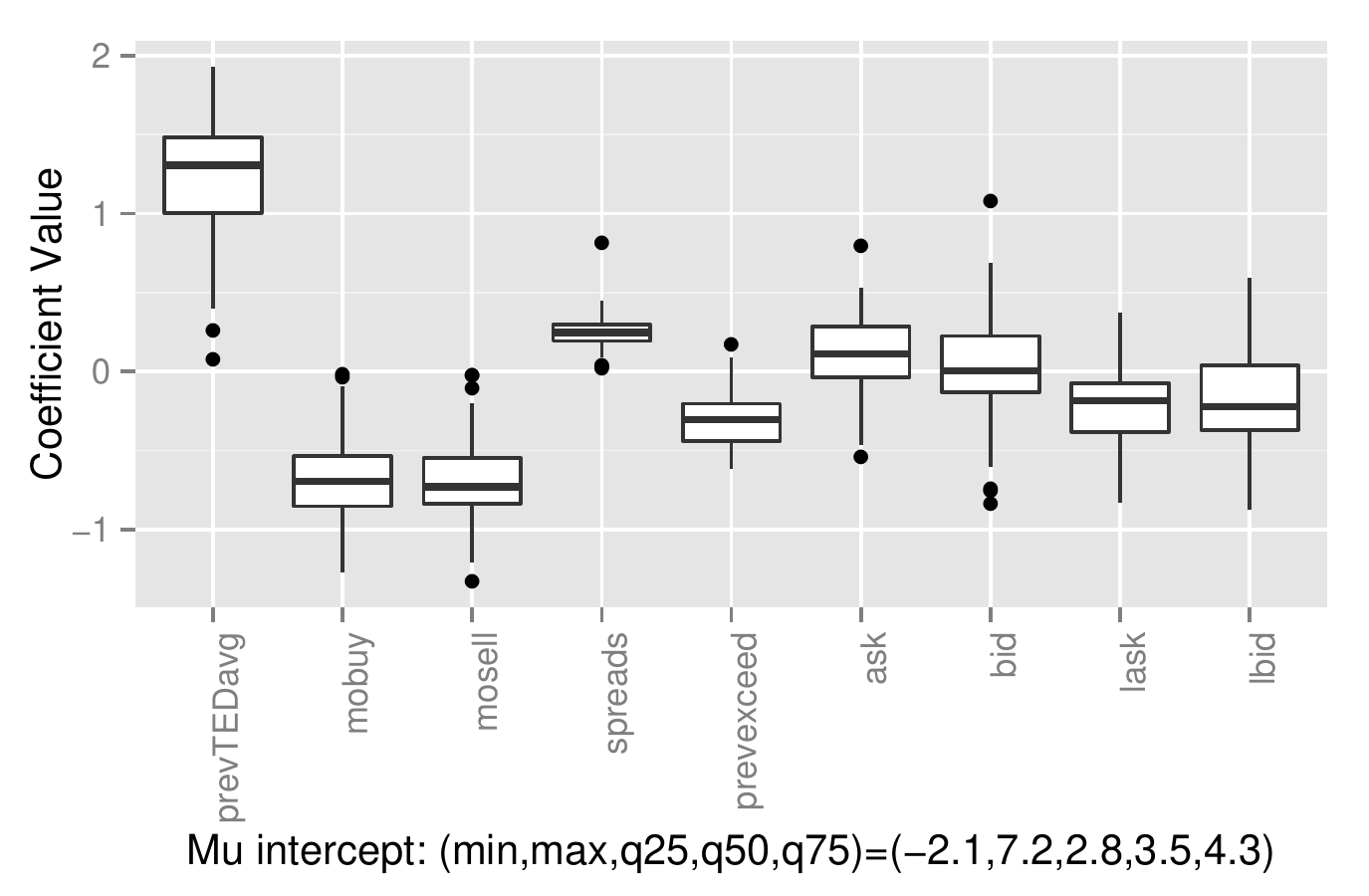}
		\includegraphics[width=0.48\textwidth]{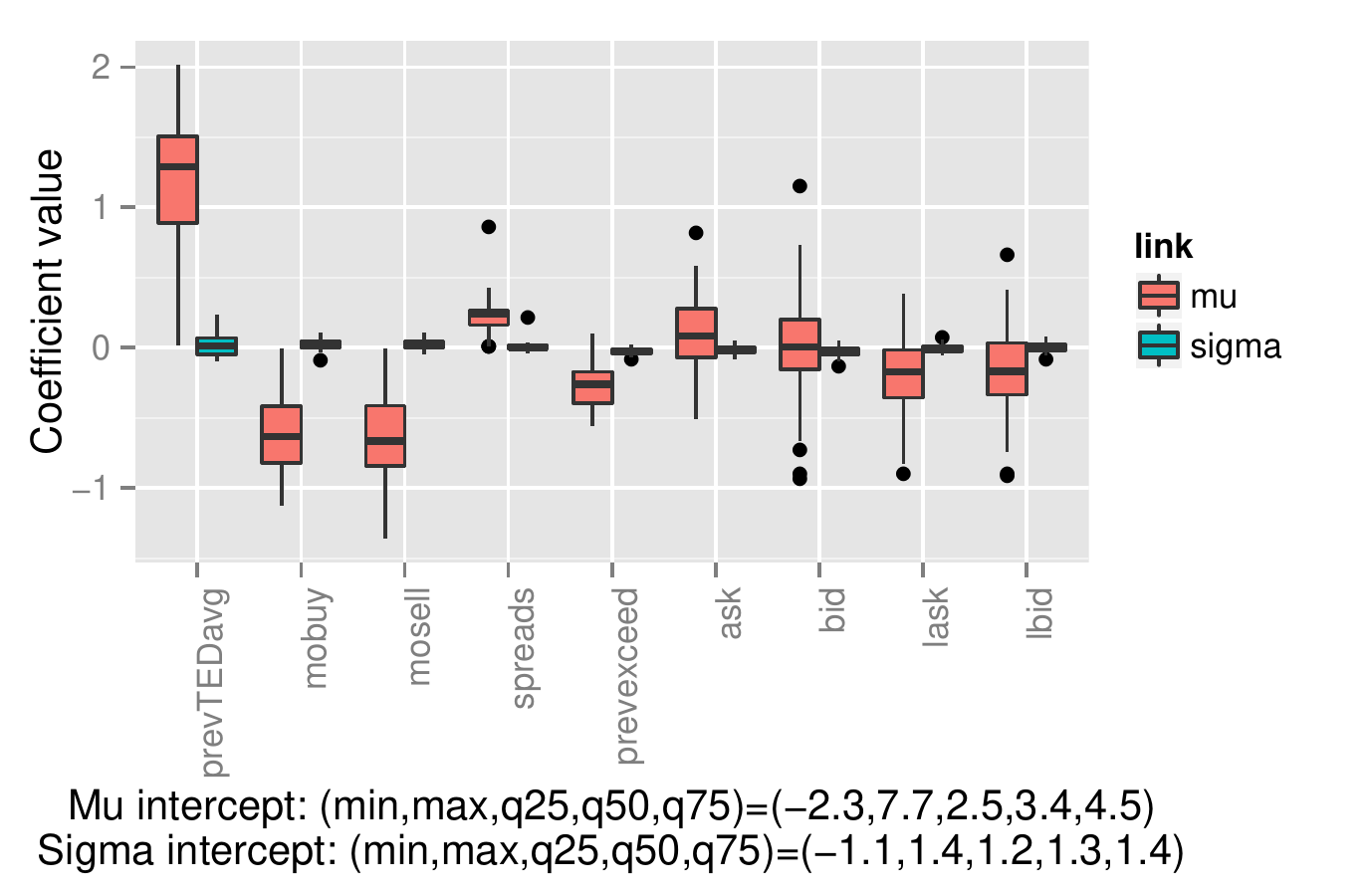}
	\caption{Boxplots of the model coefficients obtained from fitting the regression models with the various distributional assumptions separately for each day in our dataset for the threshold corresponding to the median spread for stock Deutsche Telecom (stock symbol DTEd). (Top): Lognormal. (Middle): Gamma. (Bottom): Weibull. Left: Considering link function for $\mu$ only. Right: Considering link function for $\mu$ and $\sigma$.  }
	\label{fig:boxplotallcoefsallmodels}
	\end{center}
\end{figure}

Since we have obtained model fits for every model subspace, and for every day in our dataset, we can investigate the inter-day variation of the coefficients, as well as their magnitude and sign over time. In Figure \ref{fig:boxplotallcoefsallmodels}, we summarise these results for the best fitting model on each day. The plots demonstrate for each model distribution assumption the following features: 1) the variation in each coefficient in the link functions for $\mu$ and $\sigma$ and 2) the coefficient sign, and thus its interpretation with regards to how it influences these parameters, generally related to the resilience mean and variance. 
%

We note that the signs of the coefficients generally agree for the models under the different distributional assumptions. In particular, the {\em prevTEDavg} covariate, which is an average of the last 5 log TED observations, and generally has a positive coefficient, is thus associated with a slower return to the threshold liquidity level. Thus, our model indicates that the expected TED over a particular threshold will be larger, when the duration of similar exceedances in the near past has been longer. We also find that the instantaneous spread covariate (i.e. the value of the spread at the moment when it first exceeds the threshold) appears frequently in the best model and has a positive coefficient (and would also increase the expected TED). This results matches our intuition, as the wider the spread just after an event at time $T_i+$, the longer we would expect the spread exceedance to last, on average. 

Of particular interest are the {\em mobuy} and {\em mosell} covariates, i.e. dummy variables indicating whether the exceedance resulted from a buy or sell market order respectively (if both are zero, then the exceedance was a result of a cancellation). For the majority of assets, such as Deutsche Telekom, for which results are presented in Figure \ref{fig:boxplotallcoefsallmodels}, the coefficients are generally found to be negative, indicating that exceedances from market orders are associated with an decrease in the expected TED, compared to cancellations. For a small number of assets, such as Credit Agricole we have noted that the opposite effect is found.    

\section{Liquidity drought extremes}\label{sec:quantile}

We present here an application of the model as a regulatory tool for the monitoring of liquidity. Similar to the previous setting, we would expect that regulatory bodies are interested in ensuring uninterrupted liquidity, as it is an integral part of a fair and orderly market. However, they would probably focus on the extreme liquidity levels that occur, and the durations of these extreme events. 

For this application, instead of the conditional mean response of the observation variable, we now consider conditional quantiles of the response. That is, if a TED event occurs in a (stationary) LOB regime, given covariates $\bm{x}$, we can make a prediction about the $(1-\alpha)$-th quantile, eg. the 90th quantile of the response: This is the duration of time such that there is a 90\% probability under the model that liquidity will return to the threshold level in this period. To understand the use of the model we developed in the manner of a quantile relationship, we explain briefly how to reinterpret the GAMLSS regression model, specifically in the case of the generalised gamma distribution family we adopt in this paper, as a quantile regression model structure, following the developments discussed in \cite{noufaily2013parametric}. 

In general when performing a quantile regression study, where one links the quantile behaviour of an observed response variable, in our case the TED random variables for liquidity resilience, to a set of covariates, it is achieved by either adopting a non-parametric or a parametric framework. The most common approach is to consider the non-parametric quantile regression approach, where one estimates regression coefficients without making assumptions on the distribution of the response, or equivalently the residuals. If $Y_{i}>0$ is a set of observations and $\bm x_{i}=(1,x_{i1},\dots, x_{im})$ is a vector of covariates that describe $Y_{i}$, the quantile function for the log transformed data $Y_{i}^*=\ln Y_{i} \in \Re$ is
\begin{equation}
Q_{Y^*}(u|\bm x_{i})=\alpha_{0,u}+\sum\limits_{k=1}^m \alpha_{k,u} \, x_{ik} \label{NonPar}
\end{equation}
where $u \in (0,1)$ is the quantile level, $\bm \alpha_u=(\alpha_{0,u},\dots, \alpha_{k,u})$ are the linear model coefficients for quantile level $u$ which are estimated by solving 
\begin{equation} \label{EqnLossNonParQuant}
\min_{\alpha_{0,u}, \dots, \alpha_{m,u}} \sum_{i\le I} \rho_u(\epsilon_{i})=\sum_{i \le I} \epsilon_{i} [u-I(\epsilon_{i}<0)]
\end{equation}
where $\epsilon_{i}=y_{i}^*-\alpha_{0,u}-\sum\limits_{k=1}^m \alpha_{k,u} \, x_{ik}$. Then the quantile function for the original data is $Q_{Y}(u|\bm x_{i})=\exp(Q_{Y^*}(u|\bm x_{i}))$. 

It was realised by \cite{koenker1999goodness} and \cite{yu2001bayesian} 
%
that the parameter estimates of $\bm \alpha_u$ obtained by minimizing the loss function in (\ref{EqnLossNonParQuant}) will be equivalent to the maximum likelihood estimates of $\bm \alpha_u$, when $Y_{i}^*$ follows the Asymetric Laplace (AL) proxy distribution with pdf:
\begin{equation} \label{PdfAL}
f(y_{i}^*|\mu_{i},\sigma_{i}^2,p)=\frac{p(1-p)}{\sigma_{i}}\exp\left( -\frac{(y^*_{i}-\mu^*_{i})}{\sigma_{i}}[p-I(y_{i}^* \le \mu_{i})] \right)
\end{equation}
where the location parameter or mode $\mu_{i}^*$ equals to $Q_{Y^*}(u|\bm x_{i})$ in (\ref{NonPar}), the scale parameter $\sigma_{i}>0$ and the skewness parameter $p\in (0,1)$ equals to the quantile level $u$. Since the pdf (\ref{PdfAL}) contains the loss function (\ref{EqnLossNonParQuant}), it is clear that parameter estimates which maximize (\ref{PdfAL}) will minimize (\ref{EqnLossNonParQuant}).

In this formulation the AL distribution represents the conditional distribution of the observed dependent variables (responses) given the covariates. More precisely, the location parameter $\mu_{i}$ of the AL distribution links the coefficient vector $\bm \alpha_u$ and associated independent variable covariates in the linear regression model to the location of the AL distribution. It is also worth noting that under this representation it is straightforward to extend the quantile regression model to allow for heteroscedasticity in the response which may vary as a function of the quantile level $u$ under study. To achieve this, one can simply add a regression structure linked to the scale parameter $\sigma_{i}$ in the same manner as was done for the location parameter.

Equivalently, we assume $Y_{i}^*$ conditionally follows an AL distribution denoted by $Y_{i}^*\sim AL(\mu^*_{i}, \sigma_{i}^2,u)$. Then
\begin{equation}
Y_{i}^* =\mu^*_{i}+\epsilon^*_{i} \sigma_{i}
\end{equation}
where $\epsilon^*_{i} \sim AL(0,1,u)$, $\mu^*_{i}=\alpha_{0,u}+\sum\limits_{k=1}^{m} \alpha_{k,u} \, x_{ik}$,  $\sigma_{i}^2=\exp(\beta_{0,u}+\sum\limits_{k=1}^{\nu} \beta_{k,u} \, s_{ik})$ and $s_{ik}$ are covariates in the variance function. 

One could indeed consider the resulting ALD model as a GAMLSS model structure which is interpreted as a quantile regression model also. However, there is another sub-class of models for which one can develop a GAMLSS model that will also be associated with a quantile regression structure, not necessarily in the ALD family. In this paper we consider again the generalised gamma distribution family of GAMLSS structures and we observe that one can obtain the quantile function of this family of models in closed form, which is again a form of quantile regession since it directly relates the quantile function of the TED response to the covariates. 

In particular if we consider that the TED responses are modelled according to the GAMLSS regression structure under one of the available parameterizations discussed previously, such as the Generalised Gamma distribution
\begin{equation}
f_{\tau}(\tau; b(\bm{x}),a(\bm{x}),k(\bm{x})) = \frac{b}{\Gamma(k)}\frac{\tau^{b k - 1}}{a^{b k}}\exp\left(-\left(\frac{\tau}{a}\right)^{b} \right), k > 0, a >0, b > 0
\end{equation}
where we note that the parameters $b,a,k$ can be made to be functions of the covariates $\bm{x}$ under the GAMLSS structure. From this regression relationship, one can obtain the conditional quantile function for a given quantile level $u$, as determined by \cite{noufaily2013parametric}. Obtaining this quantile function for the conditional response given the covariates, linked to the response through the parameters, is achieved by representing the log generalized gamma distribution's quantile function in terms of a base quantile function, in this case given by a gamma distribution with specifically selected shape and scale parameters. The required transformation of the analytic closed-form quantile function of a Gamma random variable, denoted by $G^{-1}$, with shape $u$ and scale $k(\bm{x})$, then gives the conditional expression
\begin{equation} \label{EqnCondQReg}
Q\left(u;\bm{x}\right) = a(\bm{x})
\left[\left(\frac{1}{k(\bm{x})}\right)G^{-1}\left(u;u,k(\bm{x})\right)\right]^{\frac{1}{b(\bm{x})}},
\end{equation}
where $a(\bm{x})$, and we recall that this parameterisation corresponds to that of the {\em gamlss} package with $b=\nu$, $a=\mu \theta^{-\frac{1}{\nu}}$ and $k=\theta$, with $\theta=\frac{1}{\sigma^2 \nu^2}$ and where one or more of $\mu,\sigma,\nu$ may be functions of $\bm{x}$ since one or more of the parameters $k,b,a$ can be made functions of the covariates $\bm{x}$. 

\begin{figure}[ht!]
	\begin{center}
	\includegraphics[width=0.49\textwidth]{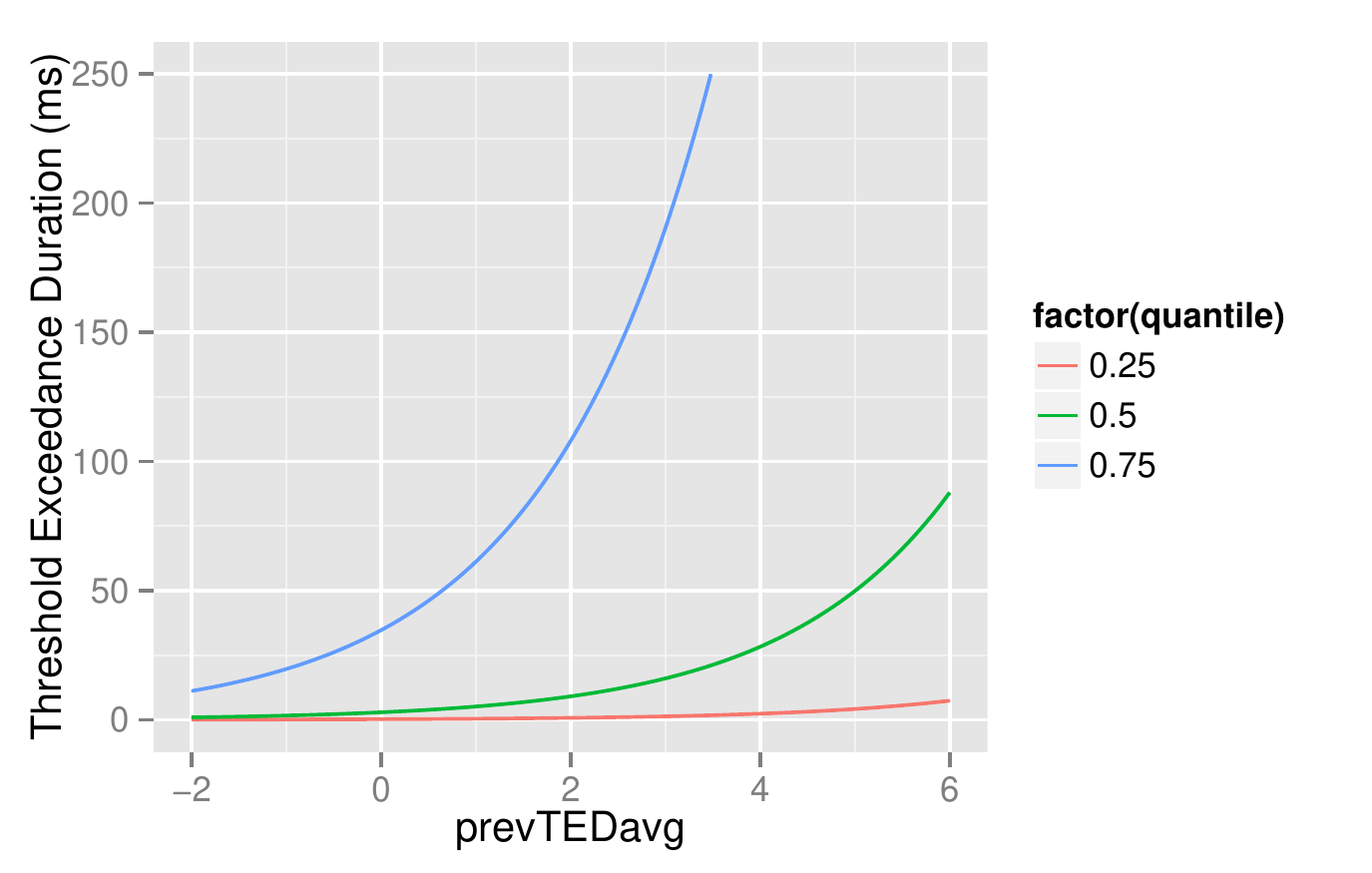}
	\includegraphics[width=0.49\textwidth]{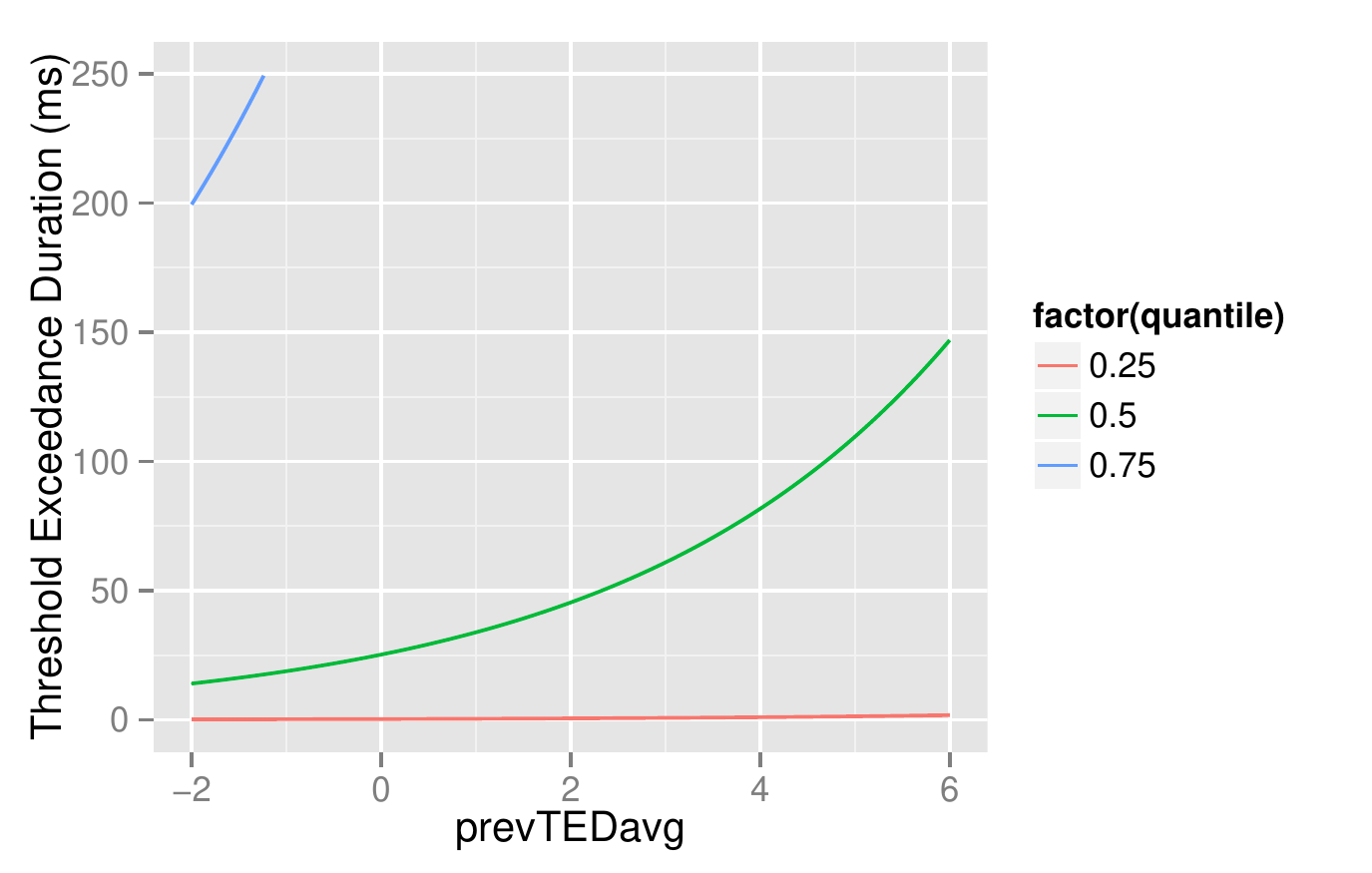}
	\includegraphics[width=0.49\textwidth]{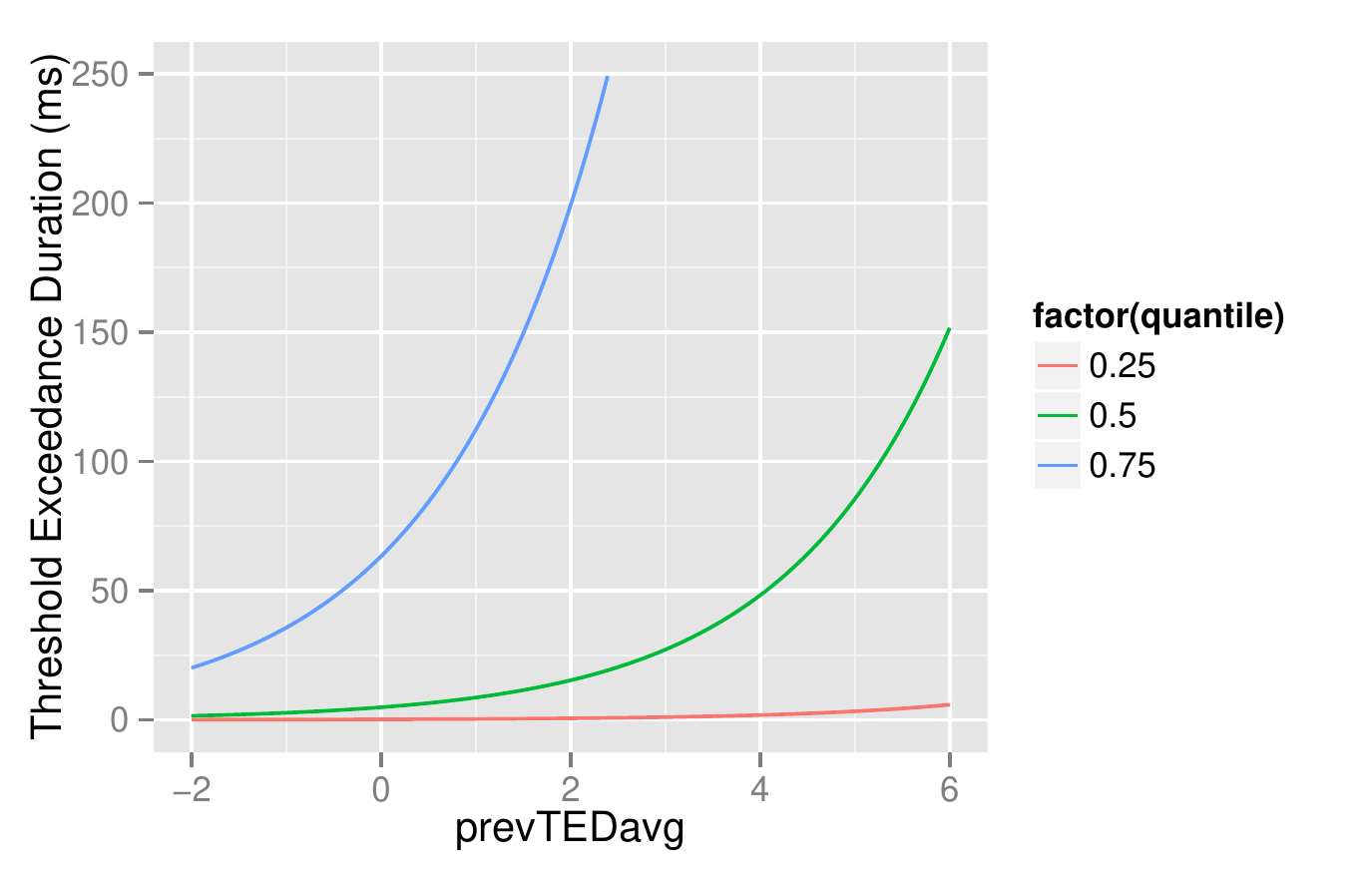}
	\includegraphics[width=0.49\textwidth]{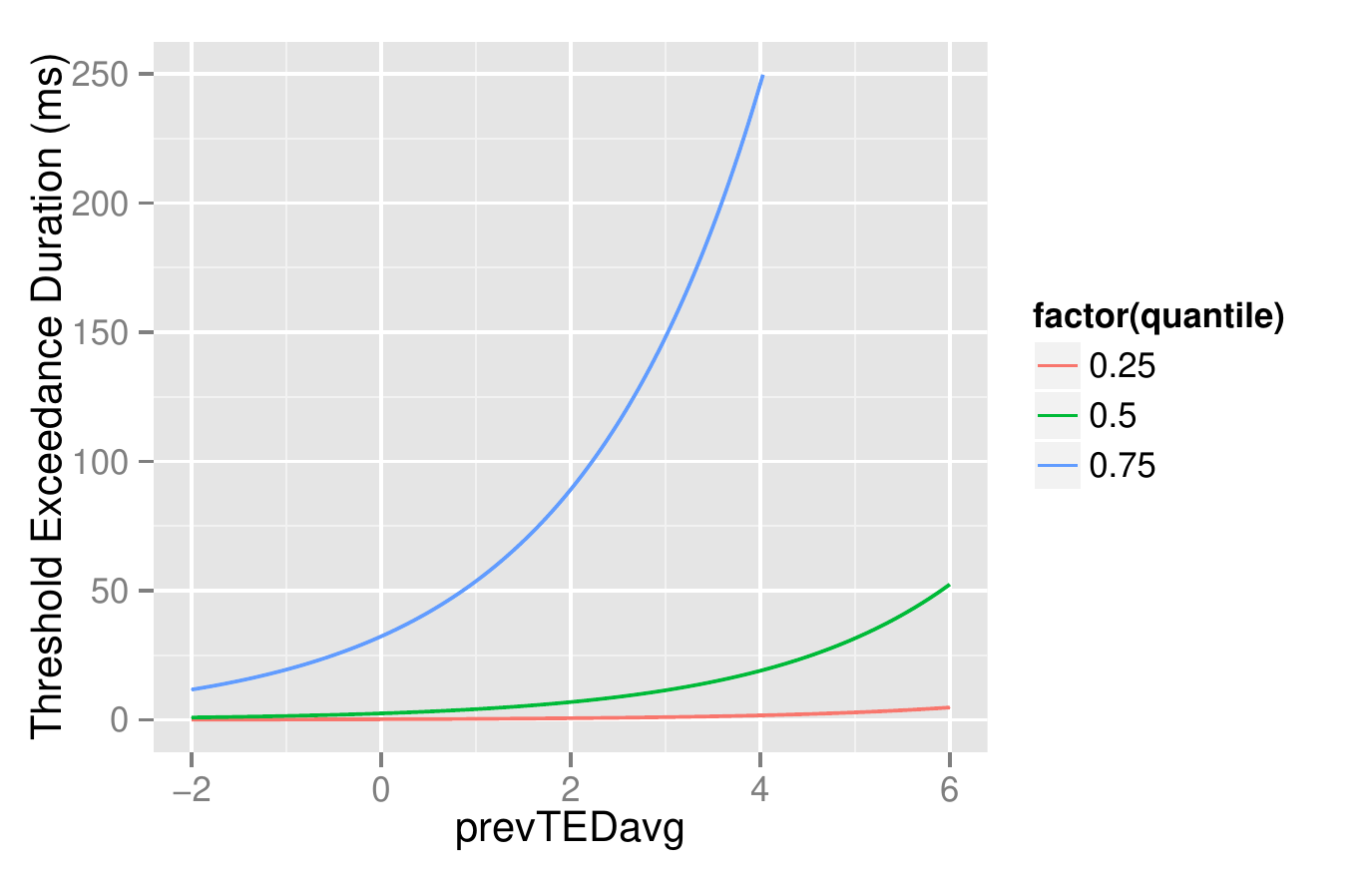}
	\caption{Quantile plots: Upper left: Lognormal. Upper right: Gamma. Lower left: Weibull. Lower right: Generalised gamma.}
	\label{fig:quantile}
	\end{center}
\end{figure}

\begin{figure}[ht!]
	\begin{center}
	\includegraphics[width=0.45\textwidth]{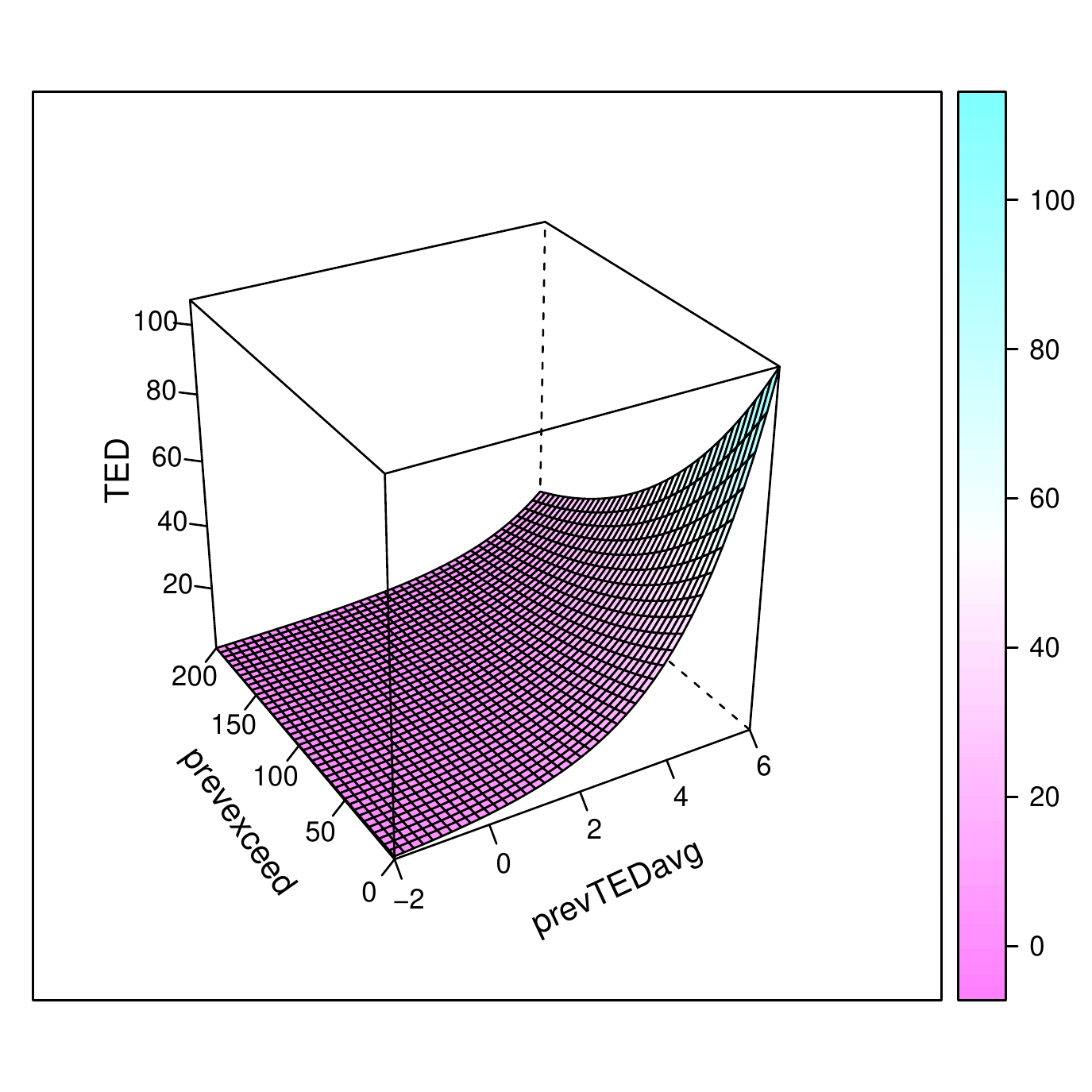}
	\includegraphics[width=0.45\textwidth]{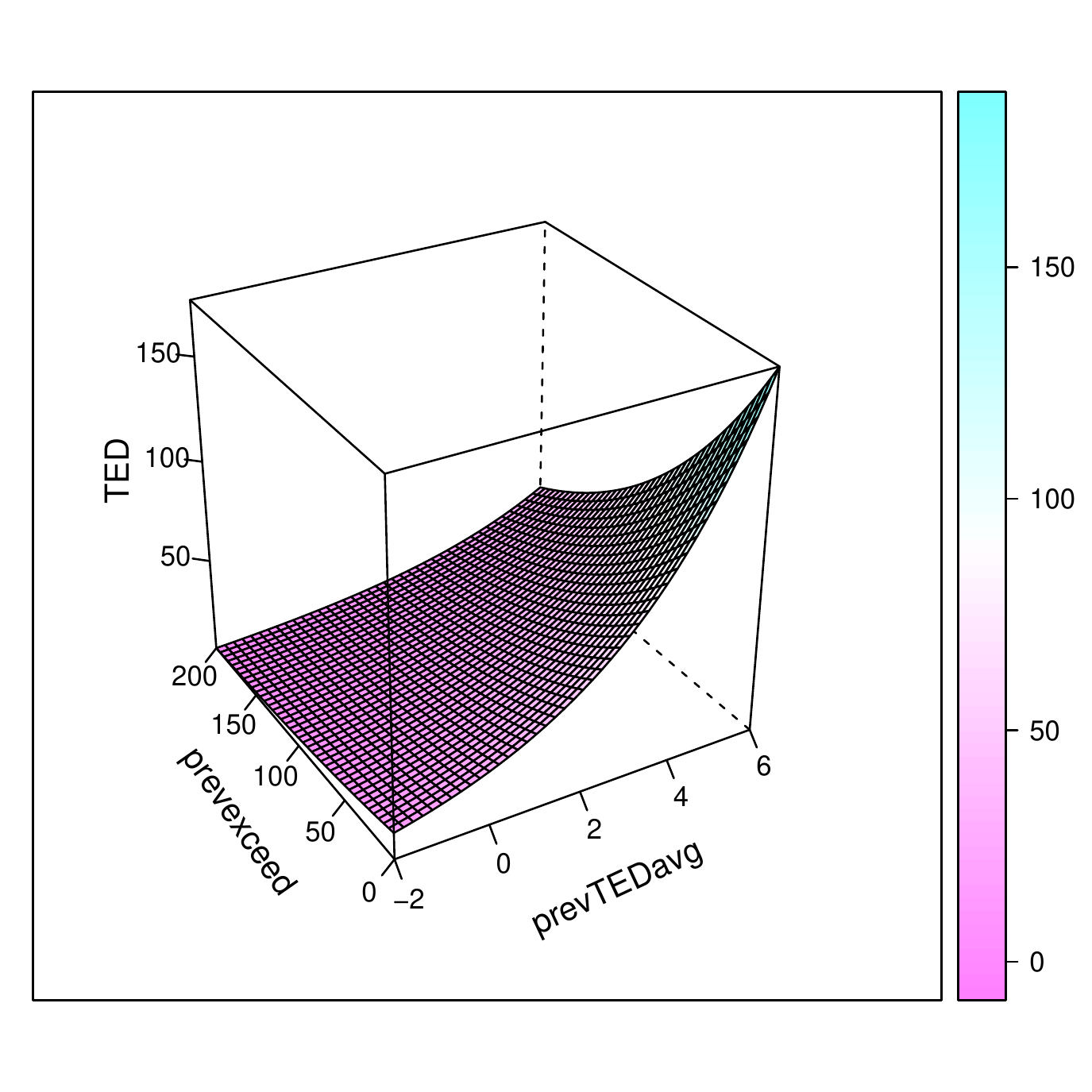}
	\includegraphics[width=0.45\textwidth]{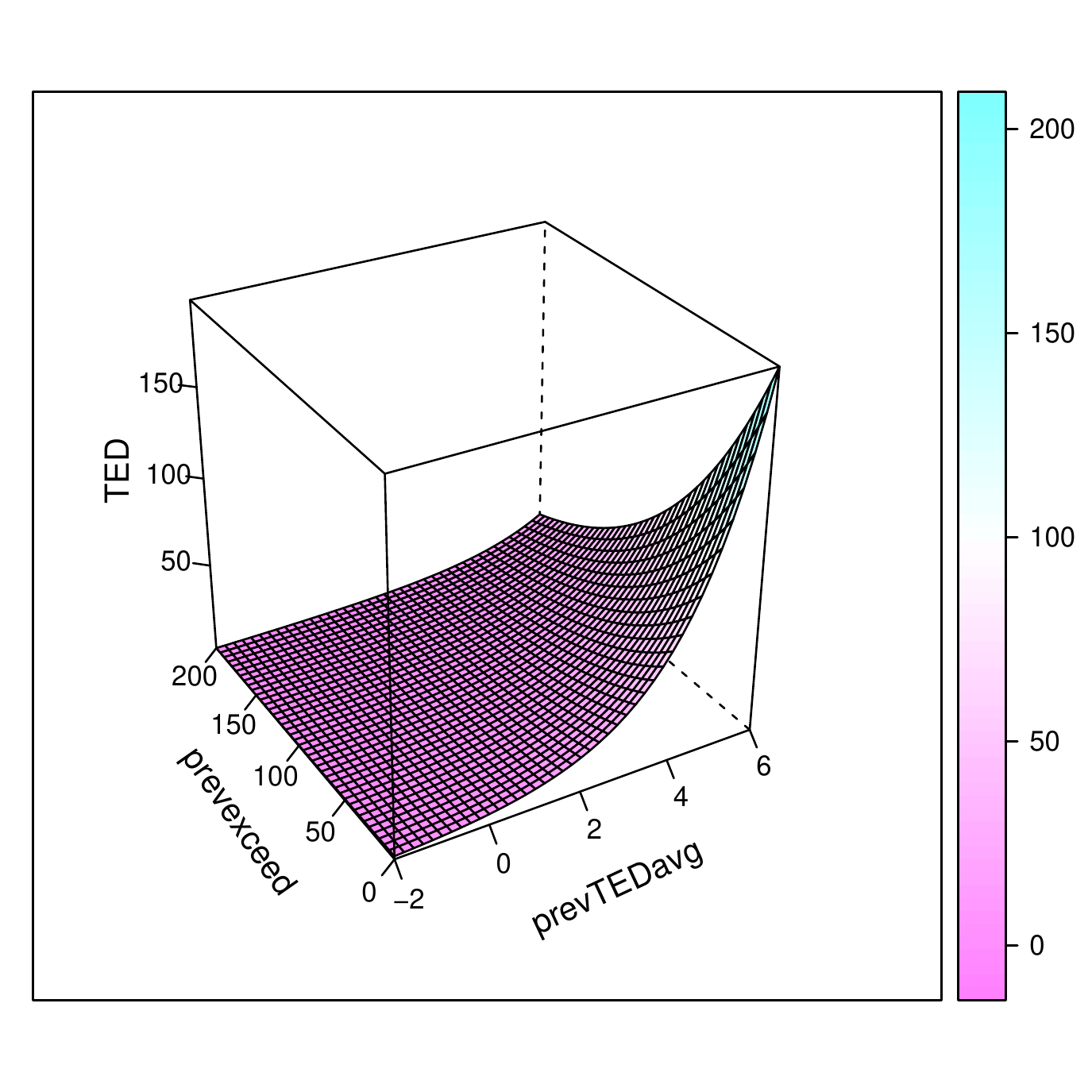}
	\includegraphics[width=0.45\textwidth]{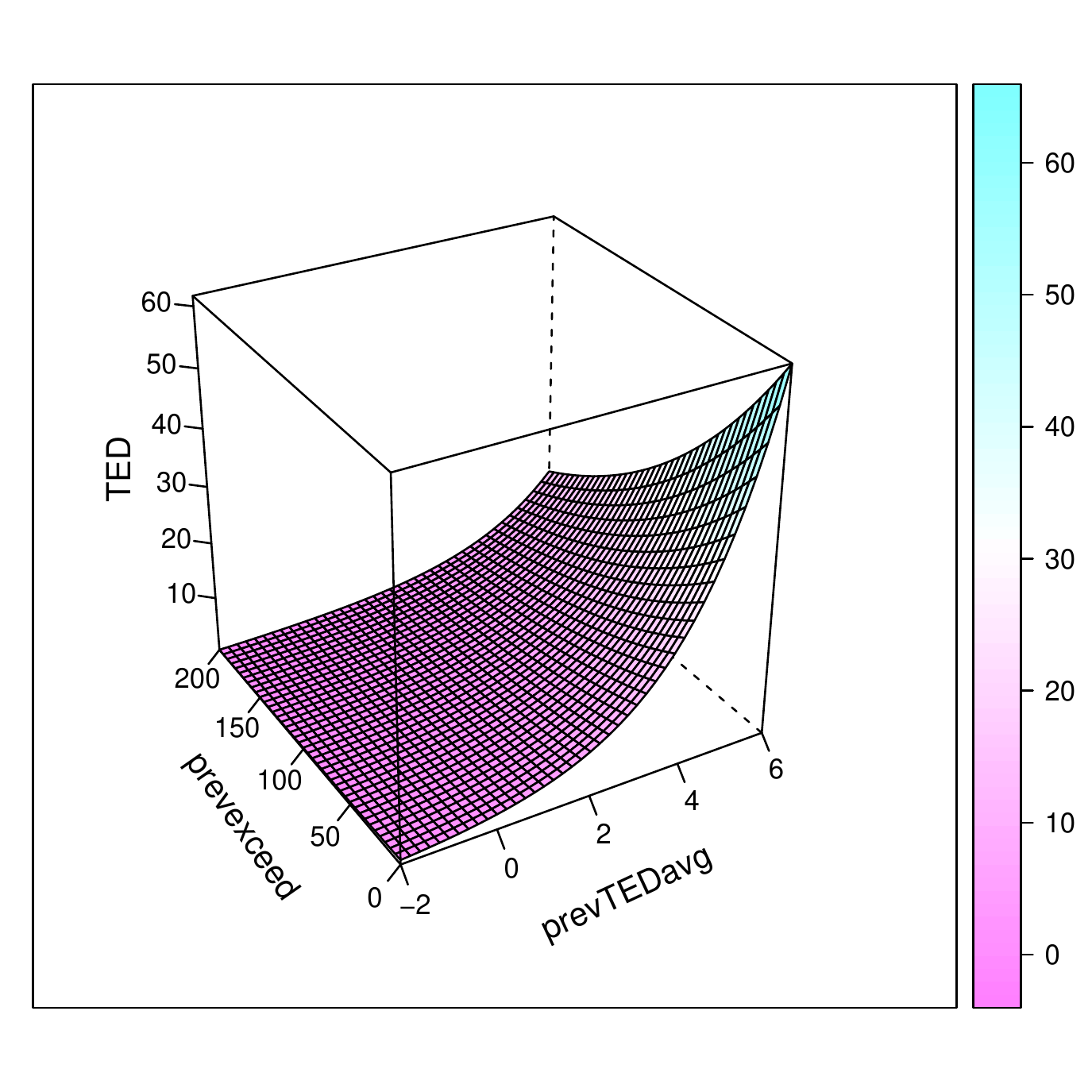}
	\caption{Quantile plots for lognormal,gamma,weibull and generalised gamma when varying 2 covariates.}
	\label{fig:quantilebivariate}
	\end{center}
\end{figure}

\begin{figure}[ht!]
	\begin{center}
	\includegraphics[width=0.45\textwidth]{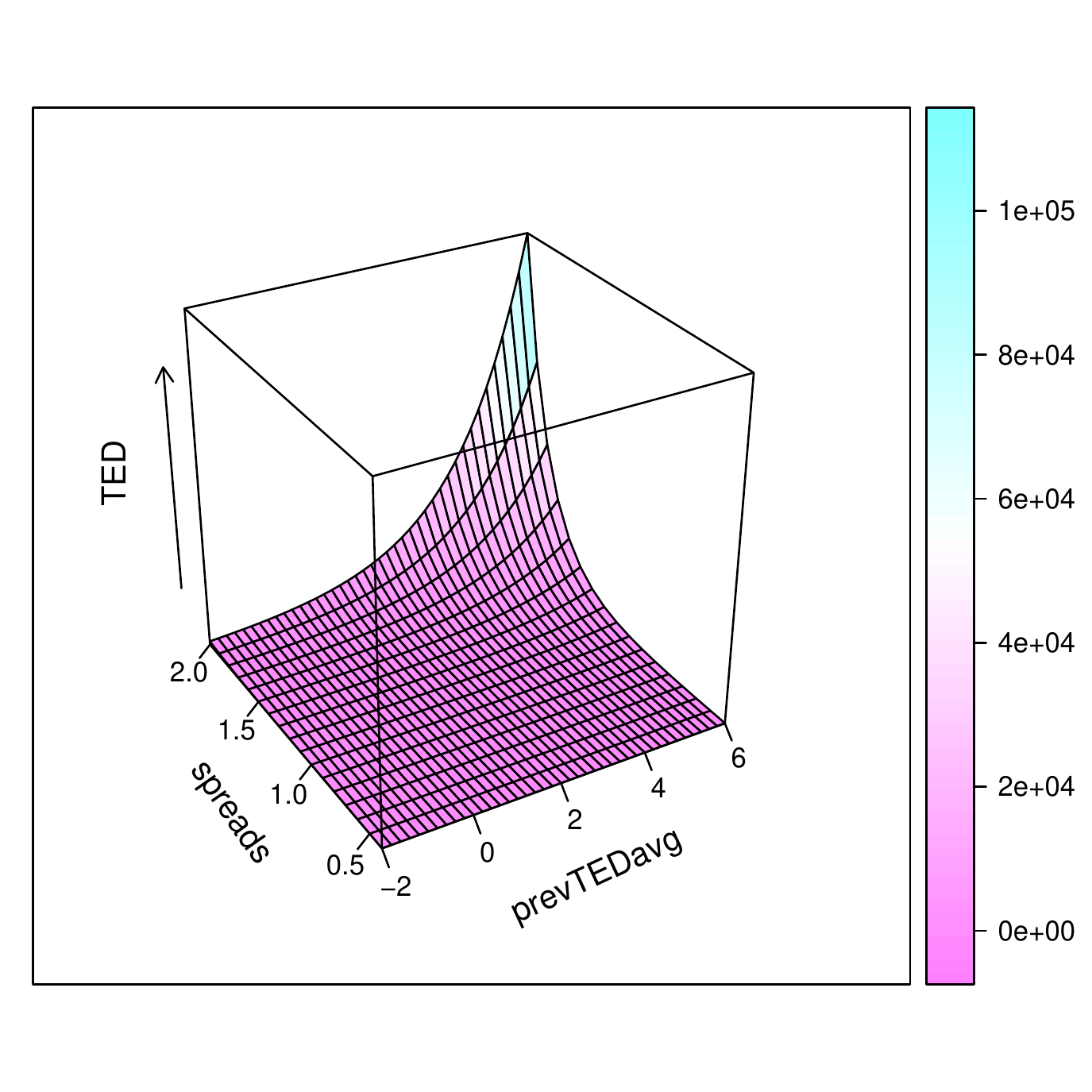}
	\includegraphics[width=0.45\textwidth]{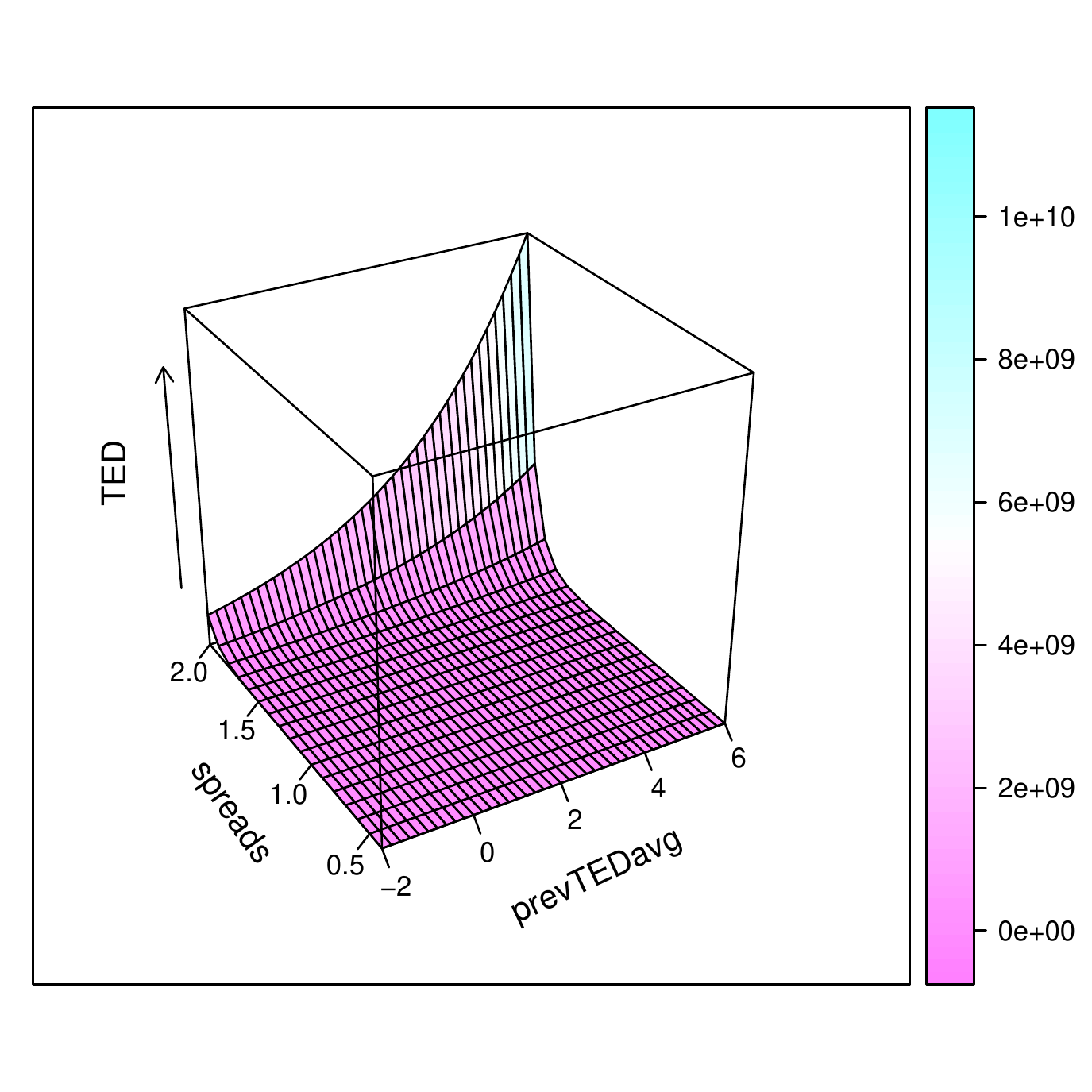}
	\includegraphics[width=0.45\textwidth]{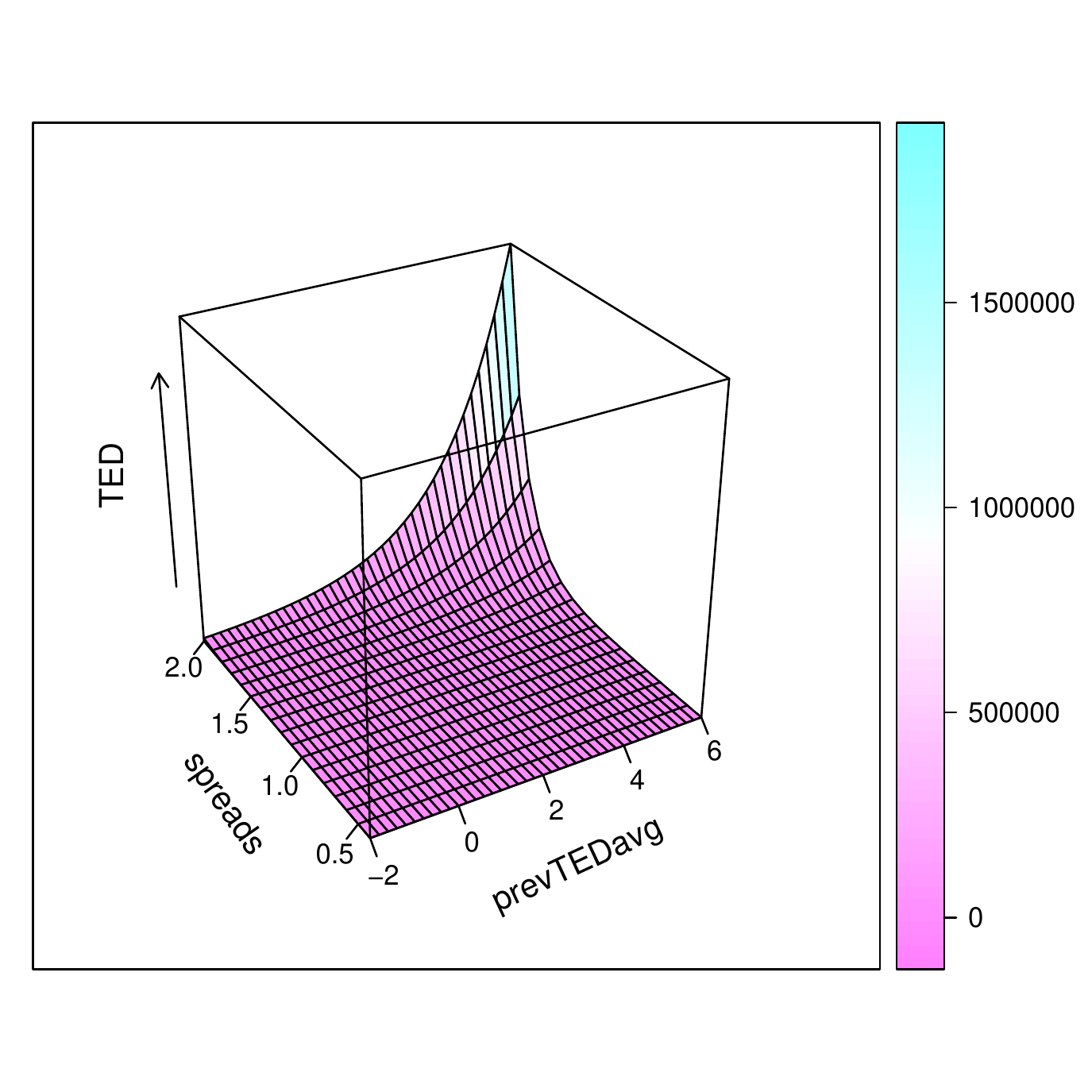}
	\includegraphics[width=0.45\textwidth]{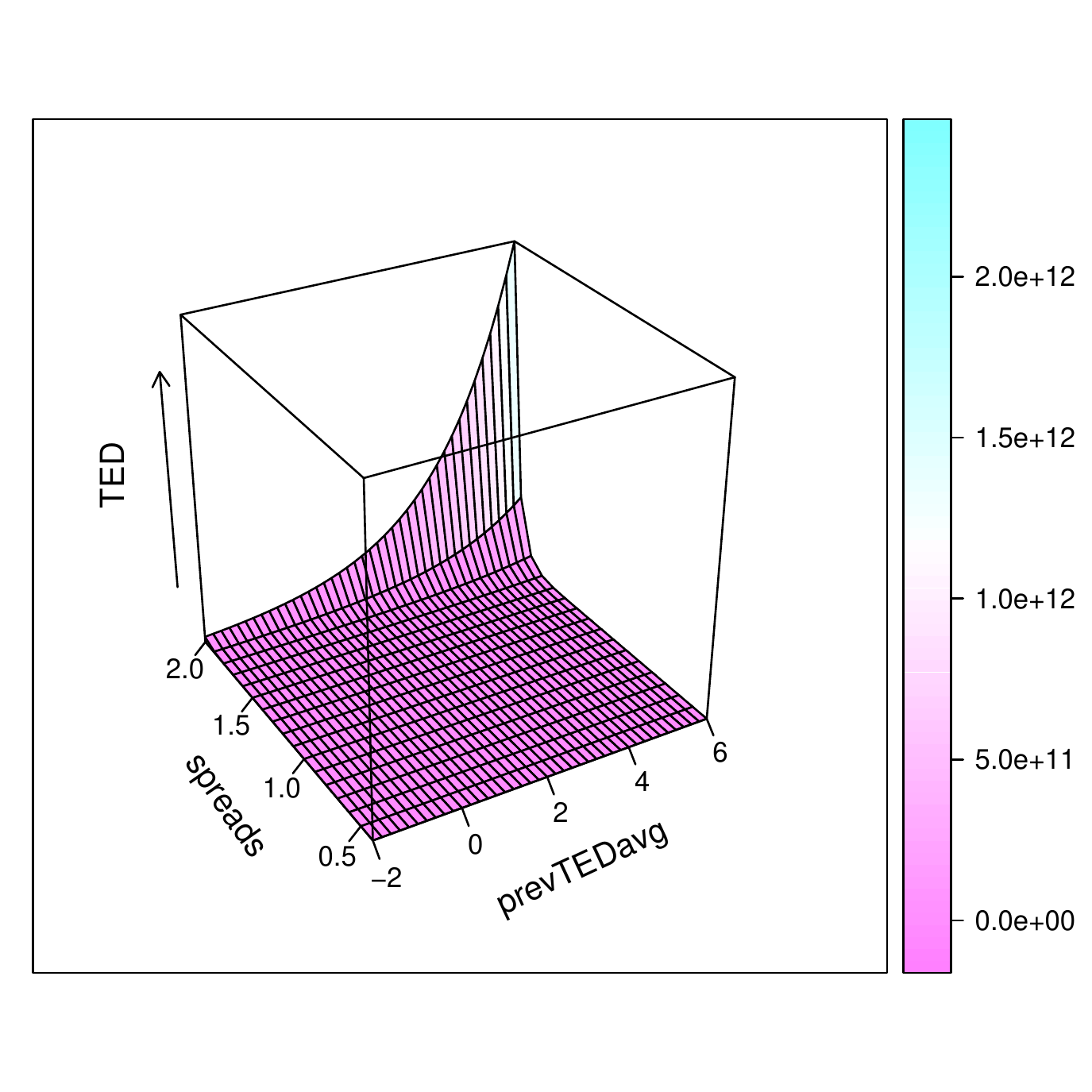}
	\caption{Quantile plots for lognormal,gamma,weibull and generalised gamma when varying 2 covariates.}
	\label{fig:quantilebivariate2}
	\end{center}
\end{figure}


To start with, we obtain the conditional quantile level of the TED, assuming that covariates take median intra-day values. The TED is defined as before, using the spread as the liquidity measure and the median of the empirical distribution as the threshold. We then allow covariate $prevTEDavg$ to vary within a range of typical intra-day values and obtain the conditional quantile levels for the four distributions we have considered in Figure \ref{fig:quantile}. We note here that these quantile functions have been obtained with model parameters from a GAMLSS structure, where we only considered the link function pertaining to the first parameter. We can see that this structure separates the effects of covariates and quantile levels on the quantile of the TED, as covariates only enter into parameter $a$ above. 


Figures \ref{fig:quantilebivariate} and \ref{fig:quantilebivariate2} show the quantile surface obtained when we allow two covariates to vary. We note from Figure \ref{fig:quantilebivariate2} that when both covariates {\em prevTEDavg} and {\em spreads} take extreme values, this leads to a vast increase in the median TED level under our model for all distributional assumptions.

Such an analysis is useful in understanding how extreme levels of particular covariates affect quantile levels of the TED. We can therefore understand how the different quantile surfaces for the TED behave for these extreme values of the spread and above. 
This enables regulators to identify which are the most important covariates associated with an increase in extreme periods of illiquidity (that is, where the liquidity measure remains above the threshold for extended period of time). In addition, to the extent that a covariate taking extreme values is considered a scenario in which the LOB is stressed, a regulator can make inferences about the duration of relative illiquidity under such stressed conditions.

In non-stressed conditions, that is, where covariates take what would be considered to be `normal' values, regulators may be interested in the range of probable values of the TED. Obtaining high quantile levels of the TED under the model could then help them identify situations which fall outside this range, which may be due to a change in the LOB regime or due to a particular event that will require their intervention. 


%
%
%
\section{Proposals and conclusion}
\label{sec:conc}
Given the intra-day variation in liquidity demand, we propose that quoting requirements of Designated Market Makers / Designated Sponsors be amended so as to include a provision for liquidity replenishment after a shock. We have shown that the Threshold Exceedance Duration (TED)\cite{panayi2014market} is a good metric for the speed of this liquidity replenishment, and indeed it has been defined so as to be able to incorporate any liquidity measure (e.g. spread, XLM) an exchange may be interested in, and any liquidity threshold that would indicate that there is sufficient liquidity in the asset. We do not suggest an explicit target for the TED for each asset, but we suggest that this should vary according to the asset's liquidity, as do current quoting requirements regarding maximum spreads and minimum posted volumes.

We have presented a comprehensive study of different regression structures that could be used to model the variation in the TED. An appropriate modelling structure would be invaluable for both the operators of the exchange and the market makers who are subject to the quoting requirements. The former, because they could use it to determine an appropriate target level of the TED, given the prevailing conditions. The latter, because when they act to replenish liquidity, it is possible that they have several options about the way in which they do it, and the model could prescribe the method that would most improve resilience in market liquidity.

In our modelling we employed various regression structures, starting from simple log-linear models to Generalised Linear Models (GLMs) and GAMLSS models. Under these approaches, we compared the explanatory power of Lognormal, Gamma, Weibull and Generalised Gamma models. We also evaluated the additional explanatory power of allowing covariates to affect each of the parameters of the distribution, as in the GAMLSS structure. 

We determined that while the Generalised Gamma model had, in most cases, the highest explanatory power of the distributions considered, its advantage over the Lognormal distribution was minimal at best. In addition, considering also a link function for a second parameter in the model increased explanatory power, but again the increase was perhaps not sufficient to justify the more involved GAMLSS modelling approach. Summarising, a simple log-linear structure is, in our opinion, the recommended approach to modelling the TED, as its estimation is very robust and its explanatory power similar to much more flexible models, such as the Generalised Gamma model.

\begin{table}[ht]
\centering
\begin{tabular}{rllll}
  \hline
 Name & Symbol & Country & Sector \\ 
  \hline
CREDIT AGRICOLE & ACAp & FRANCE & Banking Services \\ 
  ALLIANZ & ALVd & GERMANY & Insurance \\ 
  BAYER & BAYNd & GERMANY & Biotechnology / Pharmaceuticals \\ 
  BIC & BBp & FRANCE & Commercial Services / Supplies \\ 
  BMW & BMWd & GERMANY & Automobiles / Auto Parts \\ 
  DANONE & BNp & FRANCE & Food / Tobacco \\ 
  AXA & CSp & FRANCE & Insurance \\ 
  DAIMLER & DAId & GERMANY & Automobiles / Auto Parts \\ 
  DEUTSCHE BANK & DBKd & GERMANY & Banking Services \\ 
  JCDECAUX & DECp & FRANCE & Media / Publishing \\ 
  DEUTSCHE TELEKOM & DTEd & GERMANY & Telecommunications Services \\ 
  GROUPE EUROTUNNEL & GETp & FRANCE & Rails / Roads Transportation \\ 
  PUMA & PUMd & GERMANY & Textiles / Apparel \\ 
  HERMES INTL. & RMSp & FRANCE & Textiles / Apparel \\ 
  RENAULT & RNOp & FRANCE & Automobiles / Auto Parts \\ 
  SKY DEUTSCHLAND & SKYDd & GERMANY & Media / Publishing \\ 
  AXEL SPRINGER & SPRd & GERMANY & Media / Publishing \\ 
  TUI & TUI1d & GERMANY & Hotels / Entertainment Services \\ 
  UBISOFT ENTM. & UBIp & FRANCE & Leisure Products \\ 
  VOLKSWAGEN & VOWd & GERMANY & Automobiles / Auto Parts \\ 
   \hline
\end{tabular}
\caption{Information about the 20 European stocks used in the study.}
\label{tab:stockinfo}
\end{table} 

\vspace{3em}
\textbf{Declaration of interest}

The authors report no declarations of interest. 
\newpage
\bibliographystyle{plainnat}
 \bibliography{all}

\begin{thebibliography}{35}
\providecommand{\natexlab}[1]{#1}
\providecommand{\url}[1]{\texttt{#1}}
\expandafter\ifx\csname urlstyle\endcsname\relax
  \providecommand{\doi}[1]{doi: #1}\else
  \providecommand{\doi}{doi: \begingroup \urlstyle{rm}\Url}\fi

\bibitem[Abhyankar et~al.(1997)Abhyankar, Ghosh, Levin, and
  Limmack]{abhyankar1997bid}
Abhay Abhyankar, Dipak Ghosh, Eric Levin, and RJ~Limmack.
\newblock Bid-ask spreads, trading volume and volatility: Intra-day evidence
  from the {London Stock Exchange}.
\newblock \emph{Journal of Business Finance \& Accounting}, 24\penalty0
  (3):\penalty0 343--362, 1997.

\bibitem[Alfonsi et~al.(2010)Alfonsi, Fruth, and Schied]{alfonsi2010optimal}
Aur{\'e}lien Alfonsi, Antje Fruth, and Alexander Schied.
\newblock Optimal execution strategies in limit order books with general shape
  functions.
\newblock \emph{Quantitative Finance}, 10\penalty0 (2):\penalty0 143--157,
  2010.

\bibitem[Benos and Wetherilt(2012)]{benos2012role}
Evangelos Benos and Anne Wetherilt.
\newblock The role of designated market makers in the new trading landscape.
\newblock \emph{Bank of England Quarterly Bulletin}, page~Q4, 2012.

\bibitem[Brockman and Chung(1999)]{brockman1999analysis}
Paul Brockman and Dennis~Y Chung.
\newblock An analysis of depth behavior in an electronic, order-driven
  environment.
\newblock \emph{Journal of Banking \& Finance}, 23\penalty0 (12):\penalty0
  1861--1886, 1999.

\bibitem[Chan et~al.(1995)Chan, Christie, and Schultz]{chan1995market}
Kalok~C Chan, William~G Christie, and Paul~H Schultz.
\newblock Market structure and the intraday pattern of bid-ask spreads for
  {NASDAQ} securities.
\newblock \emph{Journal of Business}, pages 35--60, 1995.

\bibitem[Chlistalla et~al.(2011)Chlistalla, Speyer, Kaiser, and
  Mayer]{chlistalla2011high}
Michael Chlistalla, Bernhard Speyer, Sabine Kaiser, and Thomas Mayer.
\newblock High-frequency trading.
\newblock \emph{Deutsche Bank Research}, pages 1--19, 2011.

\bibitem[De~Jong et~al.(2008)De~Jong, Heller, et~al.]{de2008generalized}
Piet De~Jong, Gillian~Z Heller, et~al.
\newblock \emph{Generalized linear models for insurance data}, volume 136.
\newblock Cambridge University Press Cambridge, 2008.

\bibitem[Garbade(1982)]{garbade1982securities}
Kenneth~D Garbade.
\newblock \emph{Securities markets}.
\newblock McGraw-Hill New York, 1982.

\bibitem[Harris(2002)]{harris2002trading}
Larry Harris.
\newblock \emph{Trading and exchanges: Market microstructure for
  practitioners}.
\newblock Oxford University Press, USA, 2002.

\bibitem[Harvey(1976)]{harvey1976estimating}
Andrew~C Harvey.
\newblock Estimating regression models with multiplicative heteroscedasticity.
\newblock \emph{Econometrica: Journal of the Econometric Society}, pages
  461--465, 1976.

\bibitem[Hasbrouck and Seppi(2001)]{hasbrouck2001common}
Joel Hasbrouck and Duane~J. Seppi.
\newblock Common factors in prices, order flows, and liquidity.
\newblock \emph{Journal of financial Economics}, 59\penalty0 (3):\penalty0
  383--411, 2001.

\bibitem[Karolyi et~al.(2012)Karolyi, Lee, and
  Van~Dijk]{karolyi2012understanding}
G~Andrew Karolyi, Kuan-Hui Lee, and Mathijs~A Van~Dijk.
\newblock Understanding commonality in liquidity around the world.
\newblock \emph{Journal of Financial Economics}, 105\penalty0 (1):\penalty0
  82--112, 2012.

\bibitem[Kirilenko et~al.(2014)Kirilenko, Kyle, Samadi, and
  Tuzun]{kirilenko2014flash}
Andrei~A Kirilenko, Albert~S Kyle, Mehrdad Samadi, and Tugkan Tuzun.
\newblock The flash crash: The impact of high frequency trading on an
  electronic market.
\newblock 2014.

\bibitem[Koenker and Machado(1999)]{koenker1999goodness}
Roger Koenker and Jose~AF Machado.
\newblock Goodness of fit and related inference processes for quantile
  regression.
\newblock \emph{Journal of the american statistical association}, 94\penalty0
  (448):\penalty0 1296--1310, 1999.

\bibitem[Korajczyk and Sadka(2008)]{korajczyk2008pricing}
Robert~A Korajczyk and Ronnie Sadka.
\newblock Pricing the commonality across alternative measures of liquidity.
\newblock \emph{Journal of Financial Economics}, 87\penalty0 (1):\penalty0
  45--72, 2008.

\bibitem[Kyle(1985)]{kyle1985continuous}
Albert~S Kyle.
\newblock Continuous auctions and insider trading.
\newblock \emph{Econometrica: Journal of the Econometric Society}, pages
  1315--1335, 1985.

\bibitem[Lawless(1980)]{lawless1980inference}
Jerry~F Lawless.
\newblock Inference in the generalized gamma and log gamma distributions.
\newblock \emph{Technometrics}, 22\penalty0 (3):\penalty0 409--419, 1980.

\bibitem[Lipton et~al.(2013)Lipton, Pesavento, and
  Sotiropoulos]{lipton2013trade}
Alexander Lipton, Umberto Pesavento, and Michael~G Sotiropoulos.
\newblock Trade arrival dynamics and quote imbalance in a limit order book.
\newblock \emph{arXiv preprint arXiv:1312.0514}, 2013.

\bibitem[Lo et~al.(2002)Lo, MacKinlay, and Zhang]{lo2002econometric}
Andrew~W Lo, A~Craig MacKinlay, and June Zhang.
\newblock Econometric models of limit-order executions.
\newblock \emph{Journal of Financial Economics}, 65\penalty0 (1):\penalty0
  31--71, 2002.

\bibitem[Lumley(2004)]{lumley2004leaps}
Thomas Lumley.
\newblock The leaps package.
\newblock \emph{The R project for statistical computation}, 2004.

\bibitem[McCullagh and Nelder(1989)]{mccullagh1989generalized}
Peter McCullagh and John~A Nelder.
\newblock \emph{Generalized linear models}, volume~37.
\newblock CRC press, 1989.

\bibitem[Menkveld and Wang(2013)]{menkveld2013designated}
Albert~J Menkveld and Ting Wang.
\newblock How do designated market makers create value for small-caps?
\newblock \emph{Journal of Financial Markets}, 16\penalty0 (3):\penalty0
  571--603, 2013.

\bibitem[Nelder and Baker(1972)]{nelder1972generalized}
John~A Nelder and RJ~Baker.
\newblock Generalized linear models.
\newblock \emph{Encyclopedia of Statistical Sciences}, 1972.

\bibitem[Noufaily and Jones(2013)]{noufaily2013parametric}
Angela Noufaily and MC~Jones.
\newblock Parametric quantile regression based on the generalized gamma
  distribution.
\newblock \emph{Journal of the Royal Statistical Society: Series C (Applied
  Statistics)}, 62\penalty0 (5):\penalty0 723--740, 2013.

\bibitem[Obizhaeva and Wang(2012)]{obizhaeva2012optimal}
A.A. Obizhaeva and J.~Wang.
\newblock Optimal trading strategy and supply/demand dynamics.
\newblock \emph{Journal of Financial Markets}, 2012.

\bibitem[Panayi and Peters(2015)]{panayi2015liquidity}
Efstathios Panayi and Gareth~W Peters.
\newblock Liquidity commonality does not imply liquidity resilience
  commonality: A functional characterisation for ultra-high frequency
  cross-sectional lob data.
\newblock \emph{to appear, Quantitative Finance Special Issue on Big Data
  Analytics}, 2015.

\bibitem[Panayi et~al.(2014)Panayi, Peters, Danielsson, and
  Zigrand]{panayi2014market}
Efstathios Panayi, Gareth~W Peters, Jon Danielsson, and Jean-Pierre Zigrand.
\newblock Market liquidity resilience.
\newblock \emph{London School of Economics Working Paper Series}, 2014.

\bibitem[Rigby et~al.(2013)Rigby, Stasinopoulos, and
  Voudouris]{rigby2013discussion}
RA~Rigby, DM~Stasinopoulos, and V~Voudouris.
\newblock Discussion: A comparison of gamlss with quantile regression.
\newblock \emph{Statistical Modelling}, 13\penalty0 (4):\penalty0 335--348,
  2013.

\bibitem[Rigby and Stasinopoulos(2005)]{rigby2005generalized}
Robert~A Rigby and D~Mikis Stasinopoulos.
\newblock Generalized additive models for location, scale and shape.
\newblock \emph{Journal of the Royal Statistical Society: Series C (Applied
  Statistics)}, 54\penalty0 (3):\penalty0 507--554, 2005.

\bibitem[SEC(2010)]{securities2010concept}
SEC.
\newblock Concept release on equity market structure.
\newblock \emph{Federal Register}, 75\penalty0 (13):\penalty0 3594--3614, 2010.

\bibitem[Stacy(1962)]{stacy1962generalization}
Eo~W Stacy.
\newblock A generalization of the gamma distribution.
\newblock \emph{The Annals of Mathematical Statistics}, pages 1187--1192, 1962.

\bibitem[Stasinopoulos and Rigby(2007)]{stasinopoulos2007generalized}
D~Mikis Stasinopoulos and Robert~A Rigby.
\newblock Generalized additive models for location scale and shape (gamlss) in
  r.
\newblock \emph{Journal of Statistical Software}, 23\penalty0 (7):\penalty0
  1--46, 2007.

\bibitem[Venkataraman and Waisburd(2007)]{venkataraman2007value}
Kumar Venkataraman and Andrew~C Waisburd.
\newblock The value of the designated market maker.
\newblock \emph{Journal of Financial and Quantitative Analysis}, 42\penalty0
  (03):\penalty0 735--758, 2007.

\bibitem[Wood et~al.(1985)Wood, McInish, and Ord]{wood1985investigation}
Robert~A Wood, Thomas~H McInish, and J~Keith Ord.
\newblock An investigation of transactions data for {NYSE} stocks.
\newblock \emph{The Journal of Finance}, 40\penalty0 (3):\penalty0 723--739,
  1985.

\bibitem[Yu and Moyeed(2001)]{yu2001bayesian}
Keming Yu and Rana~A Moyeed.
\newblock Bayesian quantile regression.
\newblock \emph{Statistics \& Probability Letters}, 54\penalty0 (4):\penalty0
  437--447, 2001.

\end{thebibliography}

\end{document}